\documentclass[twocolumn,aps,amssymb,noshowpacs,superscriptaddress,nofootinbib,longbibliography]{revtex4-1}
\usepackage{amsmath,amssymb}
\usepackage{graphicx}
\usepackage{verbatim}
\usepackage{amsfonts}
\usepackage{dcolumn}
\usepackage{bm}
\usepackage[section]{placeins}
\usepackage{color}
\usepackage{mathrsfs}
\usepackage[colorlinks,bookmarks=true,citecolor=blue,linkcolor=red,urlcolor=blue]{hyperref}
\usepackage{afterpage}
\usepackage{ulem}
\usepackage{url} 
\usepackage[T1]{fontenc}
\usepackage{lmodern}
\usepackage[utf8]{inputenc} 
\usepackage{chemformula}
\usepackage{comment}
\usepackage{cleveref}
\usepackage{siunitx}
\crefname{equation}{Eq.}{Eqs.}
\crefname{figure}{Fig.}{Figs.}

\usepackage{longtable}
\usepackage{multirow}

\newcommand{\beq}{\begin{eqnarray} }
\newcommand{\eeq}{\end{eqnarray} }
\newcommand{\Beq}{\begin{eqnarray*} }
\newcommand{\Eeq}{\end{eqnarray*} }

\newcommand{\RNum}[1]{\uppercase\expandafter{\romannumeral #1\relax}}

\newcommand{\beginsupplement}{%
        \setcounter{table}{0}
        \renewcommand{\thetable}{S\arabic{table}}%
        \setcounter{figure}{0}
        \renewcommand{\thefigure}{S\arabic{figure}}%
        \setcounter{section}{0}
        \renewcommand{\thesection}{S\arabic{section}}%
        \setcounter{equation}{0}
        \renewcommand{\theequation}{S\arabic{equation}}%
     }

\begin{document}
\draft

\title{Enumeration of spin-space groups: Towards a complete description of symmetries of magnetic orders}
\author{Yi Jiang}
\thanks{These authors contributed equally to this study.}
\affiliation{Beijing National Laboratory for Condensed Matter Physics, and Institute of Physics, Chinese Academy of Sciences, Beijing 100190, China}
\affiliation{University of Chinese Academy of Sciences, Beijing 100049, China}
\affiliation{Donostia International Physics Center (DIPC), Paseo Manuel de Lardizábal. 20018, San Sebastián, Spain}
\author{Ziyin Song}
\thanks{These authors contributed equally to this study.}
\affiliation{Beijing National Laboratory for Condensed Matter Physics, and Institute of Physics, Chinese Academy of Sciences, Beijing 100190, China}
\affiliation{University of Chinese Academy of Sciences, Beijing 100049, China}
\author{Tiannian Zhu}
\affiliation{Beijing National Laboratory for Condensed Matter Physics, and Institute of Physics, Chinese Academy of Sciences, Beijing 100190, China}
\affiliation{University of Chinese Academy of Sciences, Beijing 100049, China}
\author{Zhong Fang}
\affiliation{Beijing National Laboratory for Condensed Matter Physics, and Institute of Physics, Chinese Academy of Sciences, Beijing 100190, China}
\author{Hongming Weng}
\affiliation{Beijing National Laboratory for Condensed Matter Physics, and Institute of Physics, Chinese Academy of Sciences, Beijing 100190, China}
\affiliation{Songshan Lake Materials Laboratory, Dongguan, Guangdong 523808, China}
\author{Zheng-Xin Liu}
\email{liuzxphys@ruc.edu.cn}
\affiliation{Department of physics, Renmin University, Beijing 100876, China}
\author{Jian Yang}
\email{yjbuptphy@gmail.com}
\affiliation{Beijing National Laboratory for Condensed Matter Physics, and Institute of Physics, Chinese Academy of Sciences, Beijing 100190, China}
\author{Chen Fang}
\email{cfang@iphy.ac.cn}
\affiliation{Beijing National Laboratory for Condensed Matter Physics, and Institute of Physics, Chinese Academy of Sciences, Beijing 100190, China}
\affiliation{Songshan Lake Materials Laboratory, Dongguan, Guangdong 523808, China}
\affiliation{Kavli Institute for Theoretical Sciences, Chinese Academy of Sciences, Beijing 100190, China}

\date{\today}
\begin{abstract}

Symmetries of three-dimensional periodic scalar fields are described by 230 space groups (SGs). Symmetries of three-dimensional periodic (pseudo-) vector fields, however, are described by the spin-space groups (SSGs), which were initially used to describe the symmetries of magnetic orders. In SSGs, the real-space and spin degrees of freedom are unlocked in the sense that an operation could have different spacial and spin rotations.
SSGs give a complete symmetry description of magnetic structures, and have natural applications in the band theory of itinerary electrons in magnetically ordered systems with weak spin-orbit coupling.
\textit{Altermagnetism}, a concept raised recently that belongs to the symmetry-compensated collinear magnetic orders but has non-relativistic spin splitting, is well described by SSGs.
Due to the vast number and complicated group structures, SSGs have not yet been systematically enumerated. In this work, we exhaust SSGs based on the invariant subgroups of SGs, with spin operations constructed from three-dimensional (3D) real representations of the quotient groups for the invariant subgroups. For collinear and coplanar magnetic orders,
the spin operations can be reduced into lower dimensional real representations.
As the number of SSGs is infinite, we only consider SSGs that describe magnetic unit cells up to 12 times crystal unit cells. We obtain 157,289 non-coplanar, 24,788 coplanar-non-collinear, and 1,421 collinear SSGs. The enumerated SSGs are stored in an online database at \url{https://cmpdc.iphy.ac.cn/ssg} with a user-friendly interface. We develop an algorithm to identify SSGs for realistic materials and find SSGs for 1,626 magnetic materials.
We also discuss several potential applications of SSGs, including the representation theory, topological states protected by SSGs, structures of spin textures, and refinement of magnetic neutron diffraction patterns using SSGs.
Our results serve as a solid starting point for further studies of symmetry and topology in magnetically ordered materials.
\end{abstract}

\maketitle

\tableofcontents

\section{Introduction}
Crystallography is a long-lived and vibrant field that studies the symmetries of crystalline materials. With the help of group theory, the symmetries of three-dimensional (3D) crystals are classified into 230 space groups (SGs)\cite{hahn1983international, Bradley2010} when combining the translational symmetries with 32 3D crystallographic point groups (PGs). Theoretically, SGs describe the symmetries of any 3D periodic scalar fields, such as the crystal potential field $V(\mathbf{r})$.

Magnetism is another century-old realm, where magnetic materials are classified according to their various magnetic orders including ferromagnetic (FM), ferrimagnetic, anti-ferromagnetic (AFM), spiral magnetic, and even more complicated orders.
In magnetic materials, the magnetic moments arrange into periodic 3D pseudo-vector fields in the spin space on top of the crystals formed by atoms in real space. The pseudo-vector field is of even parity under space inversion $\mathcal{P}$ and is odd under the time-reversal symmetry (TRS) $\mathcal{T}$.

Historically, by combining TRS with 230 SGs, 1651 (double) Shubnikov magnetic space groups (MSGs)\cite{Bradley1968, Lifshitz2004, Bradley2010, Litvin2013MagneticGT, gallego2016magndataI, gonzalez2021extension,LiuYaomsgcorep2023} were introduced in order to describe the symmetries of magnetic materials. Shubnikov MSGs are classified into four types. The 230 Type-II Shubnikov SGs describe non-magnetic materials while the rest 1421 of type-I, III, and IV describe magnetic ones. In Shubnikov MSGs, the rotations of the magnetic moments are locked with the lattice operations. For instance, a $C_{4z}$ symmetry operation stands for a $\frac{\pi}{2}$ rotation of the lattice along the $z$-axis together with the rotation of the magnetic moments for the same angle $\frac{\pi}{2}$ along the $z$-direction. In later discussion, we will call the actions on the magnetic order as ``spin operations''.

\begin{figure}[htbp]
	\centering
	\includegraphics[width=0.5\textwidth]{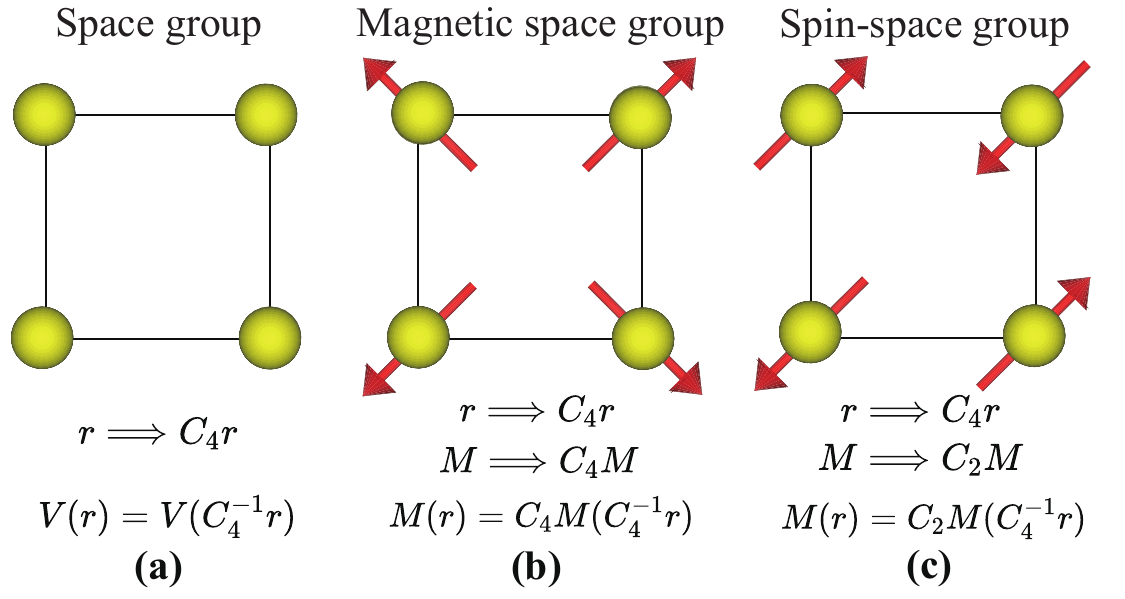}
	\caption{\label{Fig:fig1} Illustration of space groups (SGs), magnetic space groups (MSGs), and spin-space groups (SSGs).
	(a) A $C_4$-symmetric atomic configuration, generated by the SG operation $C_4$.
	(b) A magnetic order generated by the MSG operation $\{C_4||C_4\}$, where the real-space $C_4$ is accompanied by the same $C_4$ spin rotation. The notation $\{R_s||R_l\}$ denotes an operation with space rotation $R_l$ and spin rotation $R_s$.
	(c) A magnetic order generated by the SSG operation $\{C_2||C_4\}$, where the $C_4$ lattice rotation is accompanied by a $C_2$ spin rotation. MSGs, in which the spin and lattice rotation are locked, fail to give a complete symmetry description of this magnetic order.
 }
\end{figure}

MSGs, although widely used, fail to give a complete description of the symmetries of the magnetic moment fields. This is because there also exist symmetry operations that have unlocked real-space and spin rotations. Such enlarged groups were introduced as spin-space groups (SSGs)\cite{Brinkman1966, Litvin1974}, which are the natural generalizations of Shubnikov MSGs and contain Shubnikov MSGs as a subset. For instance, a real-space $C_{4z}$ rotation may accompany a $C_{2z}$ spin rotation.
In Fig.\ref{Fig:fig1}, we illustrate the difference between SGs, MSGs, and SSGs. In Fig.\ref{Fig:fig1}(a), we show an atomic configuration generated by a $C_4$ rotation, leading to a $C_4$-symmetric crystal field $V(\mathbf{r})$. In Fig.\ref{Fig:fig1}(b), a $C_4$-symmetric magnetic order is added to the atoms. We use the notation $\{R_s||R_l\}$ to denote an operation with space rotation $R_l$ (the space translation part omitted for simplicity) and spin rotation $R_s$. The MSG symmetry $\{C_4||C_4\}$ is used to describe the symmetry of the magnetic moment field $\mathbf{M}(\mathbf{r})$ in Fig.\ref{Fig:fig1}(b). In Fig.\ref{Fig:fig1}(c), a different magnetic order is shown, where the $C_4$-related atoms have $C_2$-rotated spin orientations, characterized by an SSG symmetry $\{C_2||C_4\}$, lying out of the scope of MSGs. This example demonstrates the incompleteness of MSGs in describing the symmetry of magnetic orders, and the necessity for introducing SSGs.

SGs, Shubnikov MSGs, and SSGs also have important applications in describing the symmetries of electronic structures.
Non-magnetic periodic electronic structures characterized by a single-particle Hamiltonian
$\hat{H}=\frac{\hat{\bm{p}}^2}{2m}+V({\mathbf{r}})$ have their crystal potential field $V(\mathbf{r})$ respecting certain spacial symmetries, which form a (single) SG.
The spin degree of freedom can be introduced in the Hamiltonain when the spin-orbit coupling (SOC) is present, i.e., a third term $\frac{1}{2m^{2}c^{2}} \nabla V({\mathbf{r}}) \times\hat{\bm{p}} \cdot\hat{\bm{s}}$ is added to the Hamiltonian, where $\hat{\bm{s}}$ is the electron-spin operator.
The symmetries of the (non-magnetic) spinful Hamiltonians are described by double SGs together with the TRS, i.e., type-II (double) MSGs. The SOC term locks the spacial symmetries with the corresponding spin rotations in double MSGs. Only the lattice rotations associated with corresponding spin rotations are symmetries of the Hamiltonian.
When the system is magnetically ordered, an effective ``Zeeman term'' $\mathbf{M}({\mathbf{r}})\cdot\hat{\bm{s}}$ is generated.
The effective Zeeman field may come from the mean-field decoupling of the electron interaction in the spin channel, where $\mathbf{M}(\mathbf{r})$ is the static mean-field representing the magnetization, like in the spin-density-wave state. Alternatively, this Hamiltonian also describes the motion of an electron in the background of ordered, static magnetic moments, like in the Kondo lattice.
The classifications of gapped topological states and unconventional quasiparticles protected by SGs\cite{rslager12, CLKane2012, BJYNagaosa14,cf15, watan16, Bradlyn2016, rslager17, bernevig17, Armitage2018, cf18, zhang2019catalogue, bernevig19, xgw19, cano21, song2018quantitative, song2020real} and MSGs\cite{Yang2020, gangxu16, Liang2016, Watanabe2018, Hua2018, mfvjur19, Cano2019, bernevig20,rslager21, Yang2021, wanxg2021, YiJiang2021, bernevigmtqc21, liu2021encyclopedia, zhang2021encyclopedia, lenggenhager2022triple, tang2022complete, peng2022topological, bernevig2022progress,
wanxgtangf2022prm} have been widely studied.

However, for itinerant electrons coming from light atoms or low angular momentum orbitals in magnetic materials, the SOC term is usually negligible or much smaller compared with the effective Zeeman term. Such systems are characterized by the single-particle Hamiltonian\cite{YangLiuFang2021,Liu2022prx}
\begin{equation}
    \hat{H}=\frac{\hat{\bm{p}}^2}{2m}+V({\mathbf{r}})+ \mathbf{M}({\mathbf{r}})\cdot\hat{\bm{s}}
\end{equation}
In these systems, the spin rotations are not necessarily always locked with lattice rotations due to the absence of the SOC, and the system may contain pure lattice rotation symmetries, pure spin rotation symmetries, and general symmetries with different lattice and spin rotations. Once the spin and the lattice operations are (partially) unlocked, the symmetries of the Hamiltonian form SSGs. Hence, spin-space groups not only describe the symmetries of magnetic structures, but also apply to electrons in magnetically ordered material having weak spin-orbit coupling, and magnon Hamiltonian of spin systems\cite{Corticelli2021} with weak Dzyaloshinskii-Moriya interactions.

It is worth mentioning that, SSG can always be used to describe the symmetries of the magnetic structure of a material, regardless of whether SOC is strong or not in the material. When SOC is non-negligible, the electronic structure of the material cannot be described by the SSG, and only MSG symmetries can be used. Nonetheless, SSG symmetries could serve as approximate symmetries for systems where SOC is weak or much smaller compared with the effective Zeeman term.

In 1977, Litvin tabulated 598 non-coplanar spin-point groups (SPGs)\cite{Litvin1977}. Very recently, 252 SPGs for coplanar and 90 SPGs for collinear magnetic orders have been listed\cite{Liu2022prx}.
A new concept, \textit{altermagnetism}, has also been raised recently by the authors in Ref.\cite{smejkal2021altermagnetism, TJungwirth2022, mazin2022altermagnetism}, which describes a special type of collinear symmetry-compensated magnetic order that has non-relativistic spin splitting
or
spin splitting with negligible SOC
\cite{hayami2019momentum,hayami2020prb}
in the Brillouin zone (BZ), breaking the Kramers degeneracy. The authors distinguished altermagnetic and AFM orders from symmetries using spin Laue point groups which are unlocked with the lattice point-group operations\cite{smejkal2021altermagnetism}. Therefore, altermagnetic materials\cite{reichlova2020macroscopic,feng2022anomalous,bose2022tilted,baihruo2exp2022,kshutaroruo2exp2022, ruo2aheconbs6,ruo2spinsplit2021,GBetancourtmnte2023, mazin2023altermagnetism,hariki2023xray,papaj2023andreev,ghorashi2023altermagnetic,steward2023dynamic} are naturally applicable systems of SSGs.
An increasing number of experimental and theoretical studies on magnetic materials with weak SOC, among which many adopted the concept of SSGs, have been performed, including $\mathrm{Mn}_{5}\mathrm{Si}_{3}$\cite{reichlova2020macroscopic}, $\mathrm{RuO}_{2}$\cite{feng2022anomalous,bose2022tilted,baihruo2exp2022,kshutaroruo2exp2022, ruo2aheconbs6,ruo2spinsplit2021}, $\mathrm{MnTe}$\cite{GBetancourtmnte2023, mazin2023altermagnetism},
\ch{MnTe2}\cite{liuchangmnte22023},
$\mathrm{CoNb}_3\mathrm{S}_6$\cite{liu2022chiral, Liuqhconb3s62023}, and the so-called ``low-Z'' antiferromagnetic compounds\cite{yldlowz2021}, together with many new quasi-particle types\cite{YangLiuFang2021,pjguo2021prl,pjguo2022, Liu2022prx,liu2022chiral,Liuqhconb3s62023} being theoretically predicted, which can only be realized in SSGs.

Despite the wide applications of SSGs, the systematic enumeration of SSGs is mathematically incomplete as of today. Compared with spin-point groups, the enumeration of SSGs has the following difficulties: (i) The number of symmetry operations in an SSG is infinite due to the translation group. (ii) The number of SSGs is infinite as the size of the magnetic unit cell could be arbitrary times of the original crystal unit cell, and even incommensurate magnetic orders exist. (iii) The spin rotations in SSGs could be non-crystallographic, such as a $C_n$ rotation with arbitrary integer $n$. These difficulties hinder the enumeration of SSGs in literature.

In this work, we enumerate SSGs systematically in order to give a complete symmetry description of all magnetic orders.
We first exhaust invariant subgroups of 230 SGs and then compute the corresponding quotient groups that are isomorphic to point groups. The SSGs are constructed by assigning the 3D real representations of point groups as spin rotations to quotient group operations. 2D and 1D real representations are also enumerated to construct SSGs for coplanar and collinear magnetic orders. As the number of SSGs is infinite, we restrict our enumeration of SSGs with magnetic unit cells up to 12 times the crystal unit cells. We find 157,289 non-coplanar SSGs, 24,788 coplanar-non-collinear SSGs, and 1,421 collinear SSGs.
Specially, for quotient groups isomorphic to crystallographic point groups, the size of the magnetic unit cell can only be within 12 times of the crystal unit cell and are thus exhausted in this work, which gives rise to 68,922 (non-coplanar) SSGs.
The enumerated SSGs are stored in an online database\cite{ssg_website} with a user-friendly interface.
We also develop an algorithm to identify SSGs for magnetic materials and apply the algorithm to more than 2,000 magnetic materials in \textit{Bilbao Crystallographic server}\cite{aroyo2006bilbao1,aroyo2006bilbao2,aroyo2011crystallography,gallego2016magndataI,gallego2016magndataII}, and find the corresponding SSGs for 1,626 commensurate magnetic materials without partial occupation.

The paper is organized as follows. In Sec. \ref{framework}, we give the formal definition of SSGs and the general framework for constructing SSGs. In Sec. \ref{algorithm}, we present the detailed algorithm for each step in enumerating SSGs. In Sec. \ref{Sec:results}, we summarize the enumeration results.
In Sec. \ref{example}, we use both pedagogical and realistic material examples to discuss the SSGs of magnetic structures.
In Sec. \ref{Sec:applications}, we discuss potential applications of SSGs. Finally, the paper is concluded in Sec. \ref{conclusion}.

\section{General framework}\label{framework}

\subsection{Definition of spin-space groups}
We start with 230 space groups (SGs). An SG $\mathcal{G}$ has a 3D translation group
\begin{equation}
   \bm{T}=\{\mathbf{R}_n=n_1\bm{a}_1+n_2\bm{a}_2+n_3\bm{a}_3, n_i\in\mathbb{Z}\}
\end{equation}
as an invariant subgroup, where $\bm{a}_i ~(i=1,2,3)$ are three primitive cell bases. $\mathcal{G}$ is represented as the coset decomposition
\begin{equation}
    \mathcal{G}=\bigcup_{i=1}^{n} \{R_i|\bm{\tau}_i\} \bm{T},
\end{equation}
where $R_i\in \text{O}(3)$ is a point group operation and $\bm{\tau}_i\in \mathbb{R}^3$ is a translation, which is zero in symmorphic SGs and fractional under the bases of $\bm{T}$ in non-symmorphic SGs. Denoting the collection of $R_i$ as $P=\{R_i\}$, which is the point group of $\mathcal{G}$, then we have the quotient group $\mathcal{G}/\bm{T}\cong P$. For symmorphic SGs, $P$ is a subgroup of $\mathcal{G}$, and $\mathcal{G}=\bm{T}\rtimes P$ is the semi-direct product of the translation group and the PG. For non-symmorphic SGs, $P$ is no longer a subgroup of $\mathcal{G}$, and $\mathcal{G}$ is generally a group extension of $\bm{T}$ by $P$.

SGs can be used to describe the symmetry of a scalar field $V(\mathbf{r})$ if it is invariant under all lattice operations, where an SG operation acts on $V(\mathbf{r})$ in forms of
\begin{equation}
    \begin{aligned}
        \{R_i|\bm{\tau}_i\} V(\mathbf{r})&= V(\{R_i|\bm{\tau}_i\}^{-1} \mathbf{r}).
    \end{aligned}
\end{equation}
However, to describe the complete symmetries of a periodic 3D pseudo-vector field (the magnetic moment field) $\mathbf{M}(\mathbf{r})$, one needs to define the operations that act on the pseudo-vector $\mathbf{M}$.
A natural generalization is including the spin operations $U_i$, which yields the type-I MSG
\begin{equation}
    \mathcal{G}^{\rm(I)}=\bigcup_{i=1}^{n} \{U_i|| R_i|\bm{\tau}_i\} \bm{T},
\end{equation}
where $U_i=\det(R_i)R_i\in \text{SO}(3)$ is the corresponding proper rotation matrix of $R_i$.
The group element $\{U || R|\bm{\tau}\}$ acts on the magnetic moment field $\mathbf{M}(\mathbf{r})$ as
\begin{equation}
    \{U|| R|\bm{\tau}\} \mathbf{M}(\mathbf{r})=
    \det(R) R \mathbf{M}(\{R|\bm{\tau}\}^{-1} \mathbf{r}).
\end{equation}

By further including anti-unitary time-reversal symmetry (TRS) $\mathcal{T}$,
one can obtain type-II, III and type-IV MSGs which have the group structure
$\mathcal G^{\rm (II,III,IV)}=\mathcal{G}+m\cdot \mathcal{G}$, where $m=\mathcal{T}, g\cdot\mathcal{T}$ and $\bm{\tau}\cdot\mathcal{T}$ in type-II, III and IV Shubnikov MSGs, respectively, with $g$ being an SG symmetry with non-trivial PG part and $\bm{\tau}$ a fractional translation. As $\mathcal{T}$ reverses $\mathbf{M}$, an anti-unitary Shubnikov MSG element $\{U|| R|\bm{\tau}\}\mathcal{T}$ acts on $\mathbf{M}(\mathbf{r})$ as
\begin{equation}
    \{U || R|\bm{\tau}\}\mathcal{T} \mathbf{M}(\mathbf{r})=
    -\det(R) R \mathbf{M}(\{R|\bm{\tau}\}^{-1} \mathbf{r}).
\end{equation}
Here and later we represent the action of time reversal $\mathcal T$ on the magnetic moment $\bf M$ as $-1$.

As last, when unlocking $U_i$ from $\det(R_i) R_i$, we obtain the spin-space groups, which are noted as
\begin{equation}
     \mathcal G^{(S)}=\bigcup_{i=1}^{n} \{U_i || R_i|\bm{\tau}_i\} \bm{T},
     \label{SSG_definition}
\end{equation}
where $U_i\in 
\text{O}(3)$.
When $\det(U_i)=-1$, it is assumed that it contains TRS $\mathcal T$ and is anti-unitary, i.e., $U_i\sim \det(U_i)U_i\cdot \mathcal{T}$. Under this assumption, we do not distinguish $U_i\in \text{SO}(3)\times Z_2^T$ and $U_i\in \text{O}(3)$, and the full operation of $g$ reads
\begin{equation}
g=\left\{
\begin{aligned}
    &\{U || R_i|\bm{\tau}_i\}, ~\text{if }\det(U)=+1\\
    &\{-U || R_i|\bm{\tau}_i\}\mathcal{T}, ~\text{if }\det(U)=-1\\
\end{aligned}
\right.
\end{equation}
and an SSG operation $\{U || R|\bm{\tau}\}$ acts on $\mathbf{M}(\mathbf{r})$ as
\begin{equation}
    \{U || R|\bm{\tau}\} \mathbf{M}(\mathbf{r})=
    U \mathbf{M}(\{R|\bm{\tau}\}^{-1} \mathbf{r}).
\end{equation}

To understand the group structure of SSG $\mathcal G^{(S)}$ defined in Eq. (\ref{SSG_definition}), it is convenient to introduce four key groups associated with $\mathcal G^{(S)}$\cite{Litvin1974}: (i) the group $\mathcal{G}$ formed by the lattice parts $\{R|\bm{\tau}\}$; (ii) the group $\mathcal{S}$ formed by the spin parts $\{U\}$; (iii) the group $H$ formed by pure-lattice symmetry operations $\{E|| R|\bm{\tau}\}\in \mathcal G^{(S)}$; (iv) the group $\mathcal{S}_0$ formed by pure-spin symmetry operations $\{U || E|\bm{0}\} \in \mathcal G^{(S)}$, which is called the `spin-only group' (If $U$ is an improper rotation with $\det U=-1$, then rigorously $\{U || E|\bm{0}\}$ should be written as $\{U || \mathcal T|\bm{0}\}$ since it acts non-trivially on the lattice wave vector. In this case, we still call $\{U || \mathcal T|\bm{0}\}$ a `spin-only' operation). From these definitions, it follows immediately that: (a) both $\mathcal{G}$ and $H$ are SGs; (b) $H$ is an invariant subgroup of $\mathcal{G}$, i.e., $H\triangleleft\mathcal{G}$; (c) $\mathcal{S}_0$ is an invariant subgroup of ${\mathcal{S}}$, i.e., $\mathcal{S}_0\triangleleft\mathcal{S}$. According to the isomorphism theorem\cite{Litvin1974}(or the Goursat's lemma\cite{Goursat}), the following quotient groups are isomorphic:
\begin{equation}
    Q=\mathcal{G}/H \cong  {\mathcal{S}}/\mathcal{S}_0.
\end{equation}

For non-magnetic systems, the spin-only group $\mathcal{S}_0=\text{O}(3)$, while for non-trivial magnetic orders, the pure-spin symmetries in $\mathcal{S}_0$ can only appear in two special types of magnetic orders\cite{Litvin1974}:
\begin{itemize}
    \item Collinear magnetic orders, where $\mathbf{M}(\mathbf{r})=(0, 0, \text{M}_z(\mathbf{r}))$ are set along $z$-direction without loss of generality. In this case, the spin-only group
    \begin{equation}
        \mathcal{S}_0=\{C_{\theta}|| E|\bm{0}\} + \{M_x C_{\theta}|| E|\bm{0}\} \cong \text{O}(2),
        \label{Eq:collinear_S0}
    \end{equation}
    where $C_{\theta}$ denotes the rotation with an arbitrary angle $\theta$ along the $z$-axis, and $M_x$ the mirror normal to the $x$-axis. $M_x C_{\theta}$ generates all mirrors with mirror planes passing the $z$-axis. 
    Note that $M_x$, being improper, actually denotes the anti-unitary operation  $C_{2x}\mathcal{T}$. This spin-only group can also be written as $\mathcal{S}_0=\text{SO}(2)\rtimes \mathbb{Z}_2^{M_x} \cong \text{O}(2)$, where $\text{SO}(2)$ represents the continuous group from the $C_{\theta}$ rotation, and $\mathbb{Z}_2^{M_x}$ is the $\mathbb{Z}_2$ group formed by the pure spin-rotation $M_x$.
    \item Coplanar magnetic orders, where $\mathbf{M}(\mathbf{r})=(\text{M}_x(\mathbf{r}), \text{M}_y(\mathbf{r}), 0)$ are set to lie on the $z=0$ plane without loss of generality. In this case,
    \begin{equation}
        \mathcal{S}_0=\mathbb{Z}_2^{M_z}=\{E, \{M_z || E|\bm{0}\}\}
        \label{Eq:coplanar_S0}
    \end{equation}
    where $M_z$ denotes the mirror along $z$-axis.
\end{itemize}
When an SSG $\mathcal G^{(S)}$ has a non-trivial spin-only group $\mathcal{S}_0$, $\mathcal G^{(S)}$ can always be decomposed into a direct product group by properly choosing group element, i.e.,
\begin{equation}
    \mathcal G^{(S)}=\mathcal G^{(S)\prime} \times \mathcal{S}_0,
    \label{eq_direct_product_SSG}
\end{equation}
where $\mathcal G^{(S)\prime}=\mathcal G^{(S)}/\mathcal{S}_0$ is `spin-only-free' and is still an SSG.
In Supplementary Material \cref{sec:app:proof_direct_prod_group}, we give rigorous proof for the direct product group structure.
For collinear and coplanar magnetic orders, the spin-operations in $\mathcal G^{(S)\prime}$ can only form uni-axial point groups, but not polyhedral point groups including $T, T_h, T_d, O$, and $O_h$\cite{Liu2022prx}, as the spin-only group $\mathcal{S}_0$ must be an invariant subgroup of $\mathcal G^{(S)}$.

We remark that one can also define 1651 single Shubnikov MSGs, where each operation only has a real-space part (combined with the TRS) but no spin part. Single Shubnikov MSGs are isomorphic to certain SSGs, as trivial spin rotations can be assigned to each spacial operation.

\begin{figure*}[htbp]
    \centering
    \includegraphics[width=1\textwidth]{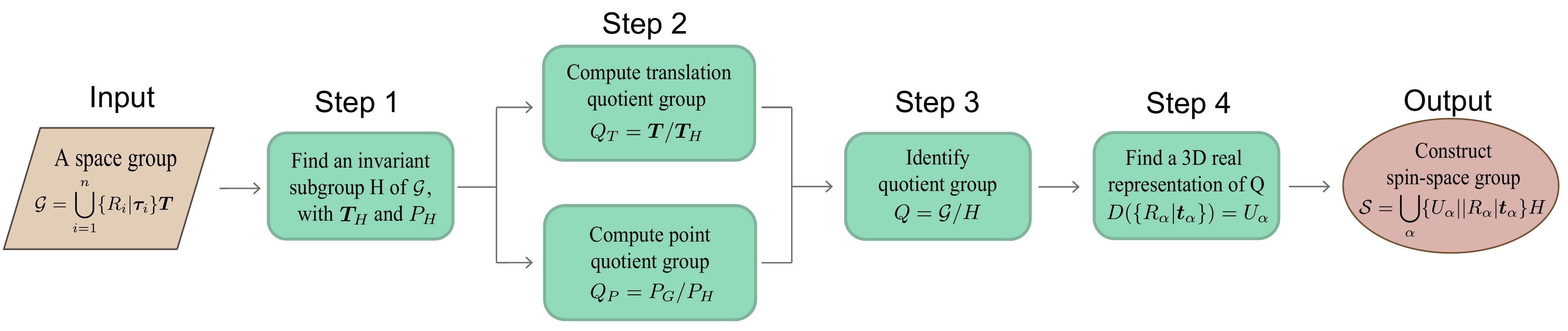}
    \caption{\label{flowchart} Flowchart of the algorithm for constructing spin-space groups. We use $\{R_\alpha|\bm{t}_\alpha\}$ to represent operations in the quotient group $Q$, where the translation part $\bm{t}_\alpha=\bm{\tau}_\alpha+\bm{R}_n$ is the original translation part $\bm{\tau}_\alpha$ associated with $R_\alpha$ in $\mathcal{G}$ plus a possible lattice translation. For collinear (coplanar) SSGs, 1D (2D) real representations are used.}
\end{figure*}

\subsection{Construction of spin-space groups}

In this work, we first construct SSGs with trivial spin-only groups $\mathcal{S}_0=\{E\}$ which describe the symmetries of general magnetic orders.
For collinear and coplanar magnetic orders, the spin-only group $\mathcal{S}_0$ has group structures shown in Eq. (\ref{Eq:collinear_S0}, \ref{Eq:coplanar_S0}) with the whole SSG being a direct product group of Eq. (\ref{eq_direct_product_SSG}), and we construct the `spin-only-free' groups $\mathcal G^{(S)'}$ for them.

To construct SSGs, first notice that the pure-lattice symmetries in an SSG $\mathcal G^{(S)}$, i.e.,
\begin{equation}
    H=\bigcup_i \{E|| R_i|\bm{\tau}_i\}\bm{T},
\end{equation}
form an invariant subgroup of $\mathcal G^{(S)}$. The quotient group $Q=\mathcal{G}/H$ is a finite group and must be isomorphic to a subgroup of $\text{O}(3)$, i.e., a point group. By assuming the spin-only group $\mathcal{S}_0$ is trivial, $Q$ is isomorphic to the group formed by spin rotations:
\begin{equation}
    Q=\mathcal G^{(S)}/H\cong \{ U_i\}.
\end{equation}
This relation gives insight into how to construct SSGs, as shown in the following.

For a given SG $\mathcal{G}$ with translation group $\bm{T}$ and PG $P$, assume $H\triangleleft \mathcal{G}$ is an invariant subgroup of $\mathcal{G}$, with $\bm{T}_H\triangleleft \bm{T}$ and $P_H\triangleleft P$. 
Denote the translational quotient group as $Q_T=\bm{T}/\bm{T}_H$, and point quotient group $Q_P=P/P_H$. The elements of $Q=\mathcal{G}/H$ are generated from the set products of $Q_T$ and $Q_P$, where the rotations from $Q_P$ have to recover their possible fractional translation parts $\bm{\tau}$ in non-symmorphic $\mathcal{G}$. As $Q_T$ is an invariant subgroup of $Q$, the following short exact sequence holds:
\begin{equation}
	\{E\}\to Q_T \stackrel{i}{\rightarrow}  Q \stackrel{\pi}{\rightarrow}  Q_P \to\{E\} ,
\end{equation}
where $i$ maps $Q_T$ to $Q$, and $\pi$ is surjection of $Q$ onto $Q_P$, which induces an isomorphism $$Q_P\cong Q/Q_T,$$ namely $Q$ is a group extension of $Q_T$ by $Q_P$.
If $Q_P$ is also a subgroup of $Q$ (which is not always the case), the group extension becomes a semi-direct product $Q=Q_T\rtimes Q_P$.

Choose the representatives of the cosets in $\mathcal G$ with respect to its invariant subgroup $H$ as $q_\alpha=\{R_\alpha|\bm{t}_\alpha\}$, such that $\mathcal{G}=\bigcup_\alpha q_\alpha H$, where $R_\alpha\in Q$ is a PG operation and $\bm{t}_\alpha=\bm{\tau}_\alpha+\bm{R}_n$ is the original translation part $\bm{\tau}_\alpha$ associated with $R_\alpha$ plus a possible lattice translation.
Assume $Q$ has a 3D real representation $D$, with representation matrix $D(q_\alpha)=U_\alpha\in \text{O}(3)$. Then an SSG $\mathcal G^{(S)}$ is constructed by
\begin{equation}
    \mathcal G^{(S)}=\bigcup_\alpha \{U_\alpha || R_\alpha|\bm{t}_\alpha \} H,
    \label{eq_SSG_construction}
\end{equation}
where $U_\alpha$ denotes the rotation in spin space, and operations in $H$ are assigned with trivial spin rotations. The algorithm is schematically illustrated in Fig. \ref{flowchart}.

For magnetic orders with non-trivial spin-only group $\mathcal{S}_0$, the general construction in Eq. (\ref{eq_SSG_construction}) is applicable but the existence of spin-only groups leads to additional equivalence relations between SSGs, and the construction of their corresponding SSGs can be simplified.

In the enumeration of SSGs, the involvement of a non-trivial spin-only group $S_0$ effectively reduces the dimensionality of the spin space to be considered due to the imposed equivalence relations.
For the three types of SSGs that have non-trivial $\mathcal{S}_0$, i.e., the non-magnetic, collinear, and coplanar, this dimensionality reduction allows us to focus solely on 0D, 1D, and 2D representations, respectively. As a result, this refinement in the enumeration process simplifies the procedure by limiting it to lower-dimensional representations, thereby reducing the breadth of groups that necessitate consideration.

For the non-magnetic case, the spin-only group $\mathcal{S}_0=\text{O}(3)$, and a `spin-only-free group' $\mathcal G^{(S)'}=\mathcal G^{(S)}/\mathcal S_0$ contains only pure-lattice operations as spin operations can be absorbed into $\mathcal{S}_0$, and is thus equivalent to one of the SG $\mathcal{G}$, hence $\mathcal G^{(S)}=\mathcal{G}\times\mathcal{S}_0$. Thus there are 230 inequivalent SSGs for non-magnetic orders, i.e., 230 SGs. Mathematically, the trivial real-representation is used to construct all spin operations in Eq. (\ref{eq_SSG_construction}), i.e., $U_\alpha\equiv E$.

For collinear magnetic orders with $\mathbf{M}(\mathbf{r})=(0, 0, \text{M}_z(\mathbf{r}))$, we have $\mathcal{S}_0 \cong \text{O}(2)$.
There are two possibilities for the group $\mathcal{G}^{(S)}$, namely $\mathcal{G}^{(S)}\cong \text{O}(2)$ or $\mathcal{G}^{(S)}\cong \text{O}(2)\times Z_2^T$, then the quotient group $Q=\mathcal{G}^{(S)}/\mathcal S_0$ is $Q\cong C_1=\{E\}$ and $Q\cong C_s=\{E,M_z\}$ (in the latter case $C_s$ has an equivalent replacement $Z_2^T= \{E, \mathcal T\}$), respectively.
A `spin-only-free group' $\mathcal G^{(S)'}=\mathcal G^{(S)}/\mathcal{S}_0$ can be constructed from Eq. (\ref{eq_SSG_construction})
where $U_\alpha$ is a 1D real representations of $Q$, i.e., $D(q_\alpha)=\pm 1$, direct-summed with two trivial representations (s.t $U_{\alpha}\in \text{O}(3)$). The SSGs for collinear magnetic orders can be further classified into three types
\begin{itemize}
    \item When $Q\cong C_1$, then $\mathcal G^{(S)'}=\mathcal{G}$, which gives 230 collinear SSGs. These SSGs describe FM orders with spin splitting.

    \item When $Q\cong \bm{T}/\bm{T}_H\cong C_s$ (or equivalently $Q\cong Z_2^T$), then  $\mathcal G^{(S)'}=H+\{M_z||\mathcal T|0,0,1\}H$ (or equivalently $\mathcal G^{(S)'}= H+ \{\mathcal T||\mathcal T|0,0,1\}H$). In this case,
    one also has $g=\{C_{2y}||E|0,0,1\}\in \mathcal G^{(S)}$ (or $\{\mathcal T||\mathcal T|0,0,1\}\in \mathcal G^{(S)}$), which leads to spin degeneracy over BZ. These SSGs describe conventional AFM with non-relativistic spin degeneracy.

    \item When $Q\cong P/P_H\cong C_s$ (or equivalently $Q\cong Z_2^T$), then  $\mathcal G^{(S)'}=H+\{M_z||R\mathcal T|\tau\}H$ (or equivalently $\mathcal G^{(S)'}= H+ \{\mathcal T||R\mathcal T|\tau \}H$).
    \begin{enumerate}
    \item[(1)]If $R=\mathcal{P}$ is the real-space inversion, then
    there exist a combined spatial inversion and time reversal symmetry $\mathcal{P}\mathcal{T}$ which leads to spin degeneracy (both non-relativistic and relativistic) in the whole BZ. These SSGs also describe conventional AFM orders.

    \item[(2)]If $R$ is not the real-space inversion, then there is no symmetry that can protect spin degeneracy over the whole BZ. These SSGs describe the so-called ``altermagnetism''\cite{smejkal2021altermagnetism, TJungwirth2022, mazin2022altermagnetism} that differ from the conventional collinear AFM orders by the non-relativistic spin splitting together with many other differences in transport properties.
    \end{enumerate}
\end{itemize}

For coplanar magnetic orders with $\mathbf{M}(\mathbf{r})=(\text{M}_x(\mathbf{r}), \text{M}_y(\mathbf{r}), 0)$, $\mathcal{S}_0=\mathbb{Z}_2^{M_z}$. The group $\mathcal{G}^{(S)}$ can be either $C_{2v}$(main axis $x$), or $C_{nh}$, $D_{nh}$(main axis $z$), with the quotient group $Q=\mathcal{G}^{(S)} /\mathcal S_0$ isomorphic to $C_s$ (main axis $x$), $C_n$, $C_{nv}$(main axis $z$) respectively. The operations in the spin-only-free group $\mathcal G^{(S)'}$ are constructed by a direct sum of a 2D real representation in the $xy$ plane and one trivial representation in the $z$ direction.

Using this algorithm, we first compute the invariant subgroups in 230 SGs and then derive the quotient groups and their 3D real representations. SSGs are constructed by assigning the 3D real representation to the quotient group elements as spin operations. In the following section, we give more detailed descriptions of each step.

\section{Detailed algorithm}\label{algorithm}

Before diving into the details of the enumeration of SSGs, it is instructive to consider how to enumerate 230 SGs. Suppose we have 32 crystallographic point groups and 14 Bravais lattices, the latter of which give the bases of translation group $\bm{T}$. Then constructing SGs is a group extension problem, i.e., an extension of the translation group $\bm{T}$ by the PG $P$.

In practice, the problem can be transformed into solving a set of linear equations. For a given PG $P$ and translation group $\bm{T}$, suppose the SG $\mathcal{G}$ constructed from them has elements $g=\{p|\bm{\tau}_{p}\}$, where $p\in P$ and $\bm{\tau}_p$ are the translational part to be solved. The following two equations must be satisfied for $\mathcal{G}$ to form a group:
\begin{equation}
\left\{
\begin{aligned}
& p\bm{T}=\bm{T},\ \forall p\in P, \\
& \bm{\tau}_{p_{1}p_{2}}=\bm{\tau}_{p_1}+p_1\bm{\tau}_{p_{2}} \bmod \bm{T},\ \forall p_{1},p_{2}\in P.
\end{aligned}
\right.
\end{equation}
The first equation means $\bm{T}$ is invariant under all $p\in P$, which is equivalent to requiring all $p\in P$ are integer $\text{O}(3)$ matrices in the primitive bases. The second equation is nothing but the group multiplication rule in $\mathcal{G}$. By iterating over 32 point groups and their compatible Bravais lattices, these equations can be solved to give all possible SGs.

However, the resultant number of SGs is larger than 230, as many of them are equivalent. To obtain inequivalent SGs, one has to define the equivalent relation between SGs. Two SGs $\mathcal{G}_1$ and $\mathcal{G}_2$ are equivalent if there exists a coordinate transformation $W=\{A|\bm{t}\}$, where $A\in \text{SL}(3, \mathbb{Z})$ and $\bm{t}\in\mathbb{R}^3$, s.t.
\begin{equation}
    \mathcal{G}_1=W \mathcal{G}_2 W^{-1}.
\end{equation}
Notice that the length of the primitive cell bases in $\bm{T}$ and the angles between them are inessential in defining equivalent SGs.
In literature\cite{hahn1983international}, there are 230 (crystallographic) space-group types if requiring $A\in \text{SL}(3, \mathbb{Z})$, and 219 affine space-group types if $A\in \text{GL}(3, \mathbb{Z})$.

\subsection{Subgroups of space groups}
Assume an SG $\mathcal{G}$ with PG $P$ and translational group $\bm{T}$ has a subgroup $H$ with sub-PG $P_H$ and sub-translational group $\bm{T}_H$. Define the t-index $I_t=|P/P_H|$ and k-index $I_k=|\bm{T}/\bm{T}_H|$, which denote the number of elements in the point and translation quotient group. Remark that although both $\bm{T}$ and $\bm{T}_H$ are infinite Abelian groups, their quotient group is a finite group when the translation basis in $\bm{T}$ and $\bm{T}_H$ are commensurate, determined by the transformation between their translation basis, i.e., original unit cell basis given by $\bm{T}$ and the supercell basis given by $\bm{T}_H$.
In literature\cite{hahn1983international}, subgroups with $I_t\ge 2$ and $I_k=1$ are called \textit{translationengleiche} subgroup (t-subgroup), and subgroups with $I_t=1$ and $I_k\ge 2$ are called \textit{klassengleiche} subgroup (k-subgroup)\footnote{\textit{Translationengleiche} means ``with the same translations'', and \textit{klassengleiche} means ``of the same (crystal) class''}.

We then consider how to enumerate subgroups of SGs. When $I_k\ge 2$, i.e., the translational bases of the subgroup are enlarged and form a supercell, the subgroup is more complicated, because an integer translation in $\mathcal{G}$ may become a fractional translation in $H$, which means the subgroup of a symmorphic SG could be a non-symmorphic SG.

Denote the elements of $H$ as $\{\tilde{p}|\tilde{\bm{\tau}}_{\tilde{p}}+\tilde{\bm{t}} \}$, where $\tilde{\bm{t}}\in \bm{T}_H$ is a lattice translation and $\tilde{\bm{\tau}}_{\tilde{p}}$ is the fractional translation associated with $\tilde{p}$ in $\mathcal{G}$, written in the bases of $\bm{T}_H$. Symbols with a tilde always represent operations in $H$.
The fractional translation in $H$ can be expressed as $\tilde{\bm{\tau}}_{\tilde{p}}=\bm{\tau}_{\tilde{p}}+\bm{t}_{\tilde{p}}$, where $\bm{\tau}_{\tilde{p}}$ is the fractional translation associated with $\tilde{p}$ in $\bm{T}$ and $\bm{t}_{\tilde{p}}$ a lattice translation in $\bm{T}$.
The following equations must be satisfied for $H$ to form a subgroup of $\mathcal{G}$:
\begin{equation}
\begin{aligned}
    &\tilde{p} \bm{T}_H = \bm{T}_H,\\
    &\bm{\tau}_{\tilde{p}_1}+ \tilde{p}_{1} \bm{\tau}_{\tilde{p}_2} - \bm{\tau}_{\tilde{p}_1\tilde{p}_2} =
    w(\tilde{p}_1, \tilde{p}_2)
    \bmod \bm{T}_H,\\
\end{aligned}
\label{subsg_eq}
\end{equation}
$\forall \tilde{p}, \tilde{p}_1, \tilde{p}_2\in P_H$, where $w(p_1, p_2) = \bm{\tau}_{p_{1} p_{2}}- \bm{\tau}_{p_{1}}- p_{1} \bm{\tau}_{p_{2}}$.
The first equation requires the translation group to be invariant under $P_H$, and the second equation is the multiplication rule that the translation part of each element must satisfy in the subgroup. The second equation can be reformulated as a modular linear equation $M\cdot v=w\bmod \bm{T}_H$, where $v$ has the dimension $3N_H$, and $M$ has dimension $3N_H N_H\times 3N_H$, with $N_H$ being the number of elements in $P_H$. The Smith normal form can be adopted to solve the modular equation, the details of which can be found in Supplementary Material \cref{SM_modular_eq}.


In practice, we first find all possible $P_H$ and $\bm{T}_H$ of given $I_t$ and $I_k$, and then solve Eq. ( \ref{subsg_eq}) to obtain all possible subgroups. $\bm{T}_H$ is generated by a supercell matrix $S=(s_1, s_2, s_3)\in \text{GL}(3, \mathbb{Z})$ with $\det=I_k$, where $s_i$ denotes a primitive basis, which consists of the integer combination coefficients of the primitive bases of $\bm{T}$.

\subsection{Invariant subgroups of space groups}
For an invariant subgroup $H\vartriangleleft \mathcal{G}$, the invariant condition, i.e., $gHg^{-1}=H$, $\forall g\in \mathcal{G}$, must be satisfied apart from Eq. (\ref{subsg_eq}).
This leads to the following equations:
\begin{equation}
\begin{aligned}
    &p \bm{T}_H = \bm{T}_H,\\
    & \tilde{p} t - t=0\bmod \bm{T}_H, \\
    &p t_{\tilde{p}}- t_{p\tilde{p}p^{-1}} =
    w(p,\tilde{p}) - w(p\tilde{p}p^{-1}, p)
    \bmod \bm{T}_H,\\
\end{aligned}
\label{inv_subsg_eq}
\end{equation}
$\forall \tilde{p}\in P_H, p\in P, t\in \bm{T}$, where
$w(p,\tilde{p})$ is defined similarly as in Eq. (\ref{subsg_eq}).

In practice, the second equation in Eq. (\ref{inv_subsg_eq}) can be transformed into another modular linear equation $M^\prime \cdot v = w^\prime \bmod \bm{T}_H$, where $v$ has dimension $3N_H$ and $M^\prime$ has dimension $3N_\mathcal{G} N_H\times 3N_H$, with $N_\mathcal{G}$ and $N_H$ being the number of elements in $P$ and $P_H$. Each subgroup is checked using Eq. (\ref{inv_subsg_eq}) to determine if it is an invariant subgroup. In this way, we obtain all invariant subgroups of SGs within a given supercell range.

\subsection{Quotient groups}

Given an invariant subgroup $H$ of $\mathcal{G}$, the quotient group $Q=\mathcal{G}/H$ can be computed by first computing the translation quotient group $Q_T=\bm{T}/\bm{T}_H$ and point quotient group $Q_P=P/P_H$.

The translation quotient group $Q_T$ is a finite Abelian group of the structure $\mathbb{Z}_{n_1}\times \mathbb{Z}_{n_2}\times \mathbb{Z}_{n_3}$, where $n_1n_2n_3=I_k$ is the supercell k-index. The point quotient group $Q_P$ is obtained from the coset representatives of $P_H$ in $P$. The whole quotient group $Q$ is a finite group, with elements being the set product of $Q_T$ and $Q_P$, where the rotations from $Q_P$ have to recover their possible fractional translation parts $\bm{\tau}$ in non-symmorphic $\mathcal{G}$. The group structure of $Q$ is determined by both $\mathcal{G}$ and $H$, which is the group extension of $Q_T$ by $Q_P$.

We compute the multiplication table of $Q$ and then identify the isomorphic abstract group. An algorithm to obtain isomorphism between finite groups is given in
Supplementary Material \cref{SM_isomorphism}.

\subsection{3D real representations of the quotient group}
The 3D real (unitary) representations of the quotient group can be used to construct SSGs. This is because the spin operations belong to $\text{O}(3)$, and the representation matrix of a 3D real representation can always be transformed into a $\text{O}(3)$ matrix. A finite group with 3D real (unitary faithful) representations must be isomorphic to a PG. Thus we consider the quotient groups that are isomorphic to point groups, either crystallographic or non-crystallographic.

Point groups are classified into the following abstract point groups that are not isomorphic to each other:
$C_n (n\in\mathbb{Z})$, $C_{nh} (n\in2\mathbb{Z})$, $D_n (n\in\mathbb{Z}, n\ge 3)$, $D_{nh} (n\in2\mathbb{Z})$, $T, O, I, T_h, O_h, I_h$.
The equivalent relations between point groups are summarized in Supplementary Material \cref{SM_eqv_PG}.
The irreducible representations (IRREPs) of 32 crystallographic point groups are tabulated in Ref.\cite{altmann1994point}, and IRREPs of non-crystallographic point groups can be found in \textit{Mathematica}\cite{Mathematica}.

IRREPs of a finite group $\mathcal{G}$ can be classified into three types, i.e., real, pseudo-real, and complex\cite{Bradley2010}.
3D real representations are constructed from IRREPs of each abstract PG, which have three possible constructions:
\begin{itemize}
    \item from three 1D IRREPs, if all three 1D IRREPs are real, or one of them is real and the other two are complex and conjugated;
    \item from a 1D and a 2D real IRREP;
    \item directly from a 3D real IRREP.
\end{itemize}
Moreover, we require that the 3D real representations are faithful, i.e., only the identity element in the PG has the identity representation matrix.
This is because otherwise, the corresponding SSG can also be constructed using a larger invariant subgroup, extended by the elements with the identity representation matrix.

\subsection{Equivalent SSGs}
To obtain the correct number of inequivalent SSGs, equivalent relation between SSGs must be defined. There are three levels of equivalence, i.e., equivalent supercells (translation subgroups), equivalent (invariant) subgroups, and equivalent 3D real representations of the quotient groups.

Before moving on to the formal definition, we first give an example to gain some insights into the problem of equivalence. Consider SG 16 $P222$, which has $C_{2x}$, $C_{2y}$, and $C_{2z}$ rotations. For a system with $P222$ symmetry, there is some arbitrariness in choosing three coordinate axes. For example, one can permute the $x, y, z$ axis to the $y, z, x$ axis, or change the $x$ and $y$ axis to $y$ and $-x$ axis, respectively. These two coordinate transformations can be represented by $C_{3, 111}$ and $C_{4z}$. Using them as two generators, a transformation group of point group $O$ can be generated. As three axes are equivalent under $C_{3, 111}$, one can choose $C_{2y}$ rotation to form a subgroup $P2_y$, and the other two subgroups $P2_x$ and $P2_z$ are equivalent to $P2_y$. Also, a $n$-fold supercell along any axis is equivalent. This example shows the importance of finding the transformations that leave the SG invariant, which can be described by the automorphism group of the SG.

For a given SG $\mathcal{G}$, its automorphism group, which is the set of coordinate transformations that leave $\mathcal{G}$ invariant, i.e., $Auto(\mathcal{G})=\{W=\{A|\bm{t}\} | A\in \text{SL}(3,\mathbb{Z}), \bm{t}\in\mathbb{R}^3\}$ s.t.
\begin{equation}
	W \mathcal{G} W^{-1}=\mathcal{G}
\end{equation}
Here $A\in \text{SL}(3,\mathbb{Z})$ is a recombination of three coordinate bases, that must have $\det(A)=1$ s.t. the volume and chirality of the unit cell is unchanged. $\bm{t}\in\mathbb{R}^3$ is a shift of the origin point, which can take an infinite number of values for a given $A$.
Notice that the ordering of operations in $\mathcal{G}$ under $W$ may change, but their matrix forms must remain unchanged. The automorphism groups of SGs are summarized in Supplementary Material \cref{SM_auto_SG}.
We remark that for the triclinic and monoclinic crystal systems, the number of automorphisms with different rotation parts is infinite, while the number for other crystal systems is finite.

The equivalence between supercells is defined as follows. For two supercells $S_1$ and $S_2$ of an SG $\mathcal{G}$, where $S_i$ is a $3\times 3$ matrix with each column being a basis vector of the supercell, they are equivalent if there exists an automorphism $W=\{A| \bm{t}\}$ of $\mathcal{G}$ and an elementary column transformation $C$ with $\det(C)=1$, s.t.
\begin{equation}
	S_1 = A S_2 C
\end{equation}
$C$ serves as a recombination of bases which is necessary.

After obtaining inequivalent supercells, we then define equivalent (invariant) subgroups under a given supercell. Two subgroups $H_1$ and $H_2$ of a given supercell $S$ of SG $\mathcal{G}$ are equivalent if there exists an automorphism $W=\{A| \bm{t}\}$ of $\mathcal{G}$ s.t.
\begin{equation}
	WH_1W^{-1}=H_2,\quad ASC=S
\end{equation}
where $C$ is an elementary column transformation with $\det(C)=1$, and the second equation means that the supercell $S$ is invariant under the automorphism, or equivalently, the automorphism $A^{-1}$ is an integer matrix in the supercell, i.e., $C=S^{-1}A^{-1}S\in \text{SL}(3,\mathbb{Z})$.

For a given invariant subgroup $H$ with supercell $T_H$ of SG $\mathcal{G}$, consider two 3D real representations $D_1$ and $D_2$ of the quotient group $Q$. We use the notation $(Q|D)$ to explicitly show the mapping from $Q$ to $D$.
$(Q|D_1)$ and $(Q|D_2)$ are equivalent if there exists an automorphism $W=\{A| \bm{t}\}$ of $\mathcal{G}$, s.t. $H$ together with $T_H$ are invariant under $W$, i.e., $WHW^{-1}=H$, $WT_HC=T_H$ ($C$ is an elementary column transformation), and
\begin{equation}
	\begin{aligned}
			(WQW^{-1}|D_1) &\cong (Q|D_2)
	\end{aligned}
\end{equation}
where $W$ only acts on the quotient group operations, and $\cong$ means the transformed representation $(WQW^{-1}|D_1)$ is equivalent to $(Q|D_2)$, which does not require their representation matrix to be the same, but only their characters to be the same.

Based on the aforementioned equivalent relations, we are able to find all inequivalent invariant subgroups and 3D real representations of the quotient group. The SSGs constructed from them are thus all inequivalent. We leave more technical details in Supplementary Material \cref{SM_algorithm_details}.

\section{Results}\label{Sec:results}

\subsection{Summary of enumeration results}
Using the algorithm introduced in the Sec.\ref{algorithm}, we obtain a vast number of SSGs. There are
\begin{itemize}
    \item $331,661$ subgroups with k-index $I_k\le 12$ and $2,443$ subgroups with $I_k=1$ (i.e., without supercell).
    \item $27,197$ invariant subgroups with k-index $I_k\le 12$ and $1,708$ invariant subgroups with $I_k=1$.
    \item $10,439$ quotient groups isomorphic to crystallographic point groups, and $68,922$ SSGs constructed by their 3D real representations.
    \item $7,994$ quotient groups isomorphic to non-crystallographic point groups with maximal operation rank less equal than 24, and $88,367$ SSGs constructed from them.
    \item The total number of inequivalent SSGs with $I_k=1$ is $8,505$.
    \item The total number of inequivalent non-coplanar SSGs is $157,289$.
    \item The number of inequivalent collinear SSGs is $1,421$, and the number of coplanar-non-collinear SSGs is $24,788$.
\end{itemize}

We use the following serial number to label (non-coplanar) SSGs:
\begin{equation}
    N_{\text{SG}}.I_k.I_t.N_{\text{3Drep}}
\end{equation}
where $N_{\text{SG}}$ is the SG number, $I_k=|T_\mathcal{G}/T_H|$ is the supercell k-index, $I_t=|P_\mathcal{G}/P_H|$ is the t-index, and $N_{\text{3Drep}}$ denotes the N-th 3D representation of given $I_k$ and $I_t$. For collinear and coplanar SSGs, we add extra $.L$ and $.P$ to the SSG label, i.e., $ N_{\text{SG}}.I_k.I_t.N_{\text{1Drep}}.L$ for collinear SSGs and $ N_{\text{SG}}.I_k.I_t.N_{\text{2Drep}}.P$ for coplanar SSGs.

In order to exhibit the vast number of SSGs, we develop a user-friendly online database\cite{ssg_website}, enabling easy searching of SSGs based on given information.
On the search page, one can specify the desired SSG in the format $N_{\text{SG}}.I_k.I_t.N_{\text{rep}}(.\text{L or .P})$ or the first three numbers $N_{\text{SG}}.I_k.I_t$ which will retrieve all eligible SSGs. A sidebar allows users to narrow down the search by selecting the space group number, SSG types, and equivalent quotient group label and specifying the range of $I_k$ and $I_t$. On the website of each SSG, we give SSG operations in $Q$, the pure-lattice operations in $H$, and other basic information about the SSG.

In the following, we compare our results with the literature and summarize some general rules for the construction of SSGs.

\subsection{Comparison with SPGs and MSGs}
Litvin enumerated $598$ spin-point groups (SPGs) in 1977\cite{Litvin1977}. SPGs are constructed using PGs, which do not have translational symmetries. The automorphism group of a PG $P$ is defined as $Auto(P)=\{A | A\in \text{SL}(3, \mathbb{R})\}$, which does not require $A$ to be an integer matrix. For example, a $\pi/4$ rotation along the $z$ axis is an automorphism of PG $D_4(422)$, but not an automorphism of SG $P422$.

For 32 symmorphic SGs that correspond to 32 PGs, we found 736 SSGs without considering supercells by restricting $I_k=1$. Compared with $598$ SPGs, there are 6 SGs, i.e., $P422$, $P4mm$, $P4/mmm$, $P622$, $P6mm$, and $P6/mmm$, that have extra SSGs, which all result from the difference in the definition of automorphism groups of SG and PG.

For 1651 Shubnikov MSGs, 230 type-II non-magnetic SGs have the time-reversal symmetry, which can be constructed using 230 SGs with spin-only groups $\mathcal{S}_0=\text{O}(3)$.

230 type-I single Shubnikov MSGs can be constructed from 230 SGs by considering $I_k=1, H=G$, and $Q=1$, i.e., all symmetries have no spin rotation part. If we consider $I_k=1, H=P1$, then $Q=P_{\mathcal{G}}$. In this case, if $\mathcal{G}$ contains no symmetry with $\det=-1$, then the corresponding SSG is isomorphic to the corresponding type-I double Shubnikov MSGs. On the other hand, if $\mathcal{G}$ contains symmetries with $\det=-1$, then the SSG is isomorphic to certain type-III double Shubnikov MSGs.

674 type-III single Shubnikov MSGs can be constructed from 230 SGs by considering all index-2 invariant subgroups of SGs with $I_t=2, I_k=1$. In this case, the quotient group $Q=\mathcal{G}/H\cong C_i$. The inversion symmetry assigned to the quotient group is equivalent to the TRS, and thus $\mathcal G^{(S)}\cong H+ g\cdot\mathcal{T}H$.

517 type-IV single Shubnikov MSGs can be constructed from 230 SGs by considering all $I_k=2, I_t=1$ invariant subgroups, which have the quotient group $Q=\mathcal{G}/H\cong C_i$. The inversion symmetry is assigned to one translation, and $\mathcal G^{(S)}\cong H + \bm{\tau}\cdot\mathcal{T} H$. The resultant number of SSGs is exactly the same number of the type-IV Shubnikov MSGs under OG setting for each SG.

\subsection{General rules for constructing quotient groups}
Despite the vast number of SSGs, many of them have quotient group structures that can be exhausted for arbitrary supercell indexes $I_k$. In the following, we list several examples of such constructions of quotient groups, and a more detailed discussion can be found in Supplementary Material \cref{SM_quotient}.

When the quotient group $Q\cong C_n$, the invariant subgroup $H$ must satisfy
$P/P_H\cong \mathbb{Z}_p$, $\bm{T}/\bm{T}_H\cong \mathbb{Z}_q$, where $pq=n$, and $p$ and $q$ are not necessarily mutually prime.
For example, for $\mathcal{G}=P4_1$, $P_H=1$, and $\bm{T}/\bm{T}_H\cong \mathbb{Z}_6$, with the supercell along $z$-direction, the quotient group $Q\cong C_{24}$ is generated by $\{C_{4z}|0,0,\frac{1}{4}+1\}$, which has rank $24$.

When $Q\cong C_{nh}$, the invariant subgroup $H$ could have the structure
$P/P_H\cong C_{2}$, $\bm{T}/\bm{T}_H\cong \mathbb{Z}_n$, where $n\in\mathbb{Z}, n\ge 2$, and $C_2$ must commute with the generator of $\bm{T}/\bm{T}_H$.
For example, for $\mathcal{G}=P2$, $P_H=1$, $P/P_H=C_2$, and $\bm{T}/\bm{T}_H\cong \mathbb{Z}_n$, with the rotation and supercell both along $y$-direction, the quotient group $Q\cong C_{nh}$ is generated by $\{C_{2y}|\bm{0}\}$ and $\{E|0,1,0\}$.

When $Q\cong D_{n}$, the invariant subgroup $H$ could also have the structure
$P/P_H\cong C_{2}$, $\bm{T}/\bm{T}_H\cong \mathbb{Z}_n$, where $n\ge 3$, but $C_2$ must not commute with the generator of $\bm{T}/\bm{T}_H$.
For example, for $\mathcal{G}=P\overline{1}$, $P_H=1$, $P/P_H=C_i$, and $\bm{T}/\bm{T}_H\cong \mathbb{Z}_n$, with the supercell along $z$-direction, the quotient group $Q\cong D_{n}$ is generated by $\{E|0,0,1\}$ and $\{\mathcal{P}|\bm{0}\}$, which do not commute with each other.

\subsection{Quotient groups isomorphic to crystallographic point groups}
We claim that we exhaust all SSGs that have spin parts isomorphic to crystallographic point groups, with reasons given below.

The translation quotient group $Q_T$ is a finite Abelian group.
For abstract point groups, only the followings are Abelian: $C_n\cong \mathbb{Z}_{n}$, $C_{nh}(n/m)\cong \mathbb{Z}_{n}\times\mathbb{Z}_2$, and $D_{2h}(mmm)\cong \mathbb{Z}_2\times\mathbb{Z}_2\times\mathbb{Z}_2$.
This can be seen from the fact that a $C_n$ rotation with $n>2$ does not commute with rotations along other axes.
Among these abstract point groups, only $C_n ~(n=1,2,3,4,6)$, $C_{nh} ~(n=2,4,6)$, and $D_{2h}$ are crystallographic point groups.

As a result, to exhaust all quotient groups that are isomorphic to crystallographic point groups, we only need to consider the supercell of k-index $I_k\in\{1,2,3,4,6,8,12\}$. Their isomorphisms are tabulated in Table.\ref{trans_quot_PG}.
Supercells with other k-indexes have translation quotient groups that are not isomorphic to any crystallographic point groups, and so are the whole quotient groups. This is because the translation quotient group is an invariant subgroup of the whole quotient group, and the whole quotient group cannot be isomorphic to a crystallographic PG if it has a subgroup that is not isomorphic to any crystallographic point groups.
This means that by considering all invariant subgroups of SGs with $I_k\le 12$, we are able to find all SSGs with spin part isomorphic to crystallographic point groups.

\begin{table}[htbp]
    \begin{tabular}{c|c|c}
	\hline\hline
	k-index & Translation quotient group                                     & Isomorphic PG \\\hline
		1       & $\mathbb{Z}_1$                                     & $C_1(1)$           \\\hline
		2       & $\mathbb{Z}_2$                                     & $C_2(2)$     \\\hline
		3       & $\mathbb{Z}_3$                                     & $C_3(3)$           \\\hline
		4       & $\mathbb{Z}_4$                                     & $C_4(4)$           \\\hline
		4       & $\mathbb{Z}_2\times\mathbb{Z}_2$                   & $C_{2h}(2/m)$         \\\hline
		6       & $\mathbb{Z}_6$                                     & $C_6(6)$           \\\hline
		8       & $\mathbb{Z}_4\times\mathbb{Z}_2$                   & $C_{4h}(4/m)$         \\\hline
		8       & $\mathbb{Z}_2\times\mathbb{Z}_2\times\mathbb{Z}_2$ & $D_{2h}(mmm)$  \\\hline
		12     & $\mathbb{Z}_6\times\mathbb{Z}_2$  & $C_{6h}(6/m)$  \\
	\hline\hline
    \end{tabular}
    \caption{\label{trans_quot_PG}Translation quotient groups that are isomorphic to crystallographic point groups. }
\end{table}

\subsection{SSGs for incommensurate magnetic orders}

Our approach also applies to constructing the SSGs for incommensurate magnetic structures. For example, consider the simplest incommensurate spiral magnetic structure along $z$-direction. In this case, the translational quotient group $\bm{T}/\bm{T}_H$ is isomorphic to the non-crystallographic point group $C_{n}$ with $n$ going to infinity (i.e., $\bm{T}/\bm{T}_H\cong \mathbb{Z}$). The corresponding SSG is generated by the operation $\{C_{2\pi\alpha}||E|\bm{a}_3\}$, i.e., the $z$-directional lattice translation $\bm{a}_3$ is accompanied by an incommensurate spin rotation $C_{2\pi\alpha}$, where $\alpha$ is an irrational number. We can approximate $\alpha$ as a rational number, i.e., $\alpha \approx \frac{m}{n}$, which leads to $|\bm{T}/\bm{T}|\approx \mathbb{Z}_n$. $n$ will go to infinity when $\alpha$ is approximated more accurately.

An interesting question arises concerning why non-crystallographic operations could appear in periodic systems. The key lies in understanding that the non-crystallographic rotations are confined only to the spin part in SSGs. These rotations do not influence the $\bm{r}$ or $\bm{k}$, but exclusively act on the magnetic moments $\bm{M}(\bm{r})$ or spinor Bloch states $\psi_{\sigma}(\bm{k})$. This distinction allows for the incorporation of non-crystallographic symmetries within the periodic framework of SSGs.

We also discuss briefly the relation of SSGs with superspace groups\cite{janner1980symmetry, van2007incommensurate}.
Superspace groups are conventionally employed in the realm of incommensurate crystallography and are generalized later to magnetic systems\cite{perez2012magnetic}.
In superspace groups, a basic (periodic) structure is described by an SG or an MSG, together with a modulation function $A_{\mu\bm{R}}$ defined on atom $\mu$ in unit cell $\bm{R}$, which characterizes the atomic displacements (e.g., charge density waves),
the magnetic moments (e.g., spin density waves), fractional occupancy of atoms, or other possible local physical quantities.
The modulation function $A_{\mu\bm{R}}(x_4)$ has a variable $x_4=\bm{q}\cdot (\bm{r}_\mu+\bm{R})$ with $\bm{q}$ being the propagation vector, either commensurate or incommensurate. $x_4$ is introduced as a higher dimension besides the three-dimensional real space and describes the aperiodicity of the system. Multiple propagation vectors are also supported in superspace groups.

Compared with (magnetic) superspace groups, SSGs are more powerful in describing complicated magnetic structures, where symmetries with unlocked spin and real-space rotations are allowed. This type of symmetry is uniquely described in SSGs.
It is worth mentioning that SSGs inherently restrict to a single principal axis around which non-crystallographic spin rotations occur.
Multiple principal axes of the non-crystallographic spin operations are not allowed in SSGs. This is because the non-crystallographic spin rotations are generated by the translation quotient group $\bm{T}/\bm{T}_H=\mathbb{Z}_{n_1}\times\mathbb{Z}_{n_2}\times\mathbb{Z}_{n_3}$, which is an abelian group. Thus the non-crystallographic spin operations assigned to the elements in $\bm{T}/\bm{T}_H$ must also commute, which means they share the same principal axis. For example, consider two SSG operations $g_1=\{C_{mx}||E|100\}$ and $g_2=\{C_{ny}||E|010\}$, with $m,n>2$. A system with $g_{1,2}$ is enforced to be non-magnetic because $g_1g_2\neq g_2g_1$, which will generate a non-trivial spin-only group that enforces zero magnetization.

However, a single principal axis in the spin part does not necessarily indicate a single propagation vector in SSGs.
An extra propagation vector could be given by the twofold spin operations perpendicular to the principal axis. For example, a SSG generated by $\{C_{nz}||E|0,0,1\}$ and $\{M_z||E|0,1,0\}$ ($n>6$ is even) has propagation vector $Q_1=(0,0,\frac{1}{n}), Q_2=(0,\frac{1}{2},0)$. Note that if $n$ is odd, $C_{nh}$ is isomorphic to $C_{2n}$, and a single propagation vector describes the corresponding magnetic order.

When the magnetic order is generated by two spin rotations along the same axis in two real-space directions, e.g., $\{C_{mz}||E|1,0,0\}$ and $\{C_{nz}||E|0,1,0\}$, they are described by a single propagation vector. This is because (i) when $m$ and $n$ are coprime, one can adopt a single generator with spin rotation $\{C_{mn,z}||E|a,b,0\}$, together with a pure translation $\{E||E|m,n,0\}$, where $an-bm=1$ s.t. the unit cell volume is maintained; (ii) when $m$ and $n$ are not coprime but have a greatest common divider $w$, one can still find a new generator $\{C_{\frac{mn}{w}, z}||E|a,b,0\}$ together with $\{E||E|\frac{m}{w}, \frac{n}{w}, 0\}$, where $an - bm=w$ (i.e., the Bézout's identity), s.t. the unit cell volume is maintained.
However, we remark that if there are multiple propagation vectors along the same direction, the magnetic moments in general cannot have the same magnitude. SSGs cannot describe this type of magnetic structure because the spin rotations in SSG are O(3) matrices that maintain the length of magnetic moments.

For incommensurate magnetic structures, they can be approximated by commensurate magnetic structures, and the discussion above still applies.
Thus, we conclude that SSGs can effectively describe incommensurate magnetic structures using non-crystallographic spin operations, which only have a single principal axis resulting from the group structure of the non-crystallographic point groups. In this case, multiple propagation vectors are still allowed.

Both SSGs and superspace groups have their unique strengths and are useful in the realm of crystallography and magnetism. It is also possible to integrate SSGs in the superspace group formalism by replacing MSGs with SSGs for the basic periodic structure.

\section{SSGs of magnetic structures}\label{example}
In this section, we give several pedagogical examples to show the construction of SSGs.

\subsection{Pedagogical examples}
\begin{figure}[htbp]
	\centering
	\includegraphics[width=0.5\textwidth]{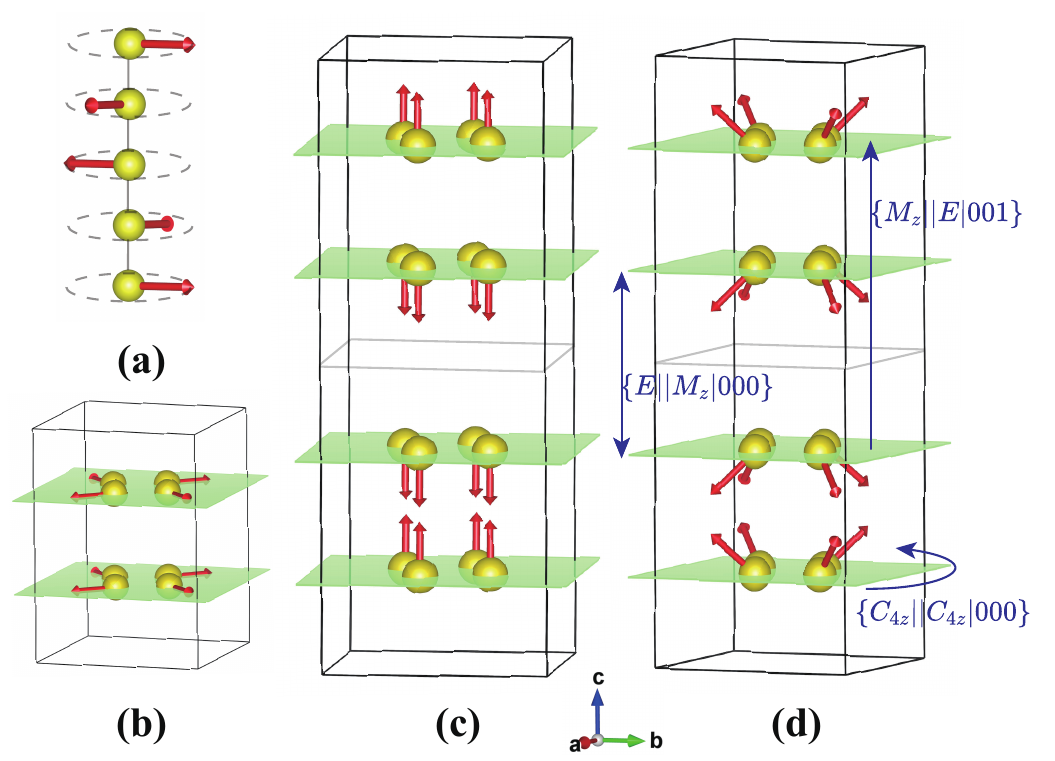}
	\caption{\label{Fig:example} Examples of magnetic orders and their spin-space groups. (a) A schematic show of spiral magnetic phase, where a lattice translation along the $z$-axis is associated with a $C_{4z}$ rotation in spin space, i.e., $\{C_{4z}||E|001\}$. (b)-(d) Three possible magnetic orders generated from the same space group $P4/m$, where (b) is a coplanar order, (c) a collinear order, and (d) a general non-coplanar order, yielding SSGs $83.1.4.1.P$, $83.2.1.1.L$, and $83.2.4.1$, respectively. In (d), we use blue arrows to denote three generators of the corresponding SSG $83.2.4.1$, which includes a pure lattice operation $\{E||M_z|000\}$, a spin-lattice locked operation $\{C_{4z}||C_{4z}|000\}$, and a spin-lattice unlocked operation $\{M_z||E|001\}$.
 }
\end{figure}

First, for each symmorphic SG, a special type of SSG can be constructed using invariant subgroups $H$ with $I_t=1$ and $I_k=n$, i.e.,
$P/P_H=1$, $\bm{T}/\bm{T}_H=\mathbb{Z}_n$, and thus $Q\cong\mathbb{Z}_n$. Assume the $n$-fold supercell is along the $z$-axis. Then the SSG is generated by a $n$-fold operation
\begin{equation}
    \{C_{n}|| E|001\}.
\end{equation}
where $E$ is the identity operation. This operation means a lattice translation in the $z$-axis is accompanied by a $C_n$ rotation in spin space, which can be used to describe the spiral magnetism. In Fig.\ref{Fig:example}(a), we show a possible spiral magnetic phase generated by $\{C_{4}|| E|001\}$.

Next, we show that different 3D real representations can construct different SSGs.
As an example, consider SG 10 $P2/m$ and a trivial invariant subgroup $H$ with $P_H=1, \bm{T}_H=\bm{T}$. The quotient group $Q\cong C_{2h}$ has four real IRREPs $\Gamma_1^\pm$ and $\Gamma_2^\pm$, as shown in Table.\ref{IRREP_C2h}. These IRREPs can be combined to form ten 3D real representations, as shown in Table.\ref{3D_rep_C2h}. Note that the ordering of the three 1D IRREPs in a 3D real representation is inessential, as it only changes the main axis of the equivalent PG which corresponds to the main axis in spin space and can be chosen arbitrarily.
These ten 3D real representations give ten inequivalent SSGs. For example, the SSG formed by $\Gamma_2^-\oplus\Gamma_1^-\oplus\Gamma_2^-$ has elements
\begin{equation}
    \{E|| E|\bm{0}\}, \{C_{2y}|| C_{2y}|\bm{0}\},
    \{\mathcal{P}|| \mathcal{P}|\bm{0}\}, \{M_{y}|| M_{y}|\bm{0}\},
\end{equation}
which is nothing but the double SG $P2/m$, with the real-space part and spin-space part being locked.
The SSG formed by $\Gamma_1^-\oplus\Gamma_2^+\oplus\Gamma_2^-$ has elements
\begin{equation}
    \{E|| E|\bm{0}\}, \{C_{2x}|| C_{2y}|\bm{0}\},
    \{C_{2y}|| \mathcal{P}|\bm{0}\}, \{C_{2z} || M_{y}|\bm{0}\},
    \label{ssg_P2/m_222}
\end{equation}
which is an SSG with real-space and spin-space parts unlocked.

As the third example, we exemplify the equivalent relation in SSGs. Consider SG 16 $P222$ and the trivial invariant subgroup as in the previous example, i.e., $Q\cong C_{2h}$. The ten 3D real representations, however, can not form ten SSGs, because only three of them are inequivalent, i.e., $\Gamma_1^-\oplus\Gamma_2^+\oplus\Gamma_2^-\cong D_2$, $\Gamma_1^+\oplus\Gamma_1^-\oplus\Gamma_2^+\cong C_{2v}$, and $\Gamma_1^-\oplus\Gamma_1^-\oplus\Gamma_2^+\cong C_{2h}$. This is because the three $C_2$ rotations in $P222$ are equivalent, and the 3D real representations equivalent to the same PG are thus equivalent.

\begin{table}[htbp]
\begin{tabular}{c|c|c|c|c}
\hline\hline
IRREP      & $E$ & $C_2$ & $\mathcal{P}$ & $M$ \\\hline
$\Gamma_1^+$ & 1   & 1     & 1   & 1   \\\hline
$\Gamma_1^-$ & 1   & 1     & -1  & -1  \\\hline
$\Gamma_2^+$ & 1   & -1    & 1   & -1  \\\hline
$\Gamma_2^-$ & 1   & -1    & -1  & 1  \\
\hline\hline
\end{tabular}
\caption{\label{IRREP_C2h}Irreducible representations (IRREPs) of point group $C_{2h}$.}
\end{table}

\begin{table}[htbp]
\begin{tabular}{c|c|c|c|c|c}
\hline\hline
3D real rep                & $E$ & $C_2$    & $\mathcal{P}$      & $M$      & Eqv PG \\ \hline
$\Gamma_1^-\oplus\Gamma_2^+\oplus\Gamma_2^-$ & $E$ & $C_{2x}$ & $C_{2y}$ & $C_{2z}$ & $D_2$      \\ \hline
$\Gamma_1^+\oplus\Gamma_1^-\oplus\Gamma_2^+$ & $E$ & $M_z$    & $M_y$    & $C_{2x}$ & $C_{2v}$   \\ \hline
$\Gamma_1^+\oplus\Gamma_1^-\oplus\Gamma_2^-$ & $E$ & $M_z$    & $C_{2x}$ & $M_y$    & $C_{2v}$   \\ \hline
$\Gamma_1^+\oplus\Gamma_2^+\oplus\Gamma_2^-$ & $E$ & $C_{2x}$ & $M_z$    & $M_y$    & $C_{2v}$   \\ \hline
$\Gamma_1^-\oplus\Gamma_1^-\oplus\Gamma_2^+$ & $E$ & $M_z$    & $C_{2z}$ & $\mathcal{P}$      & $C_{2h}$   \\ \hline
$\Gamma_1^-\oplus\Gamma_1^-\oplus\Gamma_2^-$ & $E$ & $M_z$    & $\mathcal{P}$      & $C_{2z}$ & $C_{2h}$   \\ \hline
$\Gamma_1^-\oplus\Gamma_2^+\oplus\Gamma_2^+$ & $E$ & $C_{2x}$ & $M_x$    & $\mathcal{P}$      & $C_{2h}$   \\ \hline
$\Gamma_2^-\oplus\Gamma_1^-\oplus\Gamma_2^-$ & $E$ & $C_{2y}$ & $\mathcal{P}$      & $M_y$    & $C_{2h}$   \\ \hline
$\Gamma_2^+\oplus\Gamma_2^+\oplus\Gamma_2^-$ & $E$ & $\mathcal{P}$      & $M_z$    & $C_{2z}$ & $C_{2h}$   \\ \hline
$\Gamma_2^+\oplus\Gamma_2^-\oplus\Gamma_2^-$ & $E$ & $\mathcal{P}$      & $C_{2x}$ & $M_x$    & $C_{2h}$   \\ \hline\hline
\end{tabular}
\caption{\label{3D_rep_C2h}3D real representations of point group $C_{2h}$, where the first column gives ten 3D real representations, the second to fifth columns give the representation matrix of each operation in $C_{2h}$, and the last column gives the equivalent PG of the 3D real representation.}
\end{table}

Lastly, we present a sophisticated example to demonstrate the construction of collinear, coplanar, and general non-coplanar SSGs from a single SG.
Consider SG 83 $P4/m$, which can be generated using PG operations $C_{4z}$ and $M_z$ together with the three lattice translations. A generic atomic configuration of $P4/m$ is shown in Fig.\ref{Fig:example}(b), which has eight atoms in the unit cell. We first consider a coplanar magnetic order as shown in Fig.\ref{Fig:example}(b), which has a pure spin reflection symmetry $\{M_z||E|\bm{0}\}$ owned by all coplanar orders. The magnetic moments of the four atoms on the same layer have $C_{4z}$-related directions, resulting in a spin-lattice locked operation $\{C_{4z}||C_{4z}|\bm{0}\}$. The atoms in two layers, however, share the same magnetic configuration, yielding a pure lattice symmetry $\{E||M_z|\bm{0}\}$. Thus the coplanar SSG for this magnetic order is identified as SSG $83.1.4.1.P$ in our database.

We then consider a collinear magnetic order in Fig.\ref{Fig:example}(c), which has a twofold supercell along the $z$-axis. The magnetic moments in two unit cells have reversed directions, leading to the $\{ M_z||E|001\}$ operation and the pure lattice operation $\{ E||M_z|\bm{0}\}$. The spin-lattice locked operation $\{C_{4z}||C_{4z}|\bm{0}\}$ still exists, but as the collinear order has spin-only operation $\{C_{\theta}||E|\bm{0}\}$ ($\theta$ being an arbitrary angle), the $C_{4z}$ spin rotation can be omitted and leads to the pure lattice operation $\{E||C_{4z}|\bm{0}\}$. The invariant subgroup $H$ formed by pure lattice operations is thus identified as $P4/m$ with a twofold supercell, leading to the collinear SSG $83.2.1.1.L $.

As last, we consider a general non-coplanar order in Fig.\ref{Fig:example}(d), where we mark three generators of the SSG:
\begin{equation}
\{E || M_z|\bm{0}\}, \{C_{4z}|| C_{4z}|\bm{0}\},
\{M_z || E|001\}.
\label{Eq:ssg_83_generators}
\end{equation}
The generators of the translation group $\bm{T}$ are modified to  $\{E|100\},\{E|010\},\{E|002\}$. The corresponding SSG is identified as $83.2.4.1$, which has the invariant subgroup $H$ of pure lattice operations $Pm$ with a twofold supercell along $z$. We emphasize that the SSG symmetries of this non-coplanar magnetic order are beyond MSGs. Within MSG, only the spin-lattice locked generator $\{C_{4z}|| C_{4z}|\bm{0}\}$ remains, while the pure-lattice operation $\{E || M_z|\bm{0}\}$ and the spin-lattice unlocked operation $\{M_z ||E|001\}$ lie out of the scope of MSGs.

\subsection{Realistic magnetic materials}\label{Sec:magnetic_mats}
In this work, we also develop an algorithm that can identify SSGs for realistic materials, with details given in Supplementary Material \cref{SM_identify_SSG}.
We apply the algorithm to more than 2,000 magnetic materials in \textit{Bilbao Crystallographic server}\cite{aroyo2006bilbao1,aroyo2006bilbao2,aroyo2011crystallography,gallego2016magndataI,gallego2016magndataII}, and find the corresponding SSGs for all 1,626 commensurate magnetic materials without partial occupation, with results summarized in Supplementary Material \cref{sec:result_magndata}.
Before starting, we would like to remark on the usage of SSG on magnetic materials: (i) SSGs serve as a fine-grained tool to describe symmetry and refine magnetic structures. (ii) SSGs describe the electronic structures of magnetic materials when SOC is negligible or weak compared to the spin splitting induced by the effective Zeeman term.

In the following, we present four examples of realistic materials with collinear, coplanar, and non-coplanar magnetism. We identify their corresponding SSGs and MSGs and show that SSGs have richer symmetries than MSGs.

\begin{figure}[htbp]
	\centering
	\includegraphics[width=0.5\textwidth]{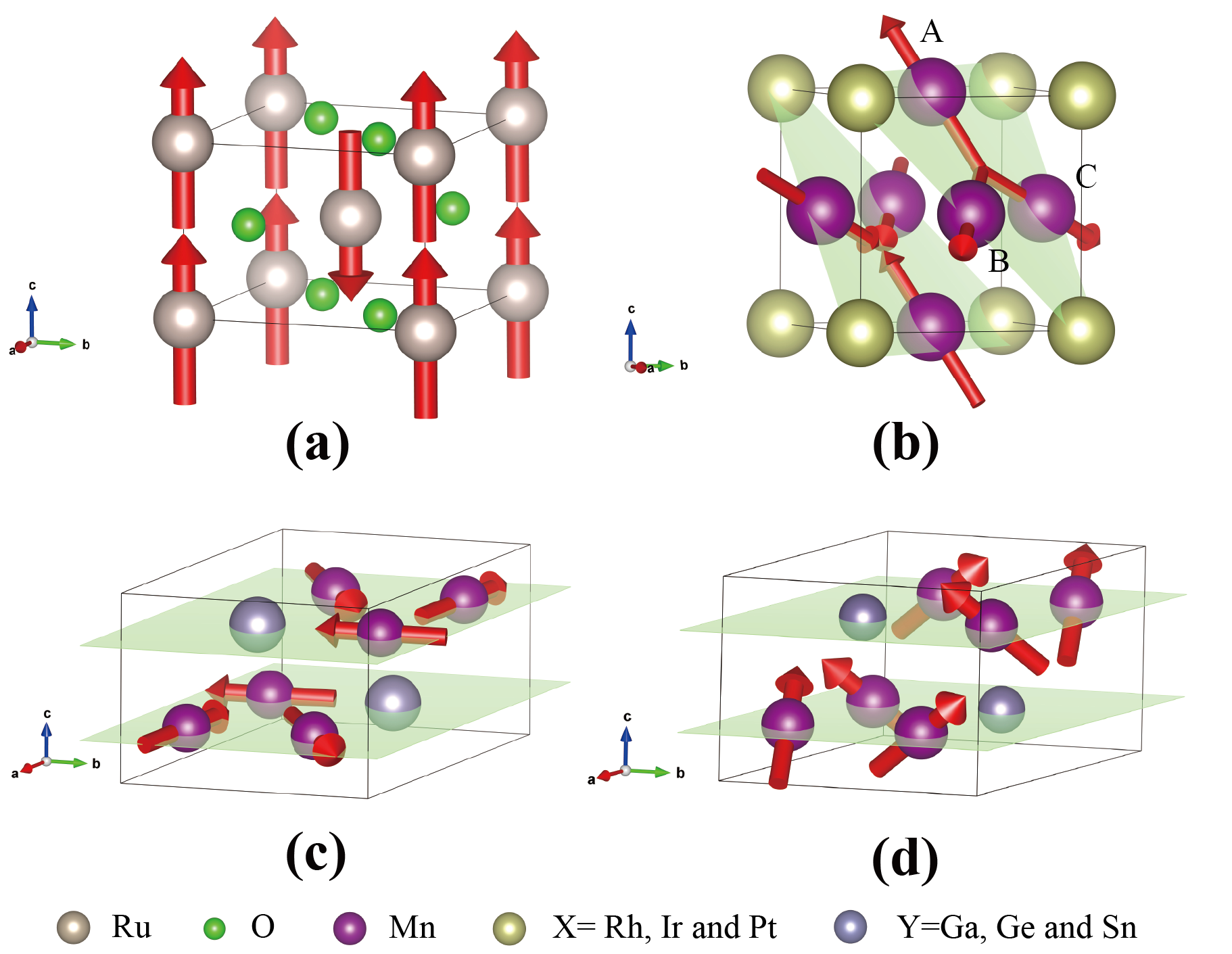}
	\caption{\label{Fig:real_material} Realistic material examples of SSGs. (a) \ch{RuO2} features a tetragonal lattice and a collinear antiferromagnetic order.  
	(b) \ch{Mn3X}   (X=Rh, Ir, and Pt) exhibits a face-centered cubic (FCC) crystal structure with a coplanar magnetic order where all magnetic moments reside on the (111) plane (highlighted in green). A, B, and C are used to label three Mn atoms located on the same plane.
	(c) \ch{Mn3Y} (Y=Ga, Ge, and Sn) has a hexagonal lattice where Mn atoms form two Kagome layers and develop a coplanar magnetic order. (d) A deformed non-coplanar magnetic structure of (c), where the magnetic moments of \ch{Mn} are rotated in $xy$ plane and acquire a $z$-directional component of equal magnitude.
 }
\end{figure}

Fig.\ref{Fig:real_material}(a) depicts the collinear magnetic structure of \ch{RuO2}, a recently proposed typical \textit{altermagnetic} material\cite{feng2022anomalous,bose2022tilted,baihruo2exp2022,kshutaroruo2exp2022, ruo2aheconbs6,ruo2spinsplit2021}. We take it as an example to demonstrate how to determine the SSG of collinear antiferromagnetic materials.
\ch{RuO2} has SG 136 $P4_2/mnm$ symmetry when magnetic moments are ignored.
The pure lattice symmetry group $H$ can be identified by considering the spin-up and spin-down atoms as different types of atoms, which is $Pmmm$ in \ch{RuO2}. The quotient group is then computed as $C_4$ with the generator $g=\{M_{001}|| 4_{001}|\frac{1}{2},\frac{1}{2},\frac{1}{2}\}$. The collinear spin-only group $\mathcal{S}_0$ defined in Eq. (\ref{Eq:collinear_S0}) is also present. The SSG is thus determined as 136.1.2.6.L in our database.
The MSG of \ch{RuO2} is 136.499 $P4_2^\prime /mnm^\prime$, which forms a subgroup of the SSG. For example, $\{4_{001}||4_{001}|\frac{1}{2},\frac{1}{2},\frac{1}{2}\}\cdot \mathcal{T}\sim \{4_{001}\cdot \mathcal{P}||4_{001}|\frac{1}{2},\frac{1}{2},\frac{1}{2}\}$ in MSG can be obtained by combining $g$ with the pure-spin operation $\{4_{001}^{-1}||E|\bm{0}\}$.

Fig.\ref{Fig:real_material}(b) and (c) display the coplanar magnetic structures of \ch{Mn3X}   (X=Rh, Ir, and Pt) and \ch{Mn3Y} (Y=Ga, Ge, and Sn), respectively, which have been reported to exhibit strong anisotropic anomalous Hall effects and spin Hall effects\cite{Mn3X_hall_effect,ahc_noncollinear_prl}. In the following, we determine their corresponding SSGs separately.

As shown in Fig.\ref{Fig:real_material}(b), \ch{Mn3X} (X=Rh, Ir, and Pt) has a face-centered cubic (FCC) crystal structure with three inequivalent face-centered Mn atoms having different magnetic orientations, but all residing on the same $(111)$ plane. The SG without considering magnetic moments is SG 221 $Pm\bar{3}m$.
The pure lattice symmetry group $H$ is determined to be $Pmmm$. The quotient group is thus isomorphic to $C_{3v}$, but the spin rotations are not straightforward to obtain.
To clarify each SSG operation, we label three \ch{Mn} atoms by A, B, and C in Fig.\ref{Fig:real_material}(b), and show the permutation of \ch{Mn} atoms by the real-space rotations in Table.\ref{tab:Mn3X_quot_op_permutation}. The spin rotations can then be determined by comparing the orientations of magnetic moments. For example, the real-space $2_{110}$ acts on \ch{Mn} atoms as
\begin{equation}
2_{110}r_A=r_A,\ 2_{110}r_B=r_C,\ 2_{110}r_C=r_B.
\end{equation}
The accompanied spin operation should also switch the magnetic moments at B and C while leaving A unchanged, which is identified as $M_{1\bar{1}0}$ with the mirror plane perpendicular to the line connecting B and C. The SSG operation $\{3_{111}||3_{111}\}$ can be identified similarly. The SSG is then determined as 221.1.6.1.P in the database, with all SG 221 operations maintained by assigning proper spin rotations.
The MSG of \ch{Mn3X} is 166.101 $R\bar{3}m^\prime$, with the main $C_3$ axis along $(111)$. Two generators of SG 221 are broken in this MSG, i.e., $4_{100}$ and $M_{100}$, with the former  having an unlocked spin rotation, i.e., $\{M_{01\bar{1}}||4_{100}\}$, and the latter being a pure lattice symmetry, i.e., $\{E||M_{100}\}$, in SSG 221.1.6.1.P.

\begin{table}[htbp]
\begin{tabular}{c|c|c|c|c}
\hline\hline
real-space operation & A & B & C & spin operation  \\\hline
$E$            & A & B & C & $E$            \\\hline
$3_{111}$            & B & C & A & $3_{111}$           \\\hline
$3^{-1}_{111}$      & C & A & B & $3^{-1}_{111}$          \\\hline
$2_{110}$            & A & C & B & $M_{1\bar{1}0}$     \\\hline
$4_{100}$            & C & B & A & $M_{01\bar{1}}$    \\\hline
$4_{010}$            & B & A & C & $M_{10\bar{1}}$    \\
\hline\hline
\end{tabular}
\caption{\label{tab:Mn3X_quot_op_permutation} Elements of the quotient group of 221.1.6.1.P, i.e., symmetries of the magnetic structure in Fig.\ref{Fig:real_material}(b). The first column is the real-space operations, the second to fourth columns the permutation of positions A $(\frac{1}{2},\frac{1}{2},1)$, B $(1,\frac{1}{2},\frac{1}{2})$, and C $(\frac{1}{2},1,\frac{1}{2})$ under these operations, and the last column the corresponding spin operations.}
\end{table}

In Fig.\ref{Fig:real_material}(c), we show the structure of \ch{Mn3Y} (Y=Ga, Ge, and Sn), which has a hexagonal lattice with Mn atoms forming two Kagome sublayers stacked along the c-axis. The magnetic moments of Mn atoms are oriented in the $(1,1,0)$, $(-1,0,0)$ and $(0,-1,0)$ directions, respectively, written under $\bm{a}_1, \bm{a}_2, \bm{a}_3$ axes.
The SG of this structure is SG 194 $P6_3/mmc$, and the pure lattice group is $P2_1/m$, generated by the inversion $\mathcal{P}$ and $\{M_z|\frac{\bm{a}_3}{2}\}$.
The quotient group is isomorphic to the previous example, i.e., $C_{3v}$, but has different operations, with generators $\{3^-_{001}||3^+_{001}\}$ and $\{2_{010}||M_{010}\}$ (or equivalently, $\{M_{210}||M_{010}\}$).
It is worth mentioning that for the former generator, its real-space operation is a $\frac{2}{3}\pi$ rotation while the spin operation is a $-\frac{2}{3}\pi$ rotation, which is not allowed in MSG. The SSG is then identified as 194.1.6.1.P in our database, with a coplanar spin-only group given in Eq. (\ref{Eq:coplanar_S0}).
The MSG of this material is 63.463 $Cmc^\prime m^\prime$, generated by $\{E||\mathcal{P}\}$, $\{2_{010}||M_{010}\}$, and $\{2_{001}||M_{001}|\frac{\bm{a}_3}{2}\}\cdot\mathcal{T}\sim \{M_{001}||M_{001}|\frac{\bm{a}_3}{2}\}$.
This MSG can be seen a subgroup of SSG 194.1.6.1.P by breaking $\{3^-_{001}||3^+_{001}\}$ and the coplanar spin-only group.
Remark that there exist two entries of \ch{Mn3Sn} on \textit{Bilbao Crystallographic server}\cite{gallego2016magndataI,gallego2016magndataII}, which come from the same Ref.\cite{brown1990determination} and differ by an overall $C_{4z}$ spin rotation. These two magnetic structures share the same SSG but with different conventions of spin rotations.

At last, we introduce a non-coplanar magnetic structure of \ch{Mn3Y} (Y=Ga, Ge, and Sn) depicted in Fig.\ref{Fig:real_material}(d), which can be obtained by imposing a hydrostatic pressure up to 5 GPa to the coplanar magnetic structure in Fig.\ref{Fig:real_material}(c) according to Ref.\cite{non_coplanar_example}.
The magnetic moments of Mn atoms are along $(1,-1,m_z)$, $(1,2,m_z)$, and $(-2,-1,m_z)$ directions. The specific value of the $z$-component $m_z$ ($\ne 0$) is inessential as it does not affect the SSG.
To demonstrate how to obtain the SSG of this novel magnetic structure, we divide the transition from coplanar to non-coplanar order into two stages. In the first stage, the coplanar magnetic moments in Fig.\ref{Fig:real_material}(c) are rotated by $\frac{\pi}{2}$ in $z$-direction in spin space, which leaves the SSG (i.e., 194.1.6.1.P) unchanged, as the spin and real space are unlocked. We remark that in SSGs the axes of the spin space can be chosen arbitrarily, while in MSG can not.
In the second stage, the magnetic moments obtain a $z$-component of equal magnitude, leading to a non-coplanar order that breaks the coplanar spin-only group symmetries.
The SSG is then identified as 194.1.6.1, with generators of the quotient group being $\{3^+_{001}||3^-_{001}\}$ and $\{M_{210}||M_{010}\}$ (note that $\{2_{010}||M_{010}\}$ is broken).
This non-coplanar SSG has the same number of operations as the coplanar SSG 194.1.6.1.P by breaking the coplanar spin-only group, while the spin rotations undergo a spin space coordinate transformation of $\frac{\pi}{2}$ rotation in $z$-direction.
The MSG of this non-coplanar structure is 12.62 $C2^\prime /m^\prime$, with generators being inversion and $\{M_{100}||2_{100}\}$. It can be seen that this MSG contains a significantly smaller number of operations compared with the SSG 194.1.6.1.

We remark that in these four examples, all SG operations (obtained by ignoring magnetic moments) are maintained in the SSGs by assigning proper spin rotations, and the first number in the label of the identified SSG (i.e., $N_{\text{SG}}$ in $N_{\text{SG}}.I_k.I_t.N_{\text{3Drep}}$) is the same as the SG. However, this not necessarily holds for all magnetic structures. For example, consider a material with randomly generated magnetic moments. Then no proper spin rotation could be assigned to SG operations and thus the SSG has only the identity operation.

\section{Applications of SSGs}\label{Sec:applications}

In this section, we discuss potential applications of SSGs. It contains four main parts: (1) the representation theory in SSGs with band representations of \ch{Mn3Sn} as a concrete example; (2) topological states protected by SSG symmetries; (3) Spin texture structures under SSGs; and (4) refining neutron scattering patterns using SSG symmetries.

\subsection{Representation theory in SSGs}\label{Sec: rep_theory}

The representation theory of SGs and MSGs has been pivotal in advancing the study of materials, including the topological quantum chemistry\cite{bernevig17, bernevigmtqc21} and symmetry-based indicators\cite{po2017symmetry, song2018quantitative, peng2022topological} for diagnosing topological crystalline phases.
The introduction of SSGs significantly expands the symmetry landscape, potentially unveiling a richer array of physical phenomena in magnetic materials, particularly those characterized by weak spin-orbit coupling (SOC).
In Ref.\cite{YangLiuFang2021}, the authors (including several authors of the current manuscript) study the co-representation of non-coplanar SSGs with supercell k-index $I_k=2$, where the little co-group is $P\times Z_2^{T}$, with $P$ being one of 32 crystallographic point groups and $Z_2^T= \{E, \mathcal T\}$. A 12-fold fermion and
13 Dirac nodal lines nexus are discovered which are topological band nodes that can only be realized in SSGs.
In Ref.\cite{PhysRevX.14.031038}, the representation theory in SSG is studied using the Complete Sets of Commuting Operators (CSCO).

In the following, we give a brief introduction to the representation theory in SSGs, with a more thorough investigation left in future work. We use \ch{Mn3Sn} as an example and compute the irreducible representations (IRREPs) for its electronic bands. We show the superiority of SSGs over MSGs by correctly capturing the multiple high-dimensional degeneracy in the band structure of \ch{Mn3Sn}.

\subsubsection{Projective representations in SSGs}
We first provide a brief introduction to the projective co-representation theory in SSGs. We restrict ourselves to the spinful representations for electron systems for the little groups in the BZ. The spinless representations can also be constructed similarly.

To begin with,
denote the representation of an SSG operation as $D(g)$ if $g$ is unitary, and $D(g)\kappa$ if anti-unitary, where $D(g)$ is the representation matrix and $\kappa$ the complex conjugation operator that satisfies $\kappa^2=1$ and $\kappa u = u^* \kappa$ for any matrix $u$. Define the multiplicity relationship of the representation matrices as
\begin{equation}
D(g_1)D(g_2)^{s(g_1)}=\omega(g_1,g_2)D(g_3),
\end{equation}
where $D(g_2)^{s(g_1)}$ indicates $D(g_2)$ when $g_1$ is unitary, and $D(g_2)^*$ when $g_1$ is anti-unitary. The factor $\omega(g_1,g_2)$ is a complex number with unit modulus, and the collection of $\omega(g_1 \in G, g_2 \in G)$ constitutes the factor system of the group, which should follow the equation
\begin{equation}
    \omega(g_1, g_2) \omega(g_1g_2, g_3) = \omega(g_2,g_3)^{s(g_1)} \omega(g_1, g_2g_3).
    \label{factor anti}
\end{equation}

We consider the factor system of an SSG at a high-symmetry point $\bm{k}$ in the BZ of an SSG, which is constructed by two parts: the $\text{SU}(2)$ factor from spin rotations and the non-symmorphic translation factor.
The $\text{SU}(2)$ factor $\omega_1$ originates from the two-to-one homomorphic relationship between $\text{SU}(2)$ and $\text{SO}(3)$ matrices and can only take the values of 1 or $-1$.
The non-symmorphic translation factor $\omega_2$, however, is only non-trivial at the non-$\Gamma$ momentum of the BZ in non-symmorphic SSGs. For two group elements $g_1=\{U_1||R_1|\bm{\tau}_1\}$ and $g_2=\{U_2||R_2|\bm{\tau}_2\}$, their translation factor is chosen as
\begin{equation}
    \omega_2(g_1,g_2) = e^{-i \bm{K_1} \cdot \bm{\tau_2}},\; \bm{K_{1}} =s(g_1)( g_1^{-1} \bm{k} - \bm{k}),
\end{equation}
where $s(g_1) =1 \text{ or } -1$ when $g_1$ is unitary or anti-unitary, respectively.
It can be proven that the factor system, composed of these two parts, satisfies the combination relations required for projective representations. In the representation theory of SSGs, fixing the choice of this factor system can avoid the arbitrariness of the overall phase in the representation.

Based on the method, we construct the factor system for the little group of SSG at all high-symmetry points. The factor system is crucial for the representation of groups.
We remark that although an SSG is isomorphic to an MSG\cite{PhysRevX.14.031037} when governed by the same multiplication relations and anti-unitary parts, they do not share the same factor system. Thus the IRREPs in SSGs are in general different from those in MSGs. This subtlety can be exemplified by the difference between the single- and double-group representations of an SG, where the double group has non-trivial spin factor systems.

With the multiplication relations and factor systems of SSGs derived, the IRREPs can then be constructed. In particular, we first consider the regular anti-unitary projective corepresentations. Regular representations are in general reducible. It can be proven that such regular representations always encompass all IRREPs. Moreover, there are mathematical methods for reducing group representations, including the use of Complete Sets of Commuting Operators (CSCO)\cite{CSCO_method,PhysRevX.14.031038} and Hamiltonian methods\cite{yang_hamiltonian_2021}. By following the Hamiltonian method, we construct all the irreducible representations of the little groups at all high-symmetry points for SSGs, which we leave in future work.

\subsubsection{Band structure and IRREPs of \ch{Mn3Sn}}

\begin{figure}[htbp]
\centering
\includegraphics[width=0.48\textwidth]{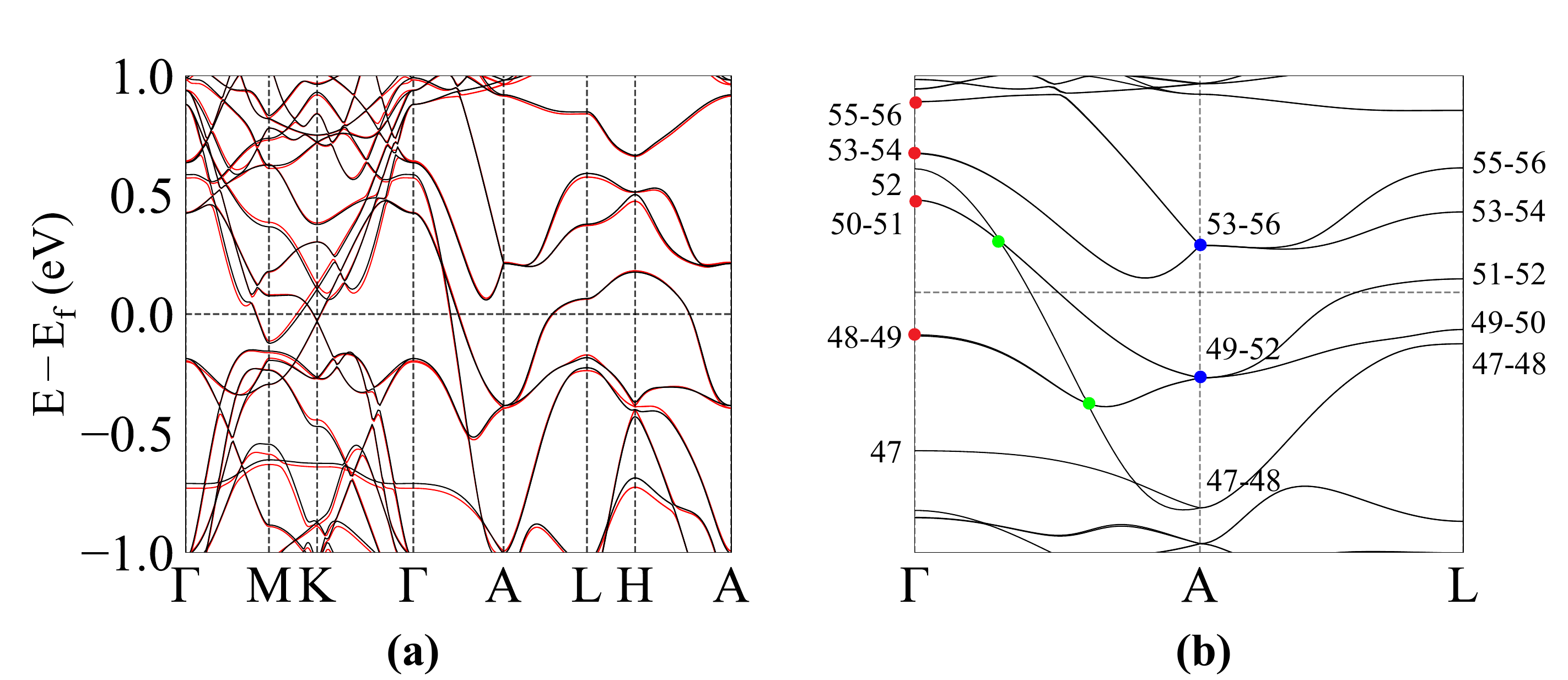}
\caption{\label{Fig:band_representations} Band structure and degeneracy of \ch{Mn3Sn}.
(a) Band structure of \ch{Mn3Sn} near the Fermi level $E_f$, where the red lines represent the bands with SOC and the black lines represent the bands without SOC. SOC only has minor effects on the band structure of \ch{Mn3Sn}.
(b) Enlarged band structure near $E_f$ of \ch{Mn3Sn} along the $\Gamma-A-L$ path for bands 47-56.
Two-fold degeneracies (red dots) at $\Gamma$ and four-fold degeneracies (blue dots) at $A$ are observed. There are also 3D crossing points (green dots) along the $\Gamma-A$ path. These high-dimensional degenerate points can only be captured by the representation theory of SSGs (see \cref{tab:band representations at gamma point of ssg} and \cref{tab:ssg character A point}), but cannot by MSG.}
\end{figure}

We use \ch{Mn3Sn} as an example to show the application of IRREPs of SSGs in the electronic band structures. We consider the coplanar-ordered \ch{Mn3Sn}. Its band structure with and without SOC is shown in \cref{Fig:band_representations}(a). SOC only has minor effects on the bands of \ch{Mn3Sn}, confirming its weak SOC nature and justifying the usage of SSGs.

In \cref{Fig:band_representations}(b), we show the enlarged bands near $E_f$ along $\Gamma-A-L$.
We first use MSG to compute the IRREPs. The coplanar-ordered \ch{Mn3Sn} has MSG 63.463 $Cmc^\prime m^\prime$, which is generated by $\{E||\mathcal{P}\}$, $\{2_{010}||M_{010}\}$, and $\{2_{001}||M_{001}|\frac{c}{2}\}\cdot\mathcal{T}$.
At $\Gamma$, MSG 63.463 only has 1D (double) IRREPs, which cannot explain the 2D degeneracies at $\Gamma$. We use \textit{symtopo}\cite{Symtopo} to identify the characters of unitary operations of these bands, with results summarized in Supplementary Material \cref{sec:app:Mn3Sn_irrep}.

Next, we consider SSG 194.1.6.1.P of coplanar \ch{Mn3Sn}. This SSG has a unitary generator $\{3^{-}_{001}||3^+_{001}\}$ and a pure lattice operation $\{E||2_{001}|\frac{z}{2}\}$ besides the three generators in MSG $Cmc^\prime m^\prime$. Utilizing the method described in the previous section for constructing anti-unitary projective representations, we obtain all the irreducible representations of this group. At the $\Gamma$ point, 194.1.6.1.P has 12 inequivalent IRREPs, including 8 one-dimensional and 4 two-dimensional IRREPs. The character table is presented in Supplementary Material \cref{sec:app:Mn3Sn_irrep}.
With the character table of the SSG, we can successfully label each group of bands using the group's irreducible representations as shown in Table~\ref{tab:band representations at gamma point of ssg}. We add a 'S' prefix to the labels of these IRREPs to distinguish them from the IRREPs of the MSG

\begin{table}[htbp]
\begin{tabular}{c|c|c|c|c|c|c}
\hline\hline
bands & \;$E$\; & \;$g_1$\; & $\;g_2\;$& $\;g_3\;$ & $\;g_4\;$ & IRREPs \\\hline
47    & 1        & 1 & i &-1&-1& $^S \bar{\Gamma}_{3}\rule{0pt}{2.6ex}$             \\\hline
48+49 &2           & -2 & 0&1&2 & $^S \bar{\Gamma}_{11}\rule{0pt}{2.6ex}$           \\\hline
50+51 &2          & 2 & 0&1&-2 & $^S \bar{\Gamma}_{10}\rule{0pt}{2.6ex}$         \\\hline
52    &1        & -1 & i&-1&1 & $^S \bar{\Gamma}_{6}\rule{0pt}{2.6ex}$     \\\hline
53+54 &2          & -2 & 0&1&2 & $^S \bar{\Gamma}_{11}\rule{0pt}{2.6ex}$        \\\hline
55+56 &2          & 2 & 0&1&2 & $^S \bar{\Gamma}_{12}\rule{0pt}{2.6ex}$        \\
\hline\hline
\end{tabular}
\caption{\label{tab:band representations at gamma point of ssg} The band representations of \ch{Mn3Sn} at $\Gamma$ using IRREPs of SSG 194.1.6.1.P. Operators $g_1 = \{E||\mathcal{P}\}$, $g_2 = \{2_{010}||M_{010}\}$, $g_3 = \{3^{-}_{001}||3^+_{001}\}$ and $g_4 = \{E||2_{001}|\frac{z}{2}\}$ are generators of SSG. The 2D degenerate points cannot be captured by IRREPs of the MSG.}
\end{table}

At the boundaries of BZ, non-symmorphic operations could lead to higher degeneracies. As can be seen in \cref{Fig:band_representations}(b), the bands at the $A$ point exhibit both two-fold and four-fold degeneracies. The MSG $Cmc^\prime m^\prime$ only has IRREPs of up to 2D at $A$, thus failing to explain the 4D degeneracy. This necessitates the introduction of SSG. Compared with the little group at $\Gamma$, the little group at $A$ has the same group elements but possesses inequivalent factor systems. The character table of IRREP at $A$ is listed in \cref{tab:ssg character A point}, showing the existence of one 2D and two 4D IRREPs. This gives a satisfying match to the DFT band structure.
Along the $A-L$ path, the SSG only has 2D irreducible representations, which is also consistent with the behavior of the bands along this path in \cref{Fig:band_representations}(b).

Moreover, along the $\Gamma-A$ path, two three-fold degeneracies are observed, protected by the symmetries of the SSG. While the MSG only has 1D representations along this path, SSG has four 1D and two 2D representations. Therefore, these two three-fold degeneracies, resulting from the crossing of doubly degenerate bands with non-degenerate bands, are uniquely captured by the SSG.

We remark that the MSG of a specific material is always a subgroup of its SSG, thus the IRREPs of the SSG can be decomposed into IRREPs of the MSG, either reducible or irreducible.
For instance, the $^S\bar{\Gamma}_{11}$ IRREP in the SSG is reduced to $\bar{\Gamma}_5, \bar{\Gamma}_6$ in the MSG.
With the IRREPs of MSG, the 2D, 3D, and 4D degeneracies in \ch{Mn3Sn} can only be explained as accidental (except for the 2D IRREPs at A). Only with SSG can we faithfully identify the high-dimensional degeneracies in the band structures of magnetic materials with weak SOC.

\begin{table}[htbp]
\begin{tabular}{c|c|c|c|c|c}
\hline\hline
IRREPs & $\;E\;$ & $\{E||\mathcal{P}\}$ & $\{2_{010}||M_{010}\}$ & $\{3^{-}_{001}||3^+_{001}\}$ & $\{E||2_{001}|\frac{z}{2}\}$ \\\hline
$^S \bar{A}_{1}\rule{0pt}{2.6ex}$            & 2 & 0 & 2i & -2 & 0             \\\hline
$^S \bar{A}_{2}\rule{0pt}{2.6ex}$           & 2 & 0 & -2i & -2 &  0           \\\hline
$^S \bar{A}_{3}\rule{0pt}{2.6ex}$           & 4 &  0 & 0 & 2 &  0  \\
\hline\hline
\end{tabular}
\caption{\label{tab:ssg character A point} Character table of SSG 194.1.6.1.P at $A$ point. In the band structure of \ch{Mn3Sn} shown in \cref{Fig:band_representations}(b), bands 47-48 have 2D $^S \bar{A}_{1}$ IRREP, while band 49-52 and 53-56 share 4D $^S \bar{A}_{3}$ IRREP.}
\end{table}

\subsection{Example of topological states protected by SSG}

In this section, we propose two new gapped topological states protected by SSG symmetries. We explicitly construct the IRREPs in these two SSGs and identify the topological surface states protected by the bulk topological states. We show that the non-trivial spin rotations in SSGs have important consequences in the IRREPs and topology.

Before starting, we discuss briefly some key aspects of the \textit{real-space recipe}\cite{song2018quantitative, song2019topological, song2020real, peng2022topological} in SSGs, which could be used to construct the complete topological classifications.
The \textit{real-space recipe} converts the problem of topological classification (i.e., exhausting all inequivalent topological states for a given symmetry group) into a ``LEGO'' puzzle, where one uses lower-dimensional topological building blocks to construct (gapped) 3D topological states. The lower-dimensional building blocks have gapless edge states, which need to be combined properly so that the resultant 3D states are gapped (i.e., the no open-edge condition). The obtained 3D states are subject to the so-called ``bubble equivalence'', a process that removes equivalent states. A simplified case in the \textit{real-space recipe} is the \textit{layer construction}\cite{song2018quantitative}, where the 3D topological states are built from 2D infinitely large layers. Generic 3D topological states are formed by small pieces of finite lower-dimensional states.

We then discuss some unique properties of the \textit{real-space recipe} in SSGs and compare those in MSGs.
First, unlike MSGs where only 2D Chern or mirror Chern insulators can be used as the lower-dimensional building blocks, SSGs could also host 2D topological insulators (TIs)\cite{Liu2022prx, PhysRevX.14.031037} protected by effective TRS symmetry $\mathcal{T}_M=\{E|| M|\bm{0}\}\cdot\mathcal{T}$, with $\mathcal{T}_M^2=-1$, where $M$ is a mirror reflection along a certain direction.
Second, there exists effective TRS on 1D lines in SSG, which could protect helical edge modes. These effective TRS include $\mathcal{T}_M=\{E||M|\bm{0}\} \cdot\mathcal{T}$ and $\mathcal{T}_{C_2}=\{E||C_2|\bm{0}\} \cdot\mathcal{T}$, which all square to -1.
Third, there exist two types of mirror symmetries in SSG, i.e., the pure mirror without spin rotation $\{E||M|\bm{0}\}$ which has eigenvalue $\pm1$, and the mirror with spin rotation $\{U_M||M|\bm{0}\}$ (where $U_M=M\cdot \mathcal{P}$ is a $C_2$ rotation), which has eigenvalue $\pm i$.
We leave a complete real-space construction of topological states in SSGs for further work.

In the following, we consider two simple but novel examples of gapped topological states in SSGs. We will start with a detailed discussion of the IRREPs of the SSG, which can be derived from the IRREPs of a corresponding MSG through a folding and shifting process of the BZ. The topological states in the SSG are then constructed using the \textit{real-space recipe}.

\subsubsection{SSG 1.4.1.2}

We first consider SSG 1.4.1.2 with generator $g=\{C_4||E|0,0,\frac{1}{4}\}\cdot\mathcal{T}$, where the translation is written under magnetic unit cell bases (i.e., $(0,0,1)$ is a lattice translation). We drive the (spinful) IRREPs in this SSG and then build a $\mathbb{Z}_2$ topological state protected by the anti-unitary operation $g$, similar to the antiferromagnetic topological insulator\cite{mong2010antiferromagnetic, fang2013topological}.

\paragraph{IRREPs.}  Based on the algorithm introduced in \cref{Sec: rep_theory}, we obtain the IRREPs at eight time-reversal-invariant momenta (TRIMs), with results summarizing in \cref{table: SSG_irreps_example}. There is only one 2D IRREP for TRIMs on the $k_z=0$ plane, while one 1D and one 2D IRREP for TRIMs on the $k_z=\pi$ plane. These IRREPs can be understood from the IRREPs of MSG 1.3 $P_S1$ with generator $g_M=\{E||E|0,0,\frac{1}{2}\}\cdot\mathcal{T}$. The unit cell of SSG 1.4.1.2 can be seen as a doubled one of MSG 1.3. In MSG 1.3, there are 2D IRREPs (i.e., Kramers' pairs) on $k_z=0$ TRIMs and 1D IRREPs on $k_z=\pi$ TRIMs. In the following, we show that the IRREPs of SSG 1.4.1.2 can be obtained from the IRREPs of MSG 1.3 through a folding process of the BZ, together with a subtle shifting of BZ in the $k_z$ direction resulting from the non-trivial spin rotation.

Define $h=\{C_2||E|0,0,\frac{1}{2}\}$ which have representation matrix satisfying $D(h)=-D(g^2)$, where the minus sign comes from $\mathcal{T}^2=-1$. This unitary operation $h$ can serve as a new translation group generator, as $h$ commutes with $g$ and all other translations. Denote the BZ defined from $h$ as mBZ. In the mBZ, the TRIMs on $k_z=0$ plane have $D(g^2)=-1$ which give Kramers' pair, while on $k_z=\pi$ plane $D(g^2)=+1$ and gives only 1D IRREPs. This is the same result as in MSG 1.3.

Then we turn to the real BZ of the SSG defined by $\{E||E|0,0,1\}$. The SSG BZ is the folded version of the mBZ. Notice that $D(h^2)=-D(\{E||E|0,0,1\})$, where the minus sign comes from the spin rotation as $U_{C_2}^2=-1$ ($U_R$ denotes the corresponding SU(2) spin rotation of $R$). We consider TRIMs on $k_z=0$ and $\pi$ planes separately:
\begin{itemize}
\item For TRIMs with $k_z=\pi$, we have $D(h^2)=+1$, as $e^{ik_z}=-1$ when $k_z=\pi$. In the mBZ, all the TRIMs on $k_z=0,\pi$ planes satisfy $D(h^2)=+1$ and are mapped to the $k_z=\pi$ TRIMs in the SSG BZ. Thus the 1D and 2D IRREPs coexist for the $k_z=\pi$ TRIMs. They are characterized by $D(h)=+1$ (for 2D IRREP) and $D(h)=-1$ (for 1D IRREP), respectively. The eigenvalue of $h$ separates the states on the $k_z=\pi$ plane into two independent sectors. The 2D IRREP sector could protect $\mathbb{Z}_2$ topological flow\cite{fu2007topological_PRL, fu2007topological_PRB} and give rise to the surface Dirac cone, while the 1D IRREP sector is an `audience' that does not hybridize with the states having the 2D IRREP.

\item For TRIMs with $k_z=0$, we have $D(h^2)=-1$ and thus $D(h)=\pm i$. This corresponds to the $k_z=\pm\frac{\pi}{2}$ planes in the mBZ, which only host 1D IRREPs. After folding into the SSG BZ, 2D IRREPs are formed at TRIMs with $k_z=0$, as $k_z=\pm\frac{\pi}{2}$ planes are related by $g$. Moreover, these 2D IRREPs on the $k_z=0$ TRIMs in the SSG BZ must be part of a nodal line, as a result of folding from mBZ.
\end{itemize}

The Dirac cone and nodal lines in the SSG BZ can also be verified from the $\bm{k}\cdot\bm{p}$ theory.
For the 2D IRREPs on TRIMs with $k_z=0$ (denote as $D_{k_z=0}^{2D}$) and $k_z=\pi$ (denote as $D_{k_z=0}^{2D}$), their representation matrix can be chosen as
\begin{equation}
D_{k_z=0}^{2D}(g)=
\left(\begin{array}{cc}
0 & 1 \\
-i & 0
\end{array}\right), \quad
D_{k_z=\pi}^{2D}(g)=
\left(\begin{array}{cc}
0 & 1 \\
-1 & 0
\end{array}\right)
\end{equation}
The $\bm{k}\cdot\bm{p}$ effective Hamilontians constructed from these representation have the form:
\begin{itemize}
\item For a TRIM on the $k_z=0$ plane, the $\bm{k}\cdot\bm{p}$ Hamiltonian has the form $(a_1 \delta k_1 + a_2 \delta k_2 + a_3  \delta k_3)\sigma_z$, where $\delta \bm{k}=\bm{k}-\bm{k}_0$ is the deviation from TRIM $\bm{k}_0$, and $a_{i=1,2,3}$ are free parameters. A nodal line is given by $a_1 \delta k_1 + a_2 \delta k_2 + a_3  \delta k_3=0$ that passes the TRIM.
\item For a TRIM on the $k_z=\pi$ plane, the $\bm{k}\cdot\bm{p}$ Hamiltonian has 9 independent terms, i.e., all combinations between $\delta k_{i=1,2,3}$ and $\sigma_{i=x,y,z}$. Thus linear Dirac crossings are allowed.
\end{itemize}

We also briefly mention the IRREPs in spinless systems, e.g., magnon bands. For spinless IRREPs in the mBZ, 2D Kramers pairs appear on the $k_z=\pi$ TRIMs, as $\mathcal{T}^2=+1$. Then $h^2=\{E||E|0,0,1\}$ in the SSG BZ. Thus on the $k_z=0$ TRIMs, there are both 2D Kramers' pairs and 1D IRREPs, while on the $k_z=\pi$ TRIMs there are 2D IRREPs from folding $k_z=\pm\frac{\pi}{2}$ in the mBZ. It can be seen that the IRREPs on $k_z=0$ and $\pi$ planes are reversed for spinless and spinful systems.



\begin{figure}[tbp]
    \centering
    \includegraphics[width=0.5\textwidth]{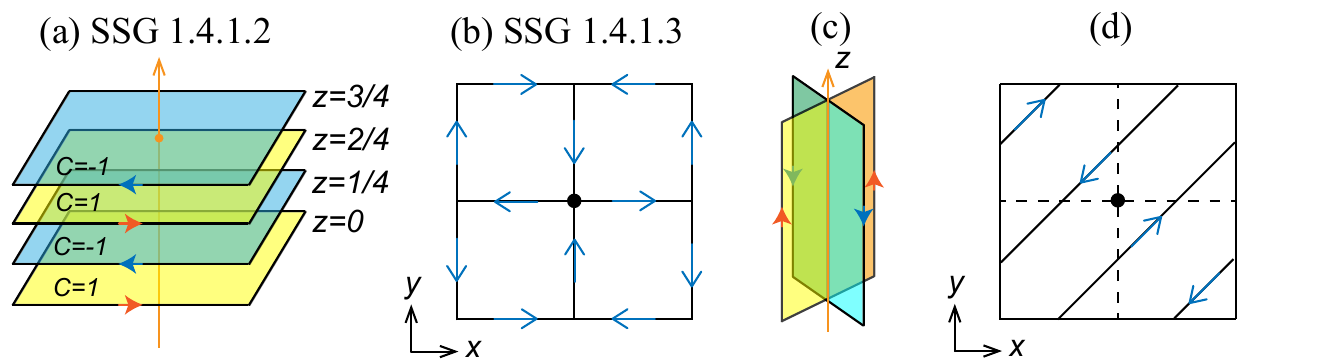}
    \caption{(a) A layer construction protected by SSG 1.4.1.2. (b)(c) A non-layer construction protected by SSG 1.4.1.3. (b) is from the top view, where arrows denote the direction of chiral edge modes when the boundary is open.
    (c) is from the side view, in which we only show the four Chern patches inside a unit cell and omit the patches on the boundary.
    (d). The layer construction deformed from the non-layer construction in (b), which has four Chern layers in a unit cell with alternating Chern numbers.
    }
    \label{fig:SSG_topo_state}
\end{figure}

\paragraph{Topological states.} This SSG protects a layer construction (LC) with a $\mathbb{Z}_2$ classification, as shown in \cref{fig:SSG_topo_state}(a). This LC (denoted as $LC_S$) consists of Chern layers with Chern number $C=1$ on $z=\frac{2n}{4}$ planes and Chern layers with $C=-1$ on $z=\frac{2n+1}{4},n\in\mathbb{Z}$ planes. Adjacent layers are related by $g$ and thus have opposite Chern number. This state is topological because the Chern layers can be deformed but cannot be eliminated as long as $g$ is maintained. Its doubled state, however, can be trivialized as Chern layers with $C=\pm 1$ can then be moved together and canceled, leading to the $\mathbb{Z}_2$ classification.

$LC_S$ can be understood from the layer construction of MSG 1.3 $P_S1$ (denoted as $LC_M$) with generator $g_M=\{E||E|0,0,\frac{1}{2}\}\cdot\mathcal{T}$. $LC_M$ has two Chern layers with Chern number $C=\pm1$ on $z=0,\frac{1}{2}$ planes in the unit cell.
In MSG 1.3, TRIMs on the $k_z=0$ plane host Kramers' pairs. For an open boundary along the $z$ direction that preserves $g_M$, a single Dirac (or more generally, an odd number) cone will appear on TRIMs with $k_z=0$ in the surface BZ, which can be verified from the 2D IRREPs on $k_z=0$ TRIMs.
$LC_S$ can be seen as the doubled $LC_M$ along the $z$ direction.
As $LC_M$ holds a single Dirac cone in the surface BZ, $LS_S$ also hosts a single Dirac cone. Doubling of the unit cell leads to the folding of the BZ in $LC_M$, which maintains the number of surface Dirac cones.

Remark that similar topological states also exist in SSGs generated by $\{C_{2n}||E|0,0,\frac{1}{2n}\}\cdot\mathcal{T}, n\in\mathbb{Z},n\ge 1$, which host a single surface Dirac cone.

\begin{table}[tbp]
\centering
\begin{tabular}{c|c|cc}
\hline\hline
SSG 1.4.1.2 & $k_z=0$ TRIMs
& \multicolumn{2}{c}{$k_z=\pi$ TRIMs}  \\ \hline
IRREP      &
\begin{tabular}[c]{@{}c@{}}one 2D \\
(nodal line)\end{tabular} &
\multicolumn{2}{c}{\begin{tabular}[c]{@{}c@{}}one 1D, \\
one 2D (Dirac point)\end{tabular}}
\\ \hline
SSG 1.4.1.3 & $(0,0)$   & \multicolumn{1}{c|}{$(\pi,\pi)$}    & $(0,\pi),(\pi,0)$      \\ \hline
IRREP  & two 1D  &
\multicolumn{1}{c|}{\begin{tabular}[c]{@{}c@{}}two 2D\\ (Dirac point)\end{tabular}} &
\begin{tabular}[c]{@{}c@{}}one 2D \\
(nodal line)\end{tabular}
\\ \hline\hline
\end{tabular}%
\caption{\label{table: SSG_irreps_example}
The summary of irreducible representations (IRREPs) at time-reversal-invariant momenta (TRIMs) in SSG 1.4.1.2 and SSG 1.4.1.3. In the table, ``Dirac point'' means the 2D IRREP forms a Dirac crossing point, while ``nodal line'' means the 2D IRREP is part of a nodal line in the BZ. In SSG 1.4.1.3, the $k_z$ is omitted in the coordinate of TRIMs. }
\end{table}

\subsubsection{SSG 1.4.1.3}
In the second example, we consider SSG 1.4.1.3, with two anti-unitary operations
\begin{equation}
    g_x\mathcal{T}=\{C_{2x}||E|\frac{1}{2},0,0\}\cdot\mathcal{T},\
    g_y\mathcal{T}=\{C_{2y}||E|0,\frac{1}{2},0\}\cdot\mathcal{T}
\end{equation}
This SSG also has unitary operations with non-trivial spin rotations:
$s_1=\{C_{2z}||E|\frac{1}{2},\frac{1}{2},0\}, s_2=\{C_{2z}||E|-\frac{1}{2},\frac{1}{2},0\}$.

\paragraph{IRREPs. }
We start from the IRREPs in SSG 1.4.1.3, with results summarizing in \cref{table: SSG_irreps_example}.
The unit cell basis in $xy$-plane of SSG 1.4.1.3 can be taken as $\bm{A}_1=(1,0), \bm{A}_2=(0,1)$. Denote the corresponding momentum in the SSG BZ as $(k_1, k_2)$. We have omitted the $z$-direction for simplicity.
In SSG 1.4.1.3, the representation of operations satisfy
\begin{equation}
\begin{aligned}
D((g_x\mathcal{T})^2) &= D(\{E||E|1,0,0\})= D(s_1)D(s_2^{-1})\\
D((g_y\mathcal{T})^2) &= D(\{E||E|0,1,0\})= -D(s_1)D(s_2)
\end{aligned}
\label{Eq: SSG1.4.1.3ops}
\end{equation}
\begin{equation}
\begin{aligned}
D(s_1^2) &=-D(\{E||E|1,1,0\})\\
D(s_2^2) &=-D(\{E||E|-1,1,0\})
\end{aligned}
\end{equation}
where minus sign comes from the spin rotation as $U_{C_{2z}}^2=-1$.
Similar to the previous SSG 1.4.1.2, we take $s_1, s_2$ as the translation group generators, and
denote the BZ thus defined as mBZ. The IRREPs of this SSG can also be built from the mBZ through a folding and shifting process.
A subtle difference here is these two operations do not commute with $g_{x,y}\mathcal{T}$ and will modify the coordinates of TRIMs and their IRREPs in the mBZ, which we show in the following.

We first identify the TRIMs in the mBZ. By definition, TRIMs are the momenta that are invariant under the anti-unitary operation (but not necessarily always have coordinate $k_i=0,\pi$). As
\begin{equation}
D(g_x\mathcal{T}) D(s_{i=1,2}) D(g_x\mathcal{T})^{-1}=-D^*(s_{i=1,2}),
\end{equation}
where the minus sign comes from the spin rotations, TRIMs must have $D(s_{1,2})=e^{i\bm{k}_m \cdot s_i}=\pm i$ (note here $s_i$ are treated as the translation group generators, and $\bm{k}_m$ is defined in the mBZ). Thus TRIMs have coordinate $\bm{k}_m=(\pm\frac{\pi}{2}, \pm\frac{\pi}{2})$ in the mBZ.
We then identify which TRIMs host Kramers' degeneracy by requiring $D((g_{x,y}\mathcal{T})^2)=-1$. Direct computation shows the TRIMs at $\pm(\frac{\pi}{2}, -\frac{\pi}{2})$ satisfy. The other two TRIMs have $D((g_{x,y}\mathcal{T})^2)=+1$ and thus only host 1D IRREP.

With IRREPs at TRIMs in the mBZ, we then construct those in the SSG BZ. As shown in \cref{Eq: SSG1.4.1.3ops}, we have
\begin{equation}
D(\bm{A}_1)=D(s_1)D(s_2^{-1}), D(\bm{A}_2)=-D(s_1)D(s_2).
\end{equation}
This relation is used to map the TRIMs between SSG BZ and the mBZ:
\begin{itemize}
\item For the $(0,0)$ TRIM, there are two 1D IRREPs. This is because at this TRIM, $D(\bm{A}_1)=D(\bm{A}_2)=e^{i\bm{k}_s\cdot \bm{A}_i}=+1$ ($\bm{k}_s$ is defined in the SSG BZ). Thus it can be mapped from the $\pm(\frac{\pi}{2}, \frac{\pi}{2})$ TRIMs in the mBZ, as $D(\bm{A}_1)=D(s_1)D(s_2^{-1})=e^{i\bm{k}_m\cdot s_1} e^{-i\bm{k}_m\cdot s_2}=1$ when $\bm{k}_m=\pm(\frac{\pi}{2}, \frac{\pi}{2})$ ($D(\bm{A}_2)$ is similar).
Since these two TRIMs in the mBZ each host one 1D IRREP, they give two 1D IRREPs at $(0,0)$ in the SSG BZ.

\item For the $(\pi, \pi)$ TRIM, there are two 2D IRREPs that protect Dirac crossings. Because at this TRIM, $D(\bm{A}_1)=D(\bm{A}_2)=-1$, and can be mapped from the $\pm(\frac{\pi}{2}, -\frac{\pi}{2})$ TRIMs in the mBZ, which host 2D Kramers' pairs.

\item For the $(0, \pi)$ TRIMs, there is one 2D IRREP coming from folding. This is because at this TRIM $D(\bm{A}_1)=+1, D(\bm{A}_2)=-1$, and can be mapped from the $(0,0), (\pi,\pi)$ in the mBZ. In the mBZ, $(0,0), (\pi,\pi)$ are generic points with only 1D trivial IRREP, and are related by $g_{x,y}\mathcal{T}$. The 2D IRREP at $(0, \pi)$ in the SSG BZ cannot support Dirac crossings but must be part of a nodal line due to the folding of BZ.

\item The $(\pi, 0)$ TRIM is similar to $(0, \pi)$, which host one 2D IRREP from folding the 1D IRREPs at the $(0,\pi),(\pi,0)$ in the mBZ.
\end{itemize}

\paragraph{Topological states. }
SSG 1.4.1.3 protects a $\mathbb{Z}_2$ non-layer construction, as shown in \cref{fig:SSG_topo_state}(b)(c). This non-layer construction is an axion insulator with a vanishing net Chern number, protected by the antiunitary SSG translations $g_{x,y}\mathcal{T}$. It can be deformed into a layer construction as shown in \cref{fig:SSG_topo_state}(d). This state is topological because the Chern layers can be deformed but cannot be removed when $g_{x,y}\mathcal{T}$ is maintained. Two copies of the states are topologically trivial as the Chern layers can then be eliminated, validating the $\mathbb{Z}_2$ classification.

We then consider the surface states of this topological state. Surface Dirac cones could only appear on surface BZ that supports the partner-switching $\mathbb{Z}_2$ topological flow of the surface Dirac cone\cite{fu2007topological_PRL, fu2007topological_PRB}, which normally requires an even number of TRIMs that can host Kramers' pair.
From the IRREPs, the $xz$ or $yz$ surface hosts two TRIMs with Kramers' pairs and thus can host a single (or odd number of) surface Dirac cone. The $xy$ surface has only one TRIM at $(\pi, \pi)$ that hosts Kramers' pair. Surprisingly, this surface can still protect the topological surface Dirac cone, as the mBZ has two TRIMs with Kramer's pair on this surface and supports the $\mathbb{Z}_2$ flow. The folding process will not change the topological nature of the flow. In the SSG BZ, the topological $\mathbb{Z}_2$-flow starts from a Dirac cone at $(\pi, \pi)$ and connects to the nodal line (which passes $(\pi,0)$ and $(0,\pi)$), and then goes bask to $(\pi, \pi)$. This surface state is unique in SSG, resulting from the modified group structure given by the spin rotations.

\subsection{Spin textures}

The spin texture $\bm{S}(\bm{k})$ is defined as the expectation value of the spin operator on a Bloch state\cite{PhysRevX.14.031037}, i.e., $\bm{S}(\bm{k})=\langle \psi_{\bm{k}}| \bm{\sigma} |\psi_{\bm{k}} \rangle $, which is usually defined on the Fermi surface. Ref.\cite{PhysRevX.14.031037} gives a detailed discussion on the properties of the spin textures based on the SSG BZ. Motivated by Ref.\cite{PhysRevX.14.031037}, we give a brief discussion on the symmetry properties of the spin textures in SSGs, focusing on the dimension of the spin textures constraint by the SSG operations.

For an SSG operation $g=\{U_R||R|\bm{\tau}\}$, the spin texture $\bm{S}(\bm{k})$ transform as
\begin{equation}
g\bm{S}(\bm{k})=
U_R \bm{S}(s_g R^{-1}\bm{k})
\end{equation}
where $s_g=\pm1$ for $\det(U_R)=\pm1$, i.e., when $\det(U_R)=-1$, $g$ is antiunitary and reverse $\bm{k}$.
We first consider the effects of spin-only group $\mathcal{S}_0$ on $\bm{S}(\bm{k})$.
\begin{itemize}
\item For collinear SSGs, the spin rotation $\{C_{\theta}||E|\bm{0}\}\in\mathcal{S}_0$ enforces $\bm{S}(\bm{k})=(0,0,S_z(\bm{k}))$ to be collinear. When the SSG also has $\mathcal{P}\mathcal{T}$ or antiunitary translation $\mathcal{T}\cdot\bm{\tau}$, spin degeneracy is guaranteed over the whole BZ, which will enforce $\bm{S}(\bm{k})=\bm{0}$. Without these two types of symmetries, spin splitting is expected and the magnetic order is described as \textit{altermagnetism}
\cite{smejkal2021altermagnetism, TJungwirth2022, mazin2022altermagnetism}
recently, which allows for non-trivial spin texture.

\item For coplanar SSGs, $\{M_z||E|\bm{0}\}\sim\{C_{2z}||E|\bm{0}\}\cdot\mathcal{T}\in\mathcal{S}_0$ enforces $\bm{S}(\bm{k})=M_z \bm{S}(-\bm{k})$, and $\bm{S}(\bm{k})$ vanishes on TRIMs.
\item  For the non-magnetic case, $\mathcal{S}_0=\text{O}(3)$ will enforce $\bm{S}(\bm{k})=\bm{0}$. Remark that in the presence of SOC, the $\text{O}(3)$ spin-only group is broken and the spin texture is in general non-vanishing.
\end{itemize}

We then consider general SSGs. The following SSG operations will reduce the dimension of $\bm{S}(\bm{k})$:
\begin{itemize}
\item When the SSG contains $\{P||P|\bm{0}\}\sim\mathcal{P}\mathcal{T}$, $\bm{S}(\bm{k})$ is enforced to be zero.
\item When the SSG contains $\{C_n||E|\bm{\tau}\}$, we have $\bm{S}(\bm{k})=C_n\bm{S}(\bm{k})$ and thus $\bm{S}(\bm{k})$ must be collinear along the $C_n$ direction.
\item When the SSG contains $\{M||P|\bm{0}\}$ where $M$ is a mirror, we have $\bm{S}(\bm{k})=M\bm{S}(\bm{k})$ and thus $\bm{S}(\bm{k})$ must be coplanar on the mirror plane.
\end{itemize}

We then argue that the non-crystallographic spin rotations in SSGs will enforce $\bm{S}(\bm{k})$ to be collinear to vanishing.
For SSGs with non-crystallographic spin rotations, there always exist spin rotations $C_p\ (p\ne 2,3,4,6)$. We argue that in such SSG $\mathcal{S}$, there exists at least one operation with the form $g=\{C_n||E|\bm{\tau}\}$, i.e., a non-crystallographic spin rotation accompanied by a pure translation. This operation enforces the collinear spin texture as discussed in the previous paragraph.
We then prove the existence of $g$. Suppose the non-crystallographic $C_p$ spin rotation in $\mathcal{S}$ has some real-space rotation $R$ and translation $\bm{\tau}$ part, i.e., $h=\{C_p||R|\bm{\tau}\}$.
As $R$ can only be crystallographic with rank $m=1,2,3,4,6$, we have $h^m=\{C_p^m||E|\bm{\tau}^\prime\}$. This operation has the same form as $g$, i.e., a non-crystallographic spin rotation $C_p^m$ with no real-space rotation but only a translation. The translation part must be nonzero because otherwise, $h^m$ belongs to the spin-only group $\mathcal{S}_0$, which contradicts the assumption that the spin-only group and the spin-only free group only have a trivial intersection.

\subsection{Refine magnetic neutron diffraction patterns with SSG}

In this section, we consider the application of SSGs to facilitate the refinement of magnetic orderings from magnetic neutron diffraction. We first review the refinement algorithm based on MSGs in the literature\cite{rodriguez2001fullprof, wills2000new} and then extend it to SSGs. We use \ch{Mn3Sn} as a realistic example to demonstrate the advantage of SSG in the refinement by reducing the number of refining parameters.

\subsubsection{Algorithm of refinement}

In literature, a commonly used software for refining neutron diffraction is \textit{FullProf}\cite{rodriguez2001fullprof}, which has a representational analysis module `simulated annealing and representational
analysis' (\textit{SARAh})\cite{wills2000new}. \textit{SARAh} could list all possible MSGs and corresponding magnetic structures for a given non-magnetic SG and propagation vector.
Our algorithm based on SSG has the same spirit and could be integrated into their workflow by extending MSGs to SSGs.

We start with the necessary experimental information that can be measured before refining magnetic structures: (i) The non-magnetic SG $\mathcal{G}$ of the material refined based on X-ray diffraction (XRD) measurements in the magnetic phase, or neutron diffraction measurement in the paramagnetic phase if assuming there is no atomic displacement during the magnetic transition. $\mathcal{G}$ gives the symmetries of the crystal structure by ignoring the magnetic moments.
(ii) The magnetic unit cell (propagation vector) obtained from new Bragg peaks shown in neutron diffraction patterns compared with the paramagnetic pattern, which could either be the same as the non-magnetic unit cell (propagation vector $\bm{q}=\bm{0}$) or an enlarged supercell (propagation vector $\bm{q}\ne \bm{0}$).
(iii) The neutron diffraction patterns used for fitting in the refinement.
(iv) The net magnetic moment and its direction. Zero net magnetic moment indicates the AFM order.
One can also distinguish collinear (together with the direction of moments), coplanar non-collinear (together with the plane of moments), and non-coplanar magnetic configurations by for example measuring anisotropic magnetization, e.g., magnetic susceptibility measurements on single crystals.

With the aforementioned experimental data, the following algorithm can be adopted for refinement based on MSG\cite{rodriguez2001fullprof, wills2000new}:
\begin{enumerate}
\item List all MSGs $\mathcal{M}$ that are compatible with the non-magnetic SG $\mathcal{G}$ and magnetic unit cell (propagation vector).
In the case of multiple propagation vectors, the magnetic unit cell can also be constructed, where different combination coefficients of propagation vectors give different symmetry groups.

\item For each $\mathcal{M}$, find the Wyckoff positions of the magnetic atoms and the symmetry-allowed components of magnetic moments. The moment could be (i) free (non-coplanar), parameterized by $(m_x,m_y,m_z)$; (ii) restricted on a certain plane (coplanar non-collinear), for example, on $(m_x, m_y, 0)$; or (iii) restricted along a certain direction (collinear), for example, along $(0,0,m_z)$.
This step can also be done mathematically by decomposing the representations of the little group $G_{\bm{q}}$ operations on the magnetic moments into IRREPs of $G_{\bm{q}}$, and the bases of these IRREPs gives symmetry-independent magnetic moments\cite{rodriguez2001fullprof}.

\item Parameterize the magnetic configuration using the symmetry-independent magnetic atoms. Assume there are $N$ such atoms. Then the number of fitting parameters is between $N$ and $3N$.
Fit the parameters to the neutron data and find the magnetic configuration with minimal error.
\item If the fitting errors are large, the previous steps are repeated by considering the subgroups of $\mathcal{G}$ and their corresponding MSGs, as magnetic order could break the symmetry.
Specially, when the propagation vector already breaks some symmetries in $\mathcal{G}$, we could start from the corresponding subgroup at the beginning.
\end{enumerate}
The refinement algorithm based on MSGs can be extended to SSGs straightforwardly. SSGs have the advantage of richer symmetry operations compared with MSGs and thus could reduce the number of fitting parameters.

\begin{figure*}[htbp]
    \centering
    \includegraphics[width=0.8\textwidth]{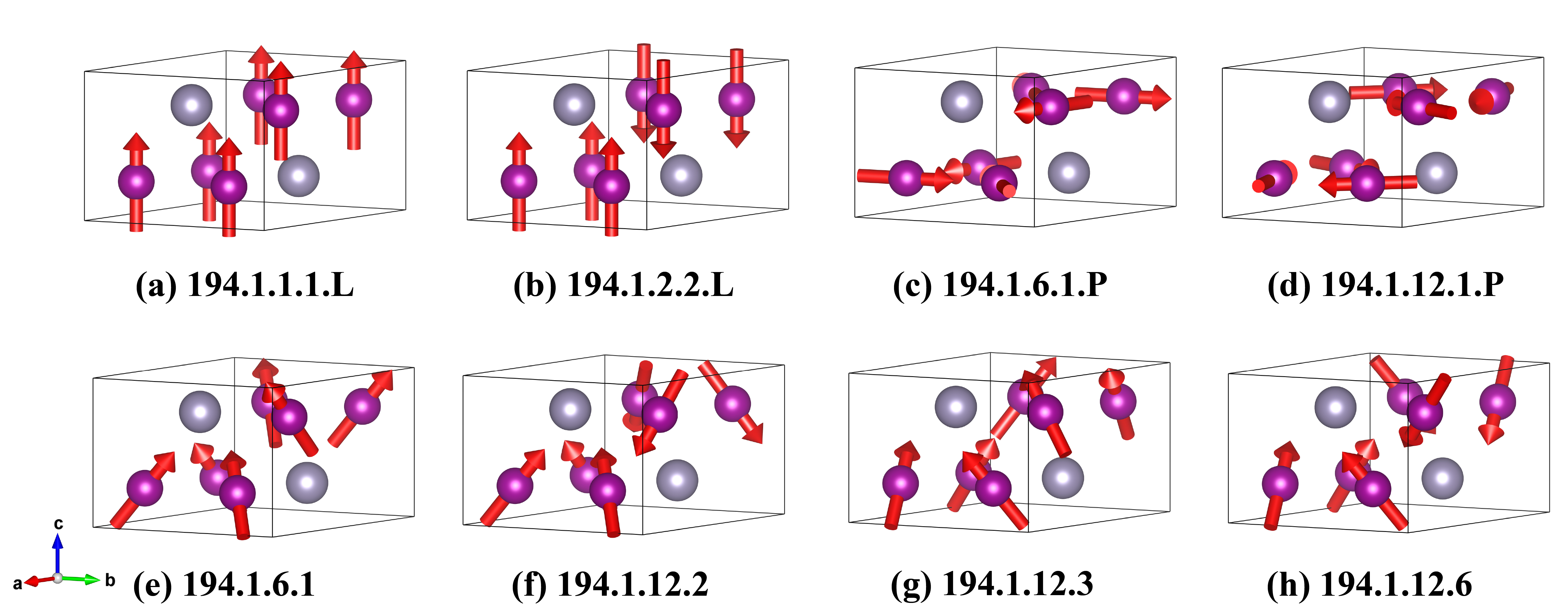}
    \caption{\label{fig: different magnetic structures of Mn3Sn from SSGs}
    The of symmetry-allowed magnetic configurations on the $6h$ Wyckoff position in eight SSGs induced from SG 194 with $I_k=1$. (a)(b) are collinear orders. (c)(d) are coplanar orders. (e)-(h) are non-coplanar orders. With SSGs, these magnetic configurations could be parametrized by only a few parameters.}
\end{figure*}

To facilitate the usage of SSGs in the refinement, we also give an algorithm to determine the symmetry-allowed directions of magnetic moments for a given SSG. Consider an SSG $\mathcal{S}$ and a Wyckoff position with representative point $\bm{s}_0$. The algorithm has the following steps:
\begin{itemize}
\item First determine the site symmetry group $G_{\bm{s}_0}$ of $\bm{s}_0$, which is formed by SSG operations that keep $\bm{s}_0$ invariant, i.e., $G_{\bm{s}_0}=\{ g=\{U||R|\tau\}\in\mathcal{S}| g\bm{s}_0 = \bm{s}_0 + \bm{R} \}$, where $\bm{R}$ is a lattice translation.
Denote the spin-only site group as $G_{\bm{s}_0}^0=\{U\}$, where $U$ is the spin rotations from all operations in $G{\bm{s}_0}$.
\item We then consider the symmetry-allowed magnetic moments $\bm{M}_0$ at $\bm{s}_0$, which should satisfy
$U_i \bm{M}_0 = \bm{M}_0,\ \forall U_i\in G_{\bm{s}_0}^0$.
In practice, this equation can be transformed into finding the intersection of eigenspaces where all $U_i$ have eigenvalues 1, or more conveniently, finding the common null space of $\bigcup_{i} (U_i - \mathbf{1}_3)$. This eigenspace is classified into the following three scenarios:
\begin{enumerate}
    \item If the dimension of the space is 0, then $\bm{M}_0=\bm{0}$. This means under SSG $\mathcal{S}$, the Wyckoff cannot host nonzero magnetic moments. In this case, when the magnetic atom is at this Wyckoff position, SSG $\mathcal{S}$ should be excluded in the refinement.
    \item If the dimension of the space is 1, then $\bm{M}_0$ is restricted to a specific direction. The magnitude of the magnetic moment is used as the sole refinement parameter.
    \item If the dimension of the space is 2, then $\bm{M}_0$ is restricted to a plane and is parameterized by two parameters, i.e., the combination coefficients of two eigenvectors.
    \item If the dimension of the space is 3, then $\bm{M}_0$ is free and has three parameters.
\end{enumerate}
\item Once $\bm{M}_0$ at $\bm{s}_0$ is determined, we use other operations of the SSG to generate the magnetic moments at other sites $\bm{s}_i$ in this Wyckoff position.
\end{itemize}
With the algorithm above, we can parameterize the magnetic moments on Wyckoff positions for a given SSG $\mathcal{S}$. These parameters are used as input for the refinement of the neutron diffraction pattern.

We remark on one possible issue with SSGs, i.e., the spin space and real space are decoupled in SSGs, and thus the spin orientations could differ by an overall rotation without breaking any symmetry. This means the symmetry-independent moments are in principle all free under SSGs. This problem could be partly overcome as the direction of moments in collinear orders and the plane of moments in coplanar orders could be pre-determined from experiments.
We also remark that in our database\cite{ssg_website}, the spin space has a fixed orientation, e.g., all collinear SSGs have the spin axis along the $z$-direction in spin space. The convention may need to be tailored for realistic materials.

\begin{table*}[htbp]
\centering
\begin{tabular}{c|c|c|c|c|c|c}
\hline\hline
SSG number   & $(-x,-2x,\frac{1}{4})$                         & $(2x,x,\frac{1}{4})$                           & $(-x,x,\frac{1}{4})$ & $(x,2x,\frac{3}{4})$                           & $(-2x,-x,\frac{3}{4})$                          & $(x,-x,\frac{3}{4})$ \\ \hline
 194.1.1.1.L\rule{0pt}{2.6ex}  & $(0,0,m_z)$                                    & $(0,0,m_z)$                                    & $(0,0,m_z)$          & $(0,0,m_z)$                                    & $(0,0,m_z)$                                     & $(0,0,m_z)$          \\ \cline{2-7}
194.1.2.2.L\rule{0pt}{2.6ex}  & $(0,0,m_z)$                                    & $(0,0,m_z)$                                    & $(0,0,m_z)$          & $(0,0,-m_z)$                                   & $(0,0,-m_z)$                                    & $(0,0,-m_z)$         \\ \hline
194.1.6.1.P\rule{0pt}{2.6ex}  & $(-\frac{m_x}{2},\frac{\sqrt{3}m_x}{2}, 0)$    & $(-\frac{m_x}{2},\frac{-\sqrt{3}m_x}{2}, 0)$   & $(m_x,0,0)$          & $(-\frac{m_x}{2},\frac{\sqrt{3}m_x}{2}, 0)$    & $(-\frac{m_x}{2},-\frac{\sqrt{3}m_x}{2}, 0)$    & $(m_x,0,0)$          \\ \cline{2-7}
194.1.12.1.P\rule{0pt}{2.6ex} & $(-\frac{\sqrt{3}m_y}{2},-\frac{m_y}{2}, 0)$   & $(\frac{\sqrt{3}m_y}{2},-\frac{m_y}{2}, 0)$    & $(0,m_y,0)$          & $(\frac{\sqrt{3}m_y}{2},\frac{m_y}{2}, 0)$     & $(-\frac{\sqrt{3}m_y}{2},\frac{m_y}{2}, 0)$     & $(0,-m_y,0)$         \\ \hline
194.1.6.1\rule{0pt}{2.6ex}    & $(-\frac{m_x}{2},\frac{\sqrt{3}m_x}{2}, m_z)$  & $(-\frac{m_x}{2},\frac{-\sqrt{3}m_x}{2}, m_z)$ & $(m_x,0,m_z)$        & $(-\frac{m_x}{2},\frac{\sqrt{3}m_x}{2}, m_z)$  & $(-\frac{m_x}{2},-\frac{\sqrt{3}m_x}{2}, m_z)$  & $(m_x,0,m_z)$        \\ \cline{2-7}
194.1.12.2\rule{0pt}{2.6ex}   & $(-\frac{m_x}{2},\frac{\sqrt{3}m_x}{2}, m_z)$  & $(-\frac{m_x}{2},\frac{-\sqrt{3}m_x}{2}, m_z)$ & $(m_x,0,m_z)$        & $(-\frac{m_x}{2},\frac{\sqrt{3}m_x}{2}, -m_z)$ & $(-\frac{m_x}{2},-\frac{\sqrt{3}m_x}{2}, -m_z)$ & $(m_x,0,-m_z)$       \\ \cline{2-7}
194.1.12.3\rule{0pt}{2.6ex}   & $(-\frac{\sqrt{3}m_y}{2},-\frac{m_y}{2},m_z)$  & $(\frac{\sqrt{3}m_y}{2},-\frac{m_y}{2},m_z)$   & $(0,m_y,m_z)$        & $(\frac{\sqrt{3}m_y}{2},\frac{m_y}{2}, m_z)$   & $(-\frac{\sqrt{3}m_y}{2},\frac{m_y}{2}, m_z)$   & $(0,-m_y,m_z)$       \\ \cline{2-7}
194.1.12.6\rule{0pt}{2.6ex}   & $(-\frac{\sqrt{3}m_y}{2},-\frac{m_y}{2}, m_z)$ & $(\frac{\sqrt{3}m_y}{2},-\frac{m_y}{2}, m_z)$  & $(0,m_y,m_z)$        & $(\frac{\sqrt{3}m_y}{2},\frac{m_y}{2}, -m_z)$  & $(-\frac{\sqrt{3}m_y}{2},\frac{m_y}{2}, -m_z)$  & $(0,-m_y,-m_z)$      \\ \hline\hline
\end{tabular}
\caption{\label{Tab: SG194_6h_allowed_magmom}
SSGs from SG 194 with $I_k=1$ and their symmetry-allowed magnetic moments on the $6h$ Wyckoff position of SG 194. There are 2 allowed collinear, 2 coplanar, and 4 non-coplanar SSGs.
SSGs have the advantage of fewer free parameters in the magnetic moments compared with MSGs and could facilitate the refinement of magnetic structures from neutron diffraction patterns.
}
\end{table*}

\subsubsection{Example of \ch{Mn3Sn}}

We take \ch{Mn3Sn}\cite{brown1990determination} as an example to exemplify the advantage of SSG in neutron diffraction refinement.
As shown in \cref{Fig:real_material}(c), \ch{Mn3Sn} has non-magnetic SG 194 $P6_3/mmc$, with magnetic Mn atoms on the kagome $6h$ Wyckoff position and Sn atoms on the $2c$ position.

We start from a general consideration of how many SSGs could support nonzero magnetic moments on the $6h$ Wyckoff position from SG 194 without involving enlarged supercells, i.e., magnetic propagation vector $\bm{q}=(0,0,0)$.
From our enumeration results, there are 8 collinear, 40 (non-collinear) coplanar, and 152 non-coplanar SSGs constructed from SG 194 with $I_k=1$. Despite the large number of SSGs, only two collinear, two coplanar, and four non-coplanar SSGs can support nonzero and inequivalent magnetic moments on the $6h$ position.
The resultant magnetic structures from these 8 SSGs are shown in Figure. \ref{fig: different magnetic structures of Mn3Sn from SSGs}, with the detailed directions of magnetic moments on each site of $6h$ tabulated in \cref{Tab: SG194_6h_allowed_magmom}.
We discuss these SSGs in more detail.
\begin{itemize}
\item For the collinear case, SSG 194.1.1.1.L and 194.1.2.2.L both correspond to FM order inplane, but FM and AFM out-of-plane, respectively. The magnetic configurations are parameterized by only one parameter $m_z$.
Remark that 194.1.2.2.L has generator $\{M_z||C_{2z}|\bm{a}_3/2\}\sim \{C_{2z}||C_{2z}|\bm{a}_3/2\}\cdot\mathcal{T}$ that connects two Mn layers, which is altermagnetic with spin-splitted electronic bands.

\item For the coplanar case, SSG 194.1.6.1.P enforces the same magnetic moments on two Mn layers, while 194.1.12.1.P has opposite directional moments. These two magnetic configurations are also parameterized by only one parameter.
The coplanar configuration described in \cref{Fig:real_material}(c) has SSG 194.1.6.1.P. We remark that in Ref.\cite{brown1990determination}, the authors did not consider the magnetic configuration given by 194.1.12.1.P, which only has a subtle difference compared with 194.1.6.1.P and might give a comparable error of refinement.

\item For the non-coplanar case, there are four SSGs: 194.1.6.1, 194.1.12.2, 194.1.12.3, and 194.1.12.6, each parameterized by 2 parameters. Among them, SSG 194.1.6.1 has two Mn layers sharing the same direction of magnetic moments, which describes the experimental structure of \ch{Mn3Sn} reported in Ref.\cite{non_coplanar_example}. The other three possible non-coplanar SSGs give two inequivalent Mn layers.
\end{itemize}

Note that among the large number of SSGs given by SG 194 with $I_k=1$, many of them only allow magnetic moments with higher symmetries on Wyckoff position $6h$, and can be omitted. For example, some non-coplanar SSGs only allow collinear orders on $6h$ and thus are equivalent to a collinear SSG.
Moreover, the directions of magnetic moments in each SSG tabulated in \cref{Tab: SG194_6h_allowed_magmom} could change by an overall rotation, due to the decoupled nature of spin and real spaces in SSGs.

For comparison, we also consider magnetic moments restricted by the MSG of \ch{Mn3Sn} in coplanar and non-coplanar order, i.e., MSG 63.463 $C mc^\prime m^\prime$ and 12.62 $C2^\prime/m^\prime$.
For MSG 63.463, the $6h$ position splits into two symmetry-independent Wyckoff positions $8g$ and $4c$, which gives 4 parameters.
For MSG 12.62, $6h$ splits into $8j$ and $4i$, which gives 5 parameters.
These two MSGs have much lower symmetry compared with the SSGs in the same table, and the $6h$ Wyckoff position splits into multiple symmetry-independent Wyckoff positions. Thus the independent components of the magnetic moments are increased and is detrimental for the neutron refinements.

\section{Conclusions}\label{conclusion}

In this work, we give an invariant subgroup-based algorithm to enumerate SSGs systematically. We implement the algorithm and find a vast number of SSGs. With the enumerated SSGs, the representation theory\cite{YangLiu,yang_hamiltonian_2021}, new types of magnetic topological states, and novel quasi-particles protected by SSG symmetries can be readily explored in the future.

In addition to the applications of SSGs discussed in our work, it is noteworthy that SSGs offer valuable insights into the transport properties of magnetic materials with weak SOC. The enhanced symmetry framework provided by SSGs is instrumental in categorizing potential responses to various external fields. For instance, SSGs offer a deeper understanding of the anomalous Hall effect (AHE) in different magnetic materials.
Collinear and coplanar magnetic materials have pure-spin operations that can be seen as effective time-reversal symmetry, which enforces the AHE to be zero when SOC is negligible\cite{smejkal_anomalous_2022, Suzuki2017cluster}. With the inclusion of SOC, the degenerate points protected by SSG symmetries are lifted and could lead to large AHE. For non-coplanar magnetic materials, AHE could exist even without SOC\cite{Shindou2001ahe}, and SSGs can be used to further classify the symmetry conditions for AHE to exist.

In conclusion, the vast number of SSGs we enumerated systematically in the work greatly enlarge the symmetries that could be used for describing magnetic materials and their band structures and could promote research in related fields.

\textit{Note added.}
Upon completing our enumeration of SSGs, we became aware of parallel studies conducted by Zhi-Da Song's group and Qihang Liu's group, which are documented in their respective works (see
Refs.\cite{PhysRevX.14.031037} and \cite{PhysRevX.14.031038}). These groups have also explored the enumeration and applications of SSGs, albeit through distinct methodologies.
In Supplementary Material \cref{sec:app:comparsion_3works}, we provide a comparative analysis of our results with those of Song's and Liu's groups. We also give a brief discussion on how to effectively map these different notational systems.
The manuscript is revised according to the referees' reports and has been published in Ref.\cite{PhysRevX.14.031039}.

\acknowledgments{Y.J. and Z.S. thanks Zhenyu Xiao, Yupeng Wang, Jie Ren, Zhen-Yuan Yang, Bingrui Peng, and Yuting Qian for helpful discussions.
C.F., H.W., and J.Y. are supported by Chinese Academy of Sciences under grant number XDB33000000, National Key Research and Development Program of China (Grant No.2022YFA1403800), and Natural Science Foundation of China (Grant No. 12188101). C.F. is supported by National Natural Science Foundation of China (NSFC) under grant number 12325404. Z.-X.L. is supported by the NSF of China (Grants No.12134020 and No. 12374166), National Key Research and Development Program of China (Grant No.2022YFA1405301, 2023YFA1406500), and the Fundamental Research Funds for the Central Universities and the Research Funds of Renmin University of China (Grant No. 19XNLG11). H.W has been supported by the New Cornerstone Science Foundation through the XPLORER PRIZE.
}

\normalem




%

\clearpage
\newpage
\onecolumngrid

\begin{center}
\textbf{\center{\large{Supplementary Material}}}
\end{center}

\beginsupplement

\section{Modular linear equations}\label{SM_modular_eq}
In this section, we introduce the Smith normal form method to solve modular linear equations of the form
\begin{equation}
	M\cdot x=w \bmod [n],
\end{equation}
where $M$ is a $m\times n$-dimensional integer matrix, $x,w$ are $n$-dimensional vectors, and $[n]$ is a simplified notation meaning that each entry of $w$ in the equation is modular $n$.

The Smith normal form of an integer matrix $M$ is
\begin{equation}
    \Lambda=LMR,
\end{equation}
where $L_{m\times m}, R_{n\times n}$ are the left and right matrix that are integral and invertible, and $\Lambda_{m\times n}=\text{Diag}(\lambda_1,..., \lambda_n)$ has the same shape as $M$ and is diagonal with integral elements.

For homogeneous equations,
\begin{equation}
	M\cdot x=0 \bmod [1],
\end{equation}
first find the smith normal form $M=L^{-1}\Lambda R^{-1}$, where $\Lambda=\text{Diag}(\lambda_1,..., \lambda_n)$ is a diagonal matrix.
Denote $v=R^{-1}\cdot x$, which satisfies a simplified equation, $\Lambda\cdot v =0\bmod [1]$, with straightforward solutions: $v_i=\frac{j}{\lambda_i}$, where $j\in\{0,1,...,\lambda_i-1\}$. Notice that if some $\lambda_i=0$, then the corresponding $v_i$ can take arbitrary values. The solution of $x$ is given by
$x=R\cdot v$.

For homogeneous equations,
\begin{equation}
	M\cdot x=0 \bmod [n],
\end{equation}
the method is similar with $v_i=n\frac{j}{\lambda_i}$.

For inhomogeneous equations,
\begin{equation}
	M\cdot x=w \bmod [1],
\end{equation}
we have $\Lambda\cdot v =L\cdot w\bmod [1]$. Denote $w^\prime=L\cdot w, v=R^{-1}\cdot x$, then $v$ is given by
$v_i=\frac{w^\prime_i}{\lambda_i}+\frac{j}{\lambda_i}$, where $j\in\{0,1,...,\lambda_i-1\}$. If $\lambda_i=0$, then $v_i$ can take arbitrary values.
The solution of $x$ is given by
$x=R\cdot v$.

For inhomogeneous equations,
\begin{equation}
	M\cdot x=w \bmod [n],
\end{equation}
the method is similar with
$v_i=\frac{w^\prime_i}{\lambda_i}+n_i\frac{j}{\lambda_i}$.

\section{Finding isomorphisms between finite groups}\label{SM_isomorphism}
For two finite groups $\mathcal{G}_1$ and $\mathcal{G}_2$ with the same number of elements $N$, the following method can be used to check if they are isomorphic\cite{group_iso}.

\begin{enumerate}
    \item First, check if $\mathcal{G}_1$ and $\mathcal{G}_2$ have the same number of elements of different ranks. If not, they must not be isomorphic. If so, proceed to the next step.
    \item  Find a set of generators of $\mathcal{G}_1$ (do not need to find the minimal set of generators). Assume there are $n$ generators.
    \item Find all subsets of $\mathcal{G}_2$ with length $n$.
    \item For each subset, exhaust all bijections to the set of generators of $\mathcal{G}_2$
    \item For each bijection, check if it can be extended to an isomorphism mapping between the $\mathcal{G}_1$ and $\mathcal{G}_2$. If true, then $\mathcal{G}_1$ and $\mathcal{G}_2$ are isomorphic.
\end{enumerate}

\section{Isomorphic point groups}\label{SM_eqv_PG}
Point groups can be enumerated into the following groups with isomorphic relations:
\begin{itemize}
	\item $C_n$, where $n\in\mathbb{N}$.
	\item $C_i, C_s$, with $C_i\cong C_m \cong C_2$.
	\item $C_{nh}$, with $C_{nh}\cong C_{2n}$, when $n\in2\mathbb{N}+1$.
	\item $S_n$, with $S_{2n}\cong C_{2n}, S_{2n+1}\cong C_{(2n+1)h}\cong C_{2(2n+1)}\cong S_{2(2n+1)}$, where $n\in\mathbb{N}$.
	\item $D_n$, where $n\in\mathbb{N}, n\ge 2$.
	\item $C_{nv}$, with $C_{nv}\cong D_{n}$, where $n\in\mathbb{N}, n\ge 2$.
	\item $D_{nh}$, with $D_{nh}\cong D_{2n}$, when $n\in2\mathbb{N}+1$.
	\item $D_{nd}$, with $D_{nd}\cong D_{2n}$, where $n\in\mathbb{N}, n\ge 2$.
	\item $T, O, T_d, T_h, O_h$, with $T_d\cong O$.
	\item $I, I_h$
\end{itemize}

Using the isomorphic relations, only the following point groups are not isomorphic to each other, which can be used as abstract point groups:
\begin{itemize}
	\item $C_n,\ n\in\mathbb{N}$
	\item $C_{nh},\ n\in2\mathbb{N}$
	\item $D_n,\ n\in\mathbb{N}, n\ge 3$
	\item $D_{nh},\ n\in2\mathbb{N}$
	\item $T, O, I, T_h, O_h, I_h$
\end{itemize}
where $\mathbb{N}$ denotes positive integers.

\section{Automorphism group of space groups}\label{SM_auto_SG}
In this section, we discuss the automorphism group $Auto(\mathcal{G})$ of each SG $\mathcal{G}$. We first enumerate the automorphism group of primitive symmorphic SGs.
\begin{itemize}
	\item For triclinic lattice, i.e., SG 1, 2, $Auto(\mathcal{G})=\{W=(A,t)|A\in \text{SL}(3,\mathbb{Z}), t\in\mathbb{R}^3\}$, where $A$ can take any $\text{SL}(3,\mathbb{Z})$ matrix.
	\item For monoclinic lattice (assuming the main axis is $b$), $A$ can take any $\text{SL}(2, \mathbb{Z})$ matrix at $a,c$ direction and identity at $b$ direction, as well as $C_{2a}, C_{2c}$.
	\item For orthorhombic lattice, $Auto(\mathcal{G})=O(432)$.
	\item For tetragonal lattice, $Auto(\mathcal{G})=O(432)$.
	\item For trigonal lattice, $Auto(\mathcal{G})=D_3(32)$.
	\item For hexagonal lattice, $Auto(\mathcal{G})=D_6(622)$.
	\item For cubic lattice, $Auto(\mathcal{G})=O(432)$.
\end{itemize}

For non-primitive Bravais lattices, the aforementioned automorphisms of the primitive Bravais lattice of the same crystal system are used as candidates. The candidates satisfying $WGW^{-1}=G$ in the corresponding primitive unit cell are automorphisms of the non-primitive SG $\mathcal{G}$.

For non-symmorphic SGs, the method is similar, and automorphisms may contain nontrivial translation parts.

\section{Details in finding inequivalent supercells, subgroups, and 3D real representations}\label{SM_algorithm_details}

For a given SG $\mathcal{G}$ with primitive cell basis $\bm{a}_{i}^{\mathcal{G}}$ and a subgroup $H$ with primitive cell basis $\bm{a}_{i}^{H}$, the supercell matrix of $H$ is defined as the transformation of primitive cell basis:
\begin{equation}
	\bm{a}_{i}^{H}=\bm{a}_{j}^{\mathcal{G}} S_{ji}, \quad
	S=
	\left(
	\begin{matrix}
		a_{11} & a_{12} & a_{13} \\
		a_{21} & a_{22} & a_{23} \\
		a_{31} & a_{32} & a_{33}
	\end{matrix}
	\right)\in \text{GL}(3,\mathbb{Z})
\end{equation}

To find inequivalent supercells of a give SG $\mathcal{G}$, we first reduce the infinite number of $\text{GL}(3,\mathbb{Z})$ supercell matrices of k-index $I_k$ to finite by considering all possible column transformations, which is the simplest equivalent relations, and simplify the supercell matrices to the form
\begin{equation}
	S=
	\left(
	\begin{matrix}
		a_{11} & 0 & 0 \\
		a_{21} & a_{22} & 0 \\
		a_{31} & a_{32} & a_{33}
	\end{matrix}
	\right)
\end{equation}
where $I_k=a_{11}a_{22}a_{33}$, $a_{ii}\in\{1,...,I_k\}$, $a_{21}\in\{0,1,...,a_{22}-1\}, a_{31},a_{32}\in\{0,1,...,a_{33}-1\}$. In this way, we reduce the number of supercells to a few hundred. We then use the automorphism group $Auto(\mathcal{G})=\{W=(A,t)\}$ to find all inequivalent supercells among them by using
\begin{equation}
	S_1=AS_2C	
\end{equation}
Note that as supercell matrices are written in the primitive basis of $\mathcal{G}$, the automorphism $W=(A,t)$ should also be written in the primitive cell.

To find inequivalent subgroups of a given SG $\mathcal{G}$ and supercell $S$, we first find all automorphisms $W=(A,t)$ that leave $S$ invariant, i.e.,
\begin{equation}
	ASC=S \Rightarrow  S^{-1}A^{-1}S\in \text{SL}(3,\mathbb{Z})
\end{equation}
For two subgroup $H_1$ and $H_2$ of supercell $S$ of $\mathcal{G}$, we check if there exist $S$-invariant $A$ s.t. it connects the point group part of $H_{1,2}$:
\begin{equation}
    A P_{H_1} = P_{H_2} A
\end{equation}
If so, we then solve the translation part of the automorphism using:
\begin{equation}
    \begin{aligned}
	(A, t)(R_1,\tau_1)&=(R_2,\tau_2)(A, t)\\
	(A^\prime, t^\prime)(R_3,\tau_3)&=(R_4,\tau_4)(A^\prime, t^\prime)\\
    \end{aligned}
\end{equation}
which is equivalent to
\begin{equation}
    \begin{aligned}
	(R_2 - 1)t=A\tau_1-\tau_2\bmod T_G\\
	(R_4-1)t^\prime = A^\prime \tau_3 - \tau_4 \bmod T_H\\
    \end{aligned}
 \label{auto_in_subsg}
\end{equation}
where $(R_{1,2},\tau_{1,2})\in G, (R_3,\tau_3)\in H_1, (R_4, \tau_4)\in H_2$ are operations written in the primitive cell of $\mathcal{G}$ and $H_{1,2}$, $(A^\prime=S^{-1}AS, t^\prime=S^{-1}t)$ is the automorphism written in the primitive cell of $H_{1,2}$, and $T_G,T_H$ denotes the primitive translation basis of $\mathcal{G}$ and $H_{1,2}$. This equation can be solved using the Smith normal form method. If there exist solutions of $t$, then subgroups $H_1$ and $H_2$ are equivalent. We remark that to solve $t$, one has to combine $\mathcal{G}$ and $H_{1,2}$.

To find inequivalent 3D real representations of a given SG $\mathcal{G}$, invariant subgroup $H$ with supercell $S$, and a quotient group $Q$, we first find automorphisms $W=(A,t)$ that leaves $S$ invariant, i.e.,
\begin{equation}
	ASC=S \Rightarrow  S^{-1}A^{-1}S\in \text{SL}(3,\mathbb{Z})
\end{equation}
We then check if the translation part of the automorphism has solutions by combining $\mathcal{G}$ and $H$:
\begin{equation}
    W \mathcal{G}=\mathcal{G} W,\quad W^\prime H=HW^\prime
    \label{eq_auto_G_H}
\end{equation}
where $W^\prime=S^{-1} W S$ is the
automorphism in the primitive cell of $H$, and is equivalent to Eq. (\ref{auto_in_subsg}) by taking $H_1=H_2=H$. In practice, we first check if the rotation part $A$ of $W$ gives an automorphism of the rotation parts of $\mathcal{G}$ and $H$, and then solve the translation part $t$ of $W$. Notice that $t$ may not be unique, and it is necessary to exhaust all possible translation parts $t$ of $W$ that give all inequivalent mappings of elements in $Q=\mathcal{G}/H$, i.e., to exhaust the automorphism group $Auto(Q)$.

For a given $A$, the solutions of $t$ are infinite, i.e., $t$ added by lattice translations of $H$ are solutions, but they can only give a finite number of inequivalent mappings of elements in $\mathcal{G}$, $H$, and $Q$. Thus by exhausting all solutions of the modular equation Eq. (\ref{eq_auto_G_H}), one obtains the automorphism group $Auto(Q)$.
We remark that for uni-axial SGs (including triclinic SGs), any arbitrary translation in the direction of the main axis is a solution of Eq. (\ref{eq_auto_G_H}), as it acts the same as identity. This means the diagonal elements of the Smith normal form of Eq. (\ref{eq_auto_G_H}) have a zero. We set such translations to zero as they do not give new automorphisms of $Q$.

For two 3D real representations $(Q|D_1)$ and $(Q|D_2)$ of $Q$, we check if there exist such $H$-invariant automorphisms that
\begin{equation}
    (WQW^{-1}|D_1) = (Q|D_2).
\end{equation}
By exhausting the elements in $Auto(Q)$ for all pairs of 3D real representations, we obtain inequivalent 3D real representations.
We remark that when $D$ is constructed by multiple 1D or 2D IRREPs of $Q$, they can be permuted, which may change the direction of the main axis of the equivalent PG of $D$, but give equivalent 3D representations.

For triclinic crystal systems, the rotation part of automorphisms $A\in\text{SL}(3,\mathbb{Z})$, while for monoclinic crystal systems, $A\in\text{SL}(2,\mathbb{Z})$ and take the form
\begin{equation}
    A=\left(
\begin{matrix}
	a_{11} & 0 & a_{13} \\
	0 & 1 & 0 \\
	a_{31} & 0 & a_{33}
\end{matrix}
\right),
\end{equation}
whose number is infinite. In practice, we reduce the number to finite by using a truncation of 24, i.e., $a_{ij}\le 24$. The step greatly reduces the number of SSGs, and the resultant SSGs are all inequivalent by increasing the truncation to 50.

\section{Quotient groups isomorphic to abstract point groups}\label{SM_quotient}
We give general rules to determine the structure of quotient groups $Q$ when they are isomorphic to abstract point groups. In Table.\ref{tab:summary_quot_PG}, we give a summary of quotient groups with $I_k\le 12$ that are isomorphic to abstract point groups.

We first list the abstract point groups that are not isomorphic to each other:
$C_n (n\in\mathbb{Z})$, $C_{nh} (n\in2\mathbb{Z})$, $D_n (n\in\mathbb{Z}, n\ge 3)$, $D_{nh} (n\in2\mathbb{Z})$, $T, O, I, T_h, O_h, I_h$.

\paragraph{$Q\cong C_n$}
There are three possible constructions:
\begin{itemize}
	\item $P/P_H\cong \mathbb{Z}_n$, $\bm{T}/\bm{T}_H\cong 1$, where $n=1,2,3,4,6$.
	\item $P/P_H\cong 1$, $\bm{T}/\bm{T}_H\cong \mathbb{Z}_n$, where $n\ge2$.
	\item $P/P_H\cong \mathbb{Z}_p$, $\bm{T}/\bm{T}_H\cong \mathbb{Z}_q$, where $pq=n$ ($p,q\ge2$). Note that $p$ and $q$ are not necessarily mutually prime.
	\begin{itemize}
		\item Example: When $G=P4_1$, $P_H=1$, and $\bm{T}/\bm{T}_H\cong \mathbb{Z}_6$, with supercell along $z$-direction, the quotient group $Q\cong C_{24}$ is generated by $\{C_{4z}|0,0,\frac{1}{4}+1\}$, which has rank $24$.
	\end{itemize}
\end{itemize}

\paragraph{$Q\cong C_{nh}\ (n\in2\mathbb{Z}, n\ge 2)$}
There are several possible constructions:
\begin{itemize}
	\item $P/P_H\cong C_{nh}$, $\bm{T}/\bm{T}_H\cong 1$, where $n=2,4,6$.
	\item $P/P_H\cong 1$, $\bm{T}/\bm{T}_H\cong \mathbb{Z}_n\times\mathbb{Z}_2$, where $n\ge2$.
	\item $P/P_H\cong C_{2}$, $\bm{T}/\bm{T}_H\cong \mathbb{Z}_n$, where $n\in\mathbb{Z}, n\ge 2$, and $C_2$ must commute with the generator of $\bm{T}/\bm{T}_H$.
	\begin{itemize}
		\item Example: When $G=P2$, $P_H=1$, $P/P_H=C_2$, and $\bm{T}/\bm{T}_H\cong \mathbb{Z}_n$, with the rotation and supercell both along $y$-direction, the quotient group $Q\cong C_{nh}$ is generated by $\{C_{2y}|0\}$ and $\{E|0,1,0\}$.
	\end{itemize}
	\item $P/P_H\cong C_{n}$, $\bm{T}/\bm{T}_H\cong \mathbb{Z}_2$, where $n=3,4,6$, and $C_n$ must commute with the generator of $\bm{T}/\bm{T}_H$.
	\begin{itemize}
		\item Example: When $G=Pn$, $P_H=1$, $P/P_H=C_n$, and $\bm{T}/\bm{T}_H\cong \mathbb{Z}_2$, with rotations and supercell both along $z$-direction, the quotient group $Q\cong C_{nh}$ is generated by $\{C_{n}|0\}$ and $\{E|0,0,1\}$.
	\end{itemize}
	\item $P/P_H\cong C_{p}$, $\bm{T}/\bm{T}_H\cong \mathbb{Z}_q\times\mathbb{Z}_2$, where $n=pq$, $p=2,3,4,6$, $q\in\mathbb{Z}$, $q\ge 2$, and $Q\cong C_{nh}$.
	\begin{itemize}
		\item Example: When $G=P4$, $P_H=1$, $P/P_H=C_4$, and $\bm{T}/\bm{T}_H\cong \mathbb{Z}_6\cong\mathbb{Z}_3\times\mathbb{Z}_2$, with rotations and supercell both along $z$-direction, the quotient group $Q\cong C_{12h}$ is generated by $\{C_{4}|0,0,2\}$ and $\{E|0,0,3\}$.
	\end{itemize}
	\item $P/P_H\cong C_{ph}$, $\bm{T}/\bm{T}_H\cong \mathbb{Z}_q$, where $n=pq$, $p=2,3,4,6$, $q\in\mathbb{Z}$, $q\ge 2$, and $Q\cong C_{nh}$.
	\begin{itemize}
		\item Example: When $G=Pmm2$, $P_H=1$, $P/P_H\cong C_{2h}$, and $\bm{T}/\bm{T}_H\cong \mathbb{Z}_q$, with $C_2$ and supercell both along $z$-direction, the quotient group $Q\cong C_{(2q)h}$ is generated by $\{C_{2z}|0,0,1\}$ and $\{M_x|0\}$.
	\end{itemize}
\end{itemize}

\paragraph{$Q\cong D_{n}\ (n\in\mathbb{Z}, n\ge 3)$}
Note that $D_n$ is not an Abelian group, in which $C_{n,z}$ rotation does not commute with the $C_{2x}$ rotation.
There are several possible constructions:
\begin{itemize}
	\item $P/P_H\cong D_{n}$, $\bm{T}/\bm{T}_H\cong 1$, where $n=3,4,6$.
	\item $P/P_H\cong C_{2}$, $\bm{T}/\bm{T}_H\cong \mathbb{Z}_n$, where $n\ge 3$, and $C_2$ must not commute with the generator of $\bm{T}/\bm{T}_H$.
	\begin{itemize}
		\item Example: When $G=P\overline{1}$, $P_H=1$, $P/P_H=C_i$, and $\bm{T}/\bm{T}_H\cong \mathbb{Z}_n$, with the supercell along $z$-direction, the quotient group $Q\cong D_{n}$ is generated by $\{E|0,0,1\}$ and $\{\mathcal{P}|0\}$, which do not commute with each other.
	\end{itemize}
	\item $P/P_H\cong D_{p}$, $\bm{T}/\bm{T}_H\cong \mathbb{Z}_q$, where $n=pq$, $p=2,3,4,6$, $q\in\mathbb{Z}$, $q\ge 2$, and $Q\cong D_{n}$.
	\begin{itemize}
		\item Example: When $G=P2_1/m$, $P/P_H\cong D_2$, and $\bm{T}/\bm{T}_H\cong \mathbb{Z}_q$, with $C_2$ and supercell both along $y$-direction, the quotient group $Q\cong D_{2q}$ is generated by $\{C_{2y}|0,\frac{1}{2}+1,0\}$ and $\{\mathcal{P}|0\}$.
	\end{itemize}
\end{itemize}

We also enumerate several constructions that can not give $Q\cong D_n$:
\begin{itemize}
	\item $P/P_H\cong 1$, $\bm{T}/\bm{T}_H\cong \mathbb{Z}_n\times\mathbb{Z}_2$. This is because $\bm{T}/\bm{T}_H$ is an Abelian group.
	\item $P/P_H\cong C_{n}$, $\bm{T}/\bm{T}_H\cong \mathbb{Z}_2$, where $n=3,4,6$. This is because if the subgroup with k-index $2$ is invariant, the doubled supercell must be along the main rotation axis, and thus the generator of $\bm{T}/\bm{T}_H$ commute with the rotation.
	\item $P/P_H\cong C_{m}$, $\bm{T}/\bm{T}_H\cong \mathbb{Z}_2\times\mathbb{Z}_2$, where $n=2m$, $m=3,4,6$. This is because if the subgroup with k-index $4$	is invariant, the supercell must be $S=\text{Diag}(2,2,1)$ (assume the main rotation axis of G is $z$), and the quotient group cannot be isomorphic to $D_n$.
\end{itemize}

\paragraph{$Q\cong D_{nh}\ (n\in2\mathbb{Z}, n\ge 2)$}
$D_{nh}=D_n\otimes\{1, \mathcal{P}\}$ can be generated by a $n$-fold rotation and two 2-fold operations. There are several possible constructions:

\begin{itemize}
	\item $P/P_H\cong D_{nh}$, $\bm{T}/\bm{T}_H\cong 1$, where $n=2,4,6$.
	\item $P/P_H\cong C_{nh}$, $\bm{T}/\bm{T}_H\cong \mathbb{Z}_2$, where only $n=2$ is allowed.
	\begin{itemize}
		\item Example: When $G=P2/m$, $P/P_H=C_{2h}$, and $\bm{T}/\bm{T}_H\cong \mathbb{Z}_2$, with the supercell along $z$-direction, the quotient group $Q\cong D_{2h}$ is generated by $\{C_{2y}|0\}$, $\{\mathcal{P}|0\}$, and $\{E|0,0,1\}$.
	\end{itemize}
	\item $P/P_H\cong C_{n}$, $\bm{T}/\bm{T}_H\cong \mathbb{Z}_2\times\mathbb{Z}_2$, where only $n=2$ is allowed.
	\begin{itemize}
		\item Example: When $G=P2$, $P/P_H\cong C_2$, and $\bm{T}/\bm{T}_H\cong \mathbb{Z}_2\times\mathbb{Z}_2$, the quotient group $Q\cong D_{2h}$ is generated by $\{C_{2y}|0\}$, $\{E|1,0,0\}$ and $\{E|0,0,1\}$.
	\end{itemize}
	\item $P/P_H\cong C_{m}$, $\bm{T}/\bm{T}_H\cong \mathbb{Z}_2\times\mathbb{Z}_2\times\mathbb{Z}_2$, where $n=2m$, and only $m=1, 2$ are allowed.
	\begin{itemize}
		\item Example: When $G=P1$, $P/P_H\cong C_1$, and $\bm{T}/\bm{T}_H\cong \mathbb{Z}_2\times\mathbb{Z}_2\times\mathbb{Z}_2$, the quotient group $Q\cong D_{2h}$ is generated by $\{E|1,0,0\}$, $\{E|0,1,0\}$, and $\{E|0,0,1\}$.
		\item Example: When $G=P4$, $P/P_H\cong C_2$, and $\bm{T}/\bm{T}_H\cong \mathbb{Z}_2\times\mathbb{Z}_2\times\mathbb{Z}_2$, the quotient group $Q\cong D_{4h}$ is generated by $\{C_{4z}|0,1,0\}$, $\{E|1,0,0\}$, and $\{E|0,0,1\}$.
	\end{itemize}
	\item $P/P_H\cong D_2\cong \mathbb{Z}_2\times\mathbb{Z}_2$, $\bm{T}/\bm{T}_H\cong \mathbb{Z}_n$, where $n\ge4$.
	\begin{itemize}
		\item Example: When $G=P222$, $P/P_H\cong D_2$, and $\bm{T}/\bm{T}_H\cong \mathbb{Z}_4$, the quotient group $Q\cong D_{4h}$.
		\item Example: When $G=Pmmm$, $P/P_H\cong D_2$, and $\bm{T}/\bm{T}_H\cong \mathbb{Z}_{12}$, the quotient group $Q\cong D_{12h}$.
	\end{itemize}
	\item $P/P_H\cong \mathbb{Z}_2$, $\bm{T}/\bm{T}_H\cong \mathbb{Z}_n\times\mathbb{Z}_2$, where $n\ge4$.
	\begin{itemize}
		\item Example: When $G=P2/m$, $P/P_H\cong C_2$, and $\bm{T}/\bm{T}_H\cong \mathbb{Z}_6\times\mathbb{Z}_2$, the quotient group $Q\cong D_{6h}$.
	\end{itemize}
	
	\item $P/P_H\cong D_{2h}\cong \mathbb{Z}_2\times\mathbb{Z}_2\times\mathbb{Z}_2$, $\bm{T}/\bm{T}_H\cong \mathbb{Z}_m$, where $n=2m$, and $m\ge2$.
	\begin{itemize}
		\item Example: When $G=Pnnn$, $P/P_H\cong D_{2h}$, and $\bm{T}/\bm{T}_H\cong \mathbb{Z}_6$, the quotient group $Q\cong D_{12h}$.
		\item Example: When $G=P4/mmm$, $P/P_H\cong D_{2h}$, and $\bm{T}/\bm{T}_H\cong \mathbb{Z}_{12}$, the quotient group $Q\cong D_{24h}$.
	\end{itemize}

	\item $P/P_H\cong D_{4}\cong C_{4v} \cong D_{2d}$, $\bm{T}/\bm{T}_H\cong \mathbb{Z}_m$, where $n=2m$, $m\ge2$.
	\begin{itemize}
		\item Example: When $G=P422$, $P/P_H= D_{4}$, and $\bm{T}/\bm{T}_H\cong \mathbb{Z}_6$, the quotient group $Q\cong D_{12h}$.
		\item Example: When $G=P4mm$, $P/P_H= C_{4v}$, and $\bm{T}/\bm{T}_H\cong \mathbb{Z}_{2}$, the quotient group $Q\cong D_{4h}$.
		\item Example: When $G=P\overline{4}m2$, $P/P_H= D_{2d}$, and $\bm{T}/\bm{T}_H\cong \mathbb{Z}_{2}$, the quotient group $Q\cong D_{4h}$.
	\end{itemize}

	\item $P/P_H\cong D_{6}\cong C_{6v}\cong D_{3h} \cong D_{3d}$, $\bm{T}/\bm{T}_H\cong \mathbb{Z}_m$, where $n=3m$, $m\ge2$.
	\begin{itemize}
		\item Example: When $G=P622$, $P/P_H= D_{6}$, and $\bm{T}/\bm{T}_H\cong \mathbb{Z}_8$, the quotient group $Q\cong D_{24h}$.
		\item Example: When $G=P6mm$, $P/P_H= C_{6v}$, and $\bm{T}/\bm{T}_H\cong \mathbb{Z}_{2}$, the quotient group $Q\cong D_{6h}$.
		\item Example: When $G=P\overline{6}m2$, $P/P_H= D_{3h}$, and $\bm{T}/\bm{T}_H\cong \mathbb{Z}_{2}$, the quotient group $Q\cong D_{6h}$.
		\item Example: When $G=P\overline{3}m$, $P/P_H= D_{3d}$, and $\bm{T}/\bm{T}_H\cong \mathbb{Z}_{2}$, the quotient group $Q\cong D_{6h}$.
	\end{itemize}

\end{itemize}

We remark that when $P/P_H\cong D_{4h}$ or $D_{6h}$, and $\bm{T}/\bm{T}_H\cong \mathbb{Z}_m$ with $m\ge2$, the quotient group cannot be isomorphic to $D_{nh}$.

\paragraph{$Q\cong T, T_h, O, O_h$}
In the following, we consider the SSGs where $Q\cong T, T_h, O, O_h$, the number of which is finite.

$T$ has 12 elements and can be generated by a $3_{111}^+$ and $2_{001}$.
There are several possible constructions:
\begin{itemize}
	\item $P/P_H\cong T$, $\bm{T}/\bm{T}_H\cong 1$.
	\item $P/P_H\cong C_{3}$, $\bm{T}/\bm{T}_H\cong \mathbb{Z}_2\times\mathbb{Z}_2$.
	\begin{itemize}
		\item Example: When $G=P3$, $P/P_H=C_3$, and $\bm{T}/\bm{T}_H\cong \mathbb{Z}_2\times\mathbb{Z}_2$, the quotient group $Q\cong T$ is generated by $\{3_{001}^+|0\}$ and $\{E|1,0,0\}$.
	\end{itemize}
\end{itemize}

$T_h$ has 24 elements and can be generated by a $3_{111}^+$, $2_{001}$, and $\mathcal{P}$.
There are several possible constructions:
\begin{itemize}
	\item $P/P_H\cong T_h$, $\bm{T}/\bm{T}_H\cong 1$.
	\item $P/P_H\cong T$, $\bm{T}/\bm{T}_H\cong \mathbb{Z}_2$.
	\item $P/P_H\cong C_6\cong S_6 \cong C_{3h}$, $\bm{T}/\bm{T}_H\cong \mathbb{Z}_2\times\mathbb{Z}_2$.
	
	\item $P/P_H\cong C_{3}$, $\bm{T}/\bm{T}_H\cong \mathbb{Z}_2\times\mathbb{Z}_2\times\mathbb{Z}_2$.
	\begin{itemize}
		\item Example: When $G=P3$, $P/P_H=C_3$, and $\bm{T}/\bm{T}_H\cong \mathbb{Z}_2\times\mathbb{Z}_2\times\mathbb{Z}_2$, the quotient group $Q\cong T_h$ can be generated by $\{3_{001}^+|0\}$, $\{E|1,0,0\}$, and $\{E|0,0,1\}$.
	\end{itemize}
\end{itemize}

$O$ has 24 elements and can be generated by a $3_{111}^+$, $2_{001}$, and $2_{110}$.
There are several possible constructions:
\begin{itemize}
	\item $P/P_H\cong O$, $\bm{T}/\bm{T}_H\cong 1$.
	
	\item $P/P_H\cong D_{3}\cong C_{3v}$, $\bm{T}/\bm{T}_H\cong \mathbb{Z}_2\times\mathbb{Z}_2$.
	\begin{itemize}
		\item Example: When $G=P321$ (SG 150), $P/P_H=D_3$, and $\bm{T}/\bm{T}_H\cong \mathbb{Z}_2\times\mathbb{Z}_2$, the quotient group $Q\cong O$ can be generated by $\{3_{001}^+|0\}$, $\{2_{110}|0\}$, and $\{E|1,0,0\}$.
	\end{itemize}
\end{itemize}

$O_h$ has 48 elements and can be generated by $3_{111}^+$, $2_{001}$, $2_{110}$ and $\mathcal{P}$.
There are several possible constructions:
\begin{itemize}
	\item $P/P_H\cong O_h$, $\bm{T}/\bm{T}_H\cong 1$.
	\item $P/P_H\cong O$, $\bm{T}/\bm{T}_H\cong \mathbb{Z}_2$.
	\item $P/P_H\cong D_{3d}\cong D_{6}\cong C_{6v}\cong D_{3h}$, $\bm{T}/\bm{T}_H\cong \mathbb{Z}_2\times\mathbb{Z}_2$.
	
	\item $P/P_H\cong D_{3}\cong C_{3v}$, $\bm{T}/\bm{T}_H\cong \mathbb{Z}_2\times\mathbb{Z}_2\times\mathbb{Z}_2$.
	\begin{itemize}
		\item Example: When $G=P321$, $P/P_H=D_3$, and $\bm{T}/\bm{T}_H\cong \mathbb{Z}_2\times\mathbb{Z}_2\times\mathbb{Z}_2$, the quotient group $Q\cong O_h$ can be generated by $\{3_{001}^+|0\}$, $\{2_{110}|0\}$, $\{E|1,0,0\}$, and $\{E|0,0,1\}$.
	\end{itemize}
\end{itemize}

\newpage
\begin{longtable}{c|c|c|c|c|c|c|c|c|c|c|c|c|c}
\caption{\label{tab:summary_quot_PG} Summary of quotient groups with $I_k\le 12$ that are isomorphic to abstract point groups.}\\
\hline\hline
PG label & Total & $I_k=$    1 & $I_k=$    2 & $I_k=$    3 & $I_k=$    4 & $I_k=$    5 &$I_k=$    6 & $I_k=$    7 & $I_k=$    8 & $I_k=$    9 & $I_k=$   10 & $I_k=$   11 & $I_k=$   12\\\hline
 $C_{1}$ &        230 &   230 &     0 &     0 &     0 &     0 &     0 &     0 &     0 &     0 &     0 &     0 &     0\\\hline
 $C_{2}$ &       1191 &   674 &   517 &     0 &     0 &     0 &     0 &     0 &     0 &     0 &     0 &     0 &     0\\\hline
 $C_{3}$ &        114 &    27 &     0 &    87 &     0 &     0 &     0 &     0 &     0 &     0 &     0 &     0 &     0\\\hline
 $C_{4}$ &        328 &    20 &   176 &     0 &   132 &     0 &     0 &     0 &     0 &     0 &     0 &     0 &     0\\\hline
 $C_{5}$ &         68 &     0 &     0 &     0 &     0 &    68 &     0 &     0 &     0 &     0 &     0 &     0 &     0\\\hline
 $C_{6}$ &        384 &    22 &    23 &   145 &     0 &     0 &   194 &     0 &     0 &     0 &     0 &     0 &     0\\\hline
 $C_{7}$ &         68 &     0 &     0 &     0 &     0 &     0 &     0 &    68 &     0 &     0 &     0 &     0 &     0\\\hline
 $C_{8}$ &        212 &     0 &     4 &     0 &    83 &     0 &     0 &     0 &   125 &     0 &     0 &     0 &     0\\\hline
 $C_{9}$ &         86 &     0 &     0 &    10 &     0 &     0 &     0 &     0 &     0 &    76 &     0 &     0 &     0\\\hline
$C_{10}$ &        304 &     0 &     0 &     0 &     0 &   129 &     0 &     0 &     0 &     0 &   175 &     0 &     0\\\hline
$C_{11}$ &         33 &     0 &     0 &     0 &     0 &     0 &     0 &     0 &     0 &     0 &     0 &    33 &     0\\\hline
$C_{12}$ &        191 &     0 &     3 &     6 &    11 &     0 &   101 &     0 &     0 &     0 &     0 &     0 &    70\\\hline
$C_{13}$ &          0 &     0 &     0 &     0 &     0 &     0 &     0 &     0 &     0 &     0 &     0 &     0 &     0\\\hline
$C_{14}$ &        129 &     0 &     0 &     0 &     0 &     0 &     0 &   129 &     0 &     0 &     0 &     0 &     0\\\hline
$C_{15}$ &         10 &     0 &     0 &     0 &     0 &    10 &     0 &     0 &     0 &     0 &     0 &     0 &     0\\\hline
$C_{16}$ &         87 &     0 &     0 &     0 &     4 &     0 &     0 &     0 &    83 &     0 &     0 &     0 &     0\\\hline
$C_{17}$ &          0 &     0 &     0 &     0 &     0 &     0 &     0 &     0 &     0 &     0 &     0 &     0 &     0\\\hline
$C_{18}$ &        147 &     0 &     0 &     4 &     0 &     0 &    10 &     0 &     0 &   133 &     0 &     0 &     0\\\hline
$C_{19}$ &          0 &     0 &     0 &     0 &     0 &     0 &     0 &     0 &     0 &     0 &     0 &     0 &     0\\\hline
$C_{20}$ &        104 &     0 &     0 &     0 &     0 &     6 &     0 &     0 &     0 &     0 &    98 &     0 &     0\\\hline
$C_{21}$ &         10 &     0 &     0 &     0 &     0 &     0 &     0 &    10 &     0 &     0 &     0 &     0 &     0\\\hline
$C_{22}$ &         62 &     0 &     0 &     0 &     0 &     0 &     0 &     0 &     0 &     0 &     0 &    62 &     0\\\hline
$C_{23}$ &          0 &     0 &     0 &     0 &     0 &     0 &     0 &     0 &     0 &     0 &     0 &     0 &     0\\\hline
$C_{24}$ &         83 &     0 &     0 &     0 &     3 &     0 &     4 &     0 &    10 &     0 &     0 &     0 &    66\\\hline
$C_{2h}$ &       1846 &   421 &  1149 &     0 &   276 &     0 &     0 &     0 &     0 &     0 &     0 &     0 &     0\\\hline
$C_{4h}$ &        527 &     6 &   161 &     0 &   264 &     0 &     0 &     0 &    96 &     0 &     0 &     0 &     0\\\hline
$C_{6h}$ &        395 &     2 &    14 &    40 &     0 &     0 &   247 &     0 &     0 &     0 &     0 &     0 &    92\\\hline
$C_{8h}$ &        254 &     0 &     0 &     0 &    44 &     0 &     0 &     0 &   210 &     0 &     0 &     0 &     0\\\hline
$C_{10h}$ &        274 &     0 &     0 &     0 &     0 &    38 &     0 &     0 &     0 &     0 &   236 &     0 &     0\\\hline
$C_{12h}$ &        186 &     0 &     0 &     0 &     3 &     0 &    56 &     0 &     0 &     0 &     0 &     0 &   127\\\hline
 $D_{3}$ &        312 &    61 &     0 &   251 &     0 &     0 &     0 &     0 &     0 &     0 &     0 &     0 &     0\\\hline
 $D_{4}$ &        964 &    74 &   366 &     0 &   524 &     0 &     0 &     0 &     0 &     0 &     0 &     0 &     0\\\hline
 $D_{5}$ &        207 &     0 &     0 &     0 &     0 &   207 &     0 &     0 &     0 &     0 &     0 &     0 &     0\\\hline
 $D_{6}$ &        958 &    42 &    35 &   367 &     0 &     0 &   514 &     0 &     0 &     0 &     0 &     0 &     0\\\hline
 $D_{7}$ &        207 &     0 &     0 &     0 &     0 &     0 &     0 &   207 &     0 &     0 &     0 &     0 &     0\\\hline
 $D_{8}$ &        556 &     0 &     6 &     0 &   199 &     0 &     0 &     0 &   351 &     0 &     0 &     0 &     0\\\hline
 $D_{9}$ &        229 &     0 &     0 &    12 &     0 &     0 &     0 &     0 &     0 &   217 &     0 &     0 &     0\\\hline
$D_{10}$ &        817 &     0 &     0 &     0 &     0 &   333 &     0 &     0 &     0 &     0 &   484 &     0 &     0\\\hline
$D_{11}$ &         44 &     0 &     0 &     0 &     0 &     0 &     0 &     0 &     0 &     0 &     0 &    44 &     0\\\hline
$D_{12}$ &        416 &     0 &     3 &    10 &    14 &     0 &   223 &     0 &     0 &     0 &     0 &     0 &   166\\\hline
$D_{13}$ &          0 &     0 &     0 &     0 &     0 &     0 &     0 &     0 &     0 &     0 &     0 &     0 &     0\\\hline
$D_{14}$ &        333 &     0 &     0 &     0 &     0 &     0 &     0 &   333 &     0 &     0 &     0 &     0 &     0\\\hline
$D_{15}$ &         13 &     0 &     0 &     0 &     0 &    13 &     0 &     0 &     0 &     0 &     0 &     0 &     0\\\hline
$D_{16}$ &        205 &     0 &     0 &     0 &     6 &     0 &     0 &     0 &   199 &     0 &     0 &     0 &     0\\\hline
$D_{17}$ &          0 &     0 &     0 &     0 &     0 &     0 &     0 &     0 &     0 &     0 &     0 &     0 &     0\\\hline
$D_{18}$ &        353 &     0 &     0 &     4 &     0 &     0 &    12 &     0 &     0 &   337 &     0 &     0 &     0\\\hline
$D_{19}$ &          0 &     0 &     0 &     0 &     0 &     0 &     0 &     0 &     0 &     0 &     0 &     0 &     0\\\hline
$D_{20}$ &        230 &     0 &     0 &     0 &     0 &    10 &     0 &     0 &     0 &     0 &   220 &     0 &     0\\\hline
$D_{21}$ &         13 &     0 &     0 &     0 &     0 &     0 &     0 &    13 &     0 &     0 &     0 &     0 &     0\\\hline
$D_{22}$ &        132 &     0 &     0 &     0 &     0 &     0 &     0 &     0 &     0 &     0 &     0 &   132 &     0\\\hline
$D_{23}$ &          0 &     0 &     0 &     0 &     0 &     0 &     0 &     0 &     0 &     0 &     0 &     0 &     0\\\hline
$D_{24}$ &        102 &     0 &     0 &     0 &     3 &     0 &     6 &     0 &    13 &     0 &     0 &     0 &    80\\\hline
$D_{2h}$ &        988 &    52 &   494 &     0 &   402 &     0 &     0 &     0 &    40 &     0 &     0 &     0 &     0\\\hline
$D_{4h}$ &       1131 &    20 &   213 &     0 &   651 &     0 &     0 &     0 &   247 &     0 &     0 &     0 &     0\\\hline
$D_{6h}$ &        815 &     4 &    16 &    86 &     0 &     0 &   516 &     0 &     0 &     0 &     0 &     0 &   193\\\hline
$D_{8h}$ &        521 &     0 &     0 &     0 &    80 &     0 &     0 &     0 &   441 &     0 &     0 &     0 &     0\\\hline
$D_{10h}$ &        584 &     0 &     0 &     0 &     0 &    82 &     0 &     0 &     0 &     0 &   502 &     0 &     0\\\hline
$D_{12h}$ &        356 &     0 &     0 &     0 &     3 &     0 &    93 &     0 &     0 &     0 &     0 &     0 &   260\\\hline
$D_{14h}$ &         82 &     0 &     0 &     0 &     0 &     0 &     0 &    82 &     0 &     0 &     0 &     0 &     0\\\hline
$D_{16h}$ &         80 &     0 &     0 &     0 &     0 &     0 &     0 &     0 &    80 &     0 &     0 &     0 &     0\\\hline
$D_{18h}$ &         84 &     0 &     0 &     0 &     0 &     0 &     2 &     0 &     0 &    82 &     0 &     0 &     0\\\hline
$D_{20h}$ &         93 &     0 &     0 &     0 &     0 &     0 &     0 &     0 &     0 &     0 &    93 &     0 &     0\\\hline
$D_{22h}$ &         12 &     0 &     0 &     0 &     0 &     0 &     0 &     0 &     0 &     0 &     0 &    12 &     0\\\hline
$D_{24h}$ &         17 &     0 &     0 &     0 &     0 &     0 &     0 &     0 &     3 &     0 &     0 &     0 &    14\\\hline
       T &         36 &    12 &     0 &     0 &    24 &     0 &     0 &     0 &     0 &     0 &     0 &     0 &     0\\\hline
       O &         77 &    24 &     0 &     0 &    53 &     0 &     0 &     0 &     0 &     0 &     0 &     0 &     0\\\hline
 $T_{h}$ &         55 &     7 &     8 &     0 &    20 &     0 &     0 &     0 &    20 &     0 &     0 &     0 &     0\\\hline
 $O_{h}$ &         88 &    10 &    10 &     0 &    38 &     0 &     0 &     0 &    30 &     0 &     0 &     0 &     0\\\hline
       I &          0 &     0 &     0 &     0 &     0 &     0 &     0 &     0 &     0 &     0 &     0 &     0 &     0\\\hline
 $I_{h}$ &          0 &     0 &     0 &     0 &     0 &     0 &     0 &     0 &     0 &     0 &     0 &     0 &     0\\\hline

Total &      18433 &  1708 &  3198 &  1022 &  2837 &   896 &  1978 &   842 &  1948 &   845 &  1808 &   283 &  1068\\
\hline\hline
\end{longtable}

\section{Proof of the direct product group structure of spin-only groups}\label{sec:app:proof_direct_prod_group}

There are three types of magnetic orders that have non-trivial spin-only groups $\mathcal{S}_0$, i.e., non-magnetic, collinear, and coplanar orders, with $\mathcal{S}_0=\text{O}(3)$, $\text{O}(2)$, and $\mathbb{Z}_2^{T}$, respectively. For each type, the full SSG can be written as the direct product group $\mathcal{G}^{(S)}=\mathcal{G}^{(S)\prime}\times \mathcal{S}_0$, where $\mathcal{G}^{(S)\prime}=\mathcal{G}^{(S)}/\mathcal{S}_0$ is the spin-only free group. In the following, We show why the direct product group structure always holds.

Mathematically, a group $\mathcal{G}=A\times B$ when (i) group $A$ and $B$ have no intersections except the identity operation; (ii) group elements in $A$ and $B$ are commutable; (iii) any elements in $\mathcal{G}$ can be expressed uniquely as a product of an element of $A$ and an element of $B$.

We then consider three cases of spin-only groups in SSGs.
\begin{itemize}
\item In the non-magnetic case, $\mathcal{S}_0=\text{O}(3)$, the direct product group structure is obvious because all elements in $\mathcal{G}^{(S)}=\mathcal{G}^{(S)\prime}$ have no spin rotation.
\item In the collinear case, $\mathcal{S}_0=\text{O}(2)$, we could choose $\mathcal{G}^{(S)\prime}$ properly such that the spin rotation can only be the identity or $M_z$, while
$\mathcal{S}_0=\{C_{\infty z}|| E|\bm{0}\} + \{M_x C_{\infty z}|| E|\bm{0}\}$ which does not contain spin $M_z$. This is always possible due to the collinear configuration (assume the moments are along the $z$ direction in spin space).
In this case, spin $M_z$ is commutable with $\mathcal{S}_0$, and they together form the $\text{O}(3)$ group. Thus $\mathcal{G}^{(S)}=\mathcal{G}^{(S)\prime}\times \mathcal{S}_0$.
\item In the coplanar case, $\mathcal{S}_0=\mathbb{Z}_2^{M_z}=\{E,\{M_z||E|\bm{0}\}\}$. We could choose $\mathcal{G}^{(S)\prime}$ properly such that it does not contain spin $M_z$, thanks to the coplanar magnetic configuration (assume the moments are in $xy$ plane in spin space). Then the direct product group structure is straightforward to verify.
\end{itemize}

\section{Band representations of \ch{Mn3Sn}}\label{sec:app:Mn3Sn_irrep}
In this section, we provide additional information about the IRREPs of the bands in \ch{Mn3Sn}.
In Table. \ref{tab:msg character gamma point}, we show the character table of IRREPs at $\Gamma$ of MSG 63.463. There only exist 1D IRREPs. In Table. \ref{tab:band representations at gamma point}, we show the IRREPs of \ch{Mn3Sn} at $\Gamma$ computed from MSG 63.463.
In Table. \ref{tab:ssg character gamma point}, we give the character table of $\Gamma$ IRREPs in SSG 194.1.6.1.P.

\begin{table}[htbp]
\begin{tabular}{c|c|c|c}
\hline\hline
MSG operations & $\{E||E\}$ & $\{E||\mathcal{P}\}$ & $\{2_{010}||M_{010}\}$  \\\hline
$\bar{\Gamma}_3\rule{0pt}{2.6ex}$            & 1 & 1 & -i             \\\hline
$\bar{\Gamma}_4\rule{0pt}{2.6ex}$          & 1 & 1 & i           \\\hline
$\bar{\Gamma}_5\rule{0pt}{2.6ex}$            & 1 & -1 & i         \\\hline
$\bar{\Gamma}_6\rule{0pt}{2.6ex}$           & 1 & -1 & -i     \\
\hline\hline
\end{tabular}
\caption{\label{tab:msg character gamma point} Character table of MSG $Cmc^\prime m^\prime$ at $\Gamma$ point.}
\end{table}

\begin{table}[htbp]
\begin{tabular}{c|c|c|c|c}
\hline\hline
bands & $\{E||E\}$ & $\{E||\mathcal{P}\}$ & $\{2_{010}||M_{010}\}$ & IRREPs \\\hline
47   & 1        & 1 & i & $\bar{\Gamma}_4\rule{0pt}{2.6ex}$           \\\hline
48+49 &2           & -2 & 0 & $\bar{\Gamma}_5, \bar{\Gamma}_6\rule{0pt}{2.6ex}$           \\\hline
50+51 &2          & 2 & 0 & $\bar{\Gamma}_3,\bar{\Gamma}_4\rule{0pt}{2.6ex}$         \\\hline
52    &1        & -1 & i & $\bar{\Gamma}_5\rule{0pt}{2.6ex}$     \\\hline
53+54 &2          & -2 & 0 & $\bar{\Gamma}_5, \bar{\Gamma}_6\rule{0pt}{2.6ex}$         \\\hline
55+56 &2          & 2 & 0 & $\bar{\Gamma}_3, \bar{\Gamma}_4\rule{0pt}{2.6ex}$         \\
\hline\hline
\end{tabular}
\caption{\label{tab:band representations at gamma point} Irreducible representations (IRREPs) for bands at $\Gamma$ of \ch{Mn3Sn}, computed with IRREPs of MSG $Cmc^\prime m^\prime$.}
\end{table}

\begin{table}[htbp]
\begin{tabular}{c|c|c|c|c|c}
\hline\hline
MSG operations & $\{E||E\}$ & $\{E||\mathcal{P}\}$ & $\{2_{010}||M_{010}\}$ & $\{3^{-}_{001}||3^+_{001}\}$ & $\{E||2_{001}|\frac{z}{2}\}$ \\\hline
$^S \bar{\Gamma}_1\rule{0pt}{2.6ex}$            & 1 & -1 & -i & -1 & -1             \\\hline
$^S \bar{\Gamma}_2\rule{0pt}{2.6ex}$            & 1 & -1 &  i & -1 & -1           \\\hline
$^S \bar{\Gamma}_3\rule{0pt}{2.6ex}$            & 1 &  1 &  i & -1 & -1         \\\hline
$^S \bar{\Gamma}_4\rule{0pt}{2.6ex}$            & 1 &  1 & -i & -1 & -1           \\\hline
$^S \bar{\Gamma}_5\rule{0pt}{2.6ex}$            & 1 & -1 & -i & -1 &  1           \\\hline
$^S \bar{\Gamma}_6\rule{0pt}{2.6ex}$            & 1 & -1 &  i & -1 &  1           \\\hline
$^S \bar{\Gamma}_7\rule{0pt}{2.6ex}$            & 1 &  1 &  i & -1 &  1           \\\hline
$^S \bar{\Gamma}_8\rule{0pt}{2.6ex}$           & 1 &  1 & -i & -1 &  1           \\\hline
$^S \bar{\Gamma}_9\rule{0pt}{2.6ex}$            & 2 & -2 & 0 & 1 & -2           \\\hline
$^S \bar{\Gamma}_{10}\rule{0pt}{2.6ex}$           & 2 &  2 & 0 & 1 & -2           \\\hline
$^S \bar{\Gamma}_{11}\rule{0pt}{2.6ex}$           & 2 & -2 & 0 & 1 &  2           \\\hline
$^S \bar{\Gamma}_{12}\rule{0pt}{2.6ex}$           & 2 &  2 & 0 & 1 &  2  \\
\hline\hline
\end{tabular}
\caption{\label{tab:ssg character gamma point} Character table of $\Gamma$ IRREPs in SSG 194.1.6.1.P.}
\end{table}


\section{Comparison of notations in two related works}\label{sec:app:comparsion_3works}

In this section, we give a brief comparison between the notations in our results, Xiao \textit{et al}\cite{PhysRevX.14.031037} and Ren \textit{et al}\cite{PhysRevX.14.031038}.

Xiao \textit{et al} construct SSGs by enumerating all inequivalent linear $\text{O}(N), N=1,2,3$ representations of space groups through induced representations from high-symmetry and generic $\bm{k}$ point in the Brillouin zone (BZ).
They obtain 1421, 9542, and 56512 SSGs for collinear ($N = 1$), coplanar
($N = 2$), and non-coplanar ($N = 3$) magnetic orders.
They classify the $\text{O}(N)$ representations into sixteen types and label an SSG with $\alpha\mathcal{I}.\mathcal{J}.\mathcal{K}$. $\alpha=L,P,N$ denotes collinear, coplanar, and non-coplanar orders, $\mathcal{I}$ from 1 to 230 denotes parent SGs, $\mathcal{J}$ denotes the representation type, and the last $\mathcal{K}$ denotes different $\text{O}(N)$ representation classes.
Their classification is complete and includes all possible supercells, as the SSGs constructed from non-high-symmetry points correspond to an infinite number of supercells, and could cover incommensurate magnetic orders.

Our method, however, starts from the real-space supercells and exhausts all invariant subgroups of an SG.
The operations in the invariant subgroup are pure lattice operations in the constructed SSG, which correspond to the SG operations with trivial representation in Xiao \textit{et al}.
The two approaches are equivalent, and the labeling can be converted.
The SSG constructed from high-symmetry points in Xiao \textit{et al} has a one-to-one mapping with our SSG, while the SSG from high-symmetry lines or generic points has a one-to-many mapping, as the $\bm{q}$ vector maps to an infinite number of supercells.

For example, consider the coplanar SSGs from SG 14 in Xiao \textit{et al}.
They construct 25 SSGs, of which 10 are derived from the representations at $\Gamma$, which corresponds to $I_k=1$ SSGs in our notation, i.e., without enlarging the unit cell. 12 SSGs originate from $Y=(0.5,0,0)$, $D=(0,0.5,0.5)$, $A=(0.5,0,0.5)$, $C=(0.5,0.5,0)$ or $Z=(0,0.5,0)$, corresponding to $I_k=2$ in our notation.
The remaining three SSGs are induced from three high-symmetry lines, i.e., $LD=(0,v,0), W=(0.5,v,0), F=(u,0,w)$.
Taking $I_k=3$ as an example. The $LD$ point yields $I_k=3$ when $v=\frac{1}{3}$ or $v=\frac{2}{3}$, corresponding to the SSG with a threefold supercell along the $y$-axis. This maps to two inequivalent SSGs $14.3.2.2.P$ and $14.3.4.2.P$ in our notation. This one-to-two mapping originates from an extra equivalence relation used in Xiao \textit{et al}, i.e., continuously varying the coordinates for non-high-symmetry points does not change the SSG. These two SSGs $14.3.2.2.P$ and $14.3.4.2.P$ have different pure-lattice groups and are considered inequivalent in our method.
The $F$ point can also lead to SSGs with a threefold extension along the $z$ direction when taking $u=0, w=\pm \frac{1}{3}$, corresponding to $14.3.2.1.P$ and $14.3.4.1.P$ in our notation. $W$ can only induce SSGs with even $I_k$. Thus we conclude the mapping in this SG 14 is fully established.

Ren \textit{et al} adopt an enumeration method similar to ours.
They also use a real-space-based method by considering supergroups of 230 SGs.
They obtain 1421 collinear, 15050 coplanar, and 75136 inequivalent non-coplanar SSGs by restricting supercell multiplicity $I_k\le 8$.
Their method differs from ours in that they consider group-supergroup pairs, while we consider group-subgroup pairs. The two methods are equivalent and have a one-to-one mapping for the SSGs.
In Ren \textit{et al}, they categorize SSGs into three types: $t$-type consists of normal subgroups obtained solely from the point group part (corresponding to $I_k=1$), $k$-type consists of normal subgroups derived only from the translation group (corresponding to $I_t=1$), and $g$-type corresponds to the all the remainings. Their labeling also utilizes the parent SG, subgroups, and supercells. This corresponds to our labeling of $\mathcal{G}$, $H$, and $I_k$. Hence, a one-to-one mapping for converting two notations can be easily derived.

\section{SSG online Database}
In this work, we develop an online database\cite{ssg_website} showing all SSGs enumerated in this work with detailed information. We give a brief introduction of the usage of the database in this section.

On the main page of our website, there is a sidebar and a search box. One can enter information about the SSG to narrow down the search. The information one can input includes the types of SSG (collinear, coplanar, or general spacial SSG), the SG number, $I_k$, $I_t$, and the quotient group label of SSG. One can also input the $N_{\mathcal{G}}, I_k, I_t$ in the search box to search for the required SSG. After clicking the search button, all SSGs that meet the conditions will be displayed with their basic information, including their quotient group label, spin part group, and the SG number of the pure lattice symmetry. Each SSG has a `Details' button. After clicking the button, the website will navigate to the next page showing all the detailed information of the chosen SSG, including the basic information, non-trivial SSG operations, and pure-lattice operations in $H$.

\section{Identifying SSG for magnetic materials}\label{SM_identify_SSG}

In this section, we describe the algorithm for identifying SSG for magnetic materials.
We apply the algorithm to 1,923 commensurate magnetic materials in \textit{Bilbao Crystallographic server}\cite{aroyo2006bilbao1,aroyo2006bilbao2,aroyo2011crystallography,gallego2016magndataI,gallego2016magndataII}, and find the corresponding SSGs for 1,626 materials. The 297 failed materials are mainly due to partial occupations or incomplete information in the structural (mcif) file. 

In the following, we first describe the algorithm and then summarize the results of materials.

\subsection{Algorithm}\label{sec:identify_SSG_algorithm}

For a given magnetic material, we first extract the structural information including the real-space unit cell basis, atomic positions, atomic number, and the magnetic moment of each atom.
The algorithm to identify SSG is divided into two parts: (i) finding the SSG operations of the magnetic structure; (ii) identifying the corresponding SSG number in our database. We then describe each step.

To find all SSG operations of the material, the main idea of our algorithm is to extract all possible real-space operations and spin rotations separately, and then loop over all possible SSG operations given by the combinations of them. In this way, the SSG operations that leave the magnetic material invariant are kept and form the SSG of the material. To be specific,
\begin{enumerate}
    \item We first identify all possible real-space operations by using the real-space atomic positions and spices but ignoring the magnetic structure, which can be readily performed using the \textit{spglib}\cite{togo2018spglib} package.
    \item All possible spin rotations are obtained by identifying the point group (PG) formed by the magnetic moments in spin space. As each magnetic moment corresponds to a point in the spin space, the collection of all magnetic moments forms an effective ``molecular'' in spin space and the corresponding molecular PG can be readily identified using packages including \textit{pymatgen}\cite{ong2013python}. We mention a few technical details in this step:
    \begin{itemize}
        \item In order to fix the origin point in spin space to avoid redundant operations, we add an extra point at the origin when identifying the PG.
        \item If the magnetic order is collinear or coplanar, we identify it at the beginning, and the corresponding spin-only group is thus know and does not need to be identified in the molecular PG. For collinear orders, the spin-only-free group could only contain the identity and a mirror operation and can be identified easily. For coplanar orders, we manually shift the plane formed by magnetic moment to transform into non-coplanar orders (as we have placed an extra point at the origin), thus exclude the spin-only group formed by the mirror operation.
    \end{itemize}
    \item With all possible real-space and spin operations, we then consider all possible combinations of them and check if they keep the material invariant. In this way, we are able to identify all SSG operations of the material.

\end{enumerate}

With all SSG operations obtained, we then extract some key information:
\begin{itemize}
    \item The SG number $N_{\mathcal{G}}$ of $\mathcal{G}$ formed by all real-space operations is obtained by ignoring the spin rotations using \textit{spglib}. In this step, \textit{spglib} will also give the coordinate transformation $(W,\tau)$ ($W$ is a rotation of coordinate axes and $\tau$ is a origin shift) that transform the SG operations into standard operations.
    \item The SG number $N_{H}$ of $H$ formed by pure-lattice operations is obtained using \textit{spglib}.
    \item The sub-PG index $I_t$ is obtained by definition $I_t=|P_\mathcal{G}/P_{H}|$.
    \item The supercell index $I_k$ is obtained using $|Q|=I_k\times I_t$ as $|Q|=|\mathcal{G}/H|$.
    \item The PG $P_S$ formed by all spin rotations, which is the equivalent PG of the 3D real representation.
\end{itemize}

With all SSG operations of the magnetic material and the key information $(N_{\mathcal{G}}, N_H, I_k, I_t, P_S)$ extracted from these operations, we then identify the corresponding SSG number in our database. To be specific:
\begin{enumerate}
    \item First, use the key information to identify all SSG candidates in the database, which could be more than one.
    \item Second, use $(W, \tau)$ to transform the SSG operations of the SSG candidates into the (real-space) coordinate system of the material, the compare if they match operations of the material. We mention some technical details in this step:
    \begin{itemize}
        \item The operations in pure-lattice group $H$ should match exactly. Note that three generators of the translation group $T_H$ given by the supercell matrix should also match.
        \item When match spin rotations in $Q$, we require the characters (i.e., trace of the rotation) should match, but do not require spin rotations to match exactly. This is because the coordinate system of the spin space is free to choose which is equivalent to a similarity transformation of the 3D real representation.
        \item The coordinate transformation $(W,\tau)$ is not unique, and the transformation $(W_0,\tau_0)$ given by \textit{spglib} may not be correct as we only use $\mathcal{G}$ when computing the transformation, and the operations in $H$ may not be maintained under the transformation. To remedy this, we exhaust all possible coordinate transformations by combining $(W_0,\tau_0)$ with all $(W,\tau)$ in the automorphism group of $\mathcal{G}$, as two difference coordinate transformations can only differ by an automorphism of $\mathcal{G}$. The number of shift $\tau$ in automorphisms $(W,\tau)$ can differ by a lattice shift and is thus infinite, and we only consider inequivalent ones. For triclinic and monoclinic lattices, i.e., from SG 1 to 15, the number of $W$ is infinite, and we set a truncation s.t. $|W_{ij}|\le 2$ which turns out to be adequate.
        \item In the previous step, an extra shift $\tau^\prime$ is sometimes needed. As there are translation operations in $Q$ given by the translational quotient group which are accompanied by non-trivial spin rotations, \textit{spglib} may mistake them with pure translations in $H$ and results in wrong $(W_0,\tau_0)$. This mismatch cannot be fixed by the automorphism group. In practice, we use a truncated set of shift candidates $\tau^\prime=\frac{1}{4}(n_1, n_2, n_3), n_i\in\{0,1,2,3\}$ which turns out to be adequate to find the correct coordinate transformation. In the following subsection, we will give an concrete example.
    \end{itemize}
\end{enumerate}

\paragraph{Example with extra origin shift $\tau^\prime$}
We give an example of material that require extra origin shift $\tau^\prime$ when identifying SSG in the database.

For material \ch{CrReO4} of SG 12 $C2/m$ and MSG 12.8.73, the SSG operations identified for the material are
\begin{equation}
    H: E, \{\mathcal{P}|\frac{1}{2}\frac{1}{2}\frac{1}{2}\},\{E|\frac{1}{2}\frac{1}{2}\frac{1}{2}\},\{\mathcal{P}|00\frac{1}{2}\},
\end{equation}
and
\begin{equation}
    Q: E, \{C_{2z}||E|00\frac{1}{2}\}, \{M_y||C_{2y}|00\frac{1}{2}\},\{M_x||C_{2y}|00\frac{1}{2}\}
\end{equation}

The identified SSG is $12.2.2.9.P$, which has pure lattice group
\begin{equation}
    H: E, \{\mathcal{P}|0\},\{E|\frac{1}{2}\frac{1}{2}0\},\{\mathcal{P}|\frac{1}{2}\frac{1}{2}0\},
\end{equation}
where $(\frac{1}{2},\frac{1}{2},0)$ is the translation given by the primitive cell basis.
The quotient group has elements:
\begin{equation}
    Q: E, \{C_{2z}||E|001\}, \{M_y||C_{2y}|0\},\{M_x||C_{2y}|001\}
\end{equation}

The coordinate transformation given by \textit{spglib} is
\begin{equation}
    W_0=\left(
\begin{matrix}
	1 & 0 & 0 \\
	0 & 1 & 0 \\
	0 & 0 & 2
\end{matrix}
\right),\quad \tau_0=\bm{0}
\end{equation}
When \textit{spglib} computes $(W_0, \tau_0)$, it only takes in the real-space operations, and thus cannot distinguish whether an operations has or has not a non-trivial spin-part. Thus the $(W_0, \tau_0)$ computed for this material cannot be used directly, as it fail to transform the operations into standard operations in the SSG database.
In the current example, there is a translation operation with non-trivial spin rotation, i.e., $\{C_{2z}||E|\frac{1}{2}\frac{1}{2}0\}$, which is mistakenly identified as the lattice translation by \textit{spglib} and lead to $\tau_0=\bm{0}$.

To fix this, we add an extra shift $\tau^\prime=(0,0,0.5)$ to $\tau_0$ and an automorphism
\begin{equation}
    W_{iso}=\left(
\begin{matrix}
	-1 & 0 & -1 \\
	0 & 1 & 0 \\
	2 & 0 & 1
\end{matrix}
\right), \tau_{iso}=\bm{0}.
\end{equation}
The final transformation is
\begin{equation}
    W=\left(
\begin{matrix}
	-1 & 0 & -2 \\
	0 & 1 & 0 \\
	2 & 0 & 2
\end{matrix}
\right),\quad \tau=(0,0,0.5).
\end{equation}

\subsection{Results of identified SSGs for commensurate magnetic materials}\label{sec:result_magndata}

Here we summarize the results of the algorithm for 1,626 commensurate materials without fractional occupancy, including their non-magnetic space groups (SGs), magnetic space groups (MSGs), and spin space groups (SSGs), as shown in Table.\ref{magdata_ssg}.
We also remark on the usage of SSG on magnetic materials: (i) SSGs can be used as a fine-grained tool to describe symmetry and refine magnetic structures compared with MSGs. (ii) SSGs describe the electronic structures of magnetic materials when SOC is negligible or weak compared to the spin splitting induced by the effective Zeeman term.

Among these 1,626 materials, there are 1,142 materials (70.2\%) whose SG number in the identified SSG is the same as the corresponding non-magnetic SG, which means all SG operations are maintained by assigning proper spin rotations. Moreover, the identified SSG is always larger than or equal to the MSG (i.e., the first number in the OG number) for each material, showing the advantage of the SSGs that contain MSGs as subgroups.

We remark that the SGs and MSGs of materials are identified using \textit{spglib} with the tolerance set to 0.01 \AA (i.e., the \textit{symprec} parameter in \textit{spglib}, which is the tolerance of two positions in Cartesian coordinate to be considered as the same, in the unit of \AA). There is a trade-off in the setting of tolerance, which in principle should be consistent with the accuracy in the structural files of magnetic materials. A too-small tolerance may fail to identify all symmetry operations, while a too-large tolerance may lead to redundant symmetries. Here we set a universal tolerance of 0.01 for simplicity for ease of high-throughput computations of SSGs, and further fine-tuning of the tolerance is possible.


As mentioned in Sec.\ref{Sec:results} in the main text, we use the following serial number to label SSGs: $N_{\text{SG}}.I_k.I_t.N_{\text{3Drep}}$,
where $N_{\text{SG}}$ is the SG number, $I_k=|T_\mathcal{G}|/|T_H|$ is the supercell k-index, $I_t=|P_\mathcal{G}|/|P_H|$ is the sub-point group t-index, and $N_{\text{3Drep}}$ denotes the N-th SSG of given $I_k$ and $I_t$. For collinear and coplanar SSGs, we add extra $.L$ and $.P$ to the SSG label, i.e., $ N_{\text{SG}}.I_k.I_t.N_{\text{1Drep}}.L$ for collinear SSGs and $ N_{\text{SG}}.I_k.I_t.N_{\text{2Drep}}.P$ for coplanar SSGs.

\begin{longtable}{c|c|c|c|c|c}
\caption{\label{magdata_ssg} Results of the space groups (SGs), magnetic space groups (MSGs) (with BNS number and OG number both listed), and spin-space groups (SSGs). We also give the label of materials in \textit{MAGNDATA}\cite{gallego2016magndataI,gallego2016magndataII} from \textit{Bilbao Crystallographic Server}.
}\\
\hline\hline
Material & MAGNDATA label & SG & MSG (BNS) & MSG (OG) & SSG \\\hline
\endfirsthead
\hline\hline
Material & MAGNDATA label & SG & MSG (BNS) & MSG (OG) & SSG \\\hline
\endhead
\hline
\endfoot
\ch{LaMnO3}&\href{http://webbdcrista1.ehu.es/magndata/index.php?this_label=0.1}{0.1} &$Pnma (62)$&62.448&62.8.509&\href{https://cmpdc.iphy.ac.cn/ssg/ssgs/62.1.2.5.L}{62.1.2.5.L} \\\hline 
\ch{Cd2Os2O7}&\href{http://webbdcrista1.ehu.es/magndata/index.php?this_label=0.2}{0.2} &$Fd-3m (227)$&227.131&227.4.1631&\href{https://cmpdc.iphy.ac.cn/ssg/ssgs/227.1.24.2}{227.1.24.2} \\\hline 
\ch{Ca3LiOsO6}&\href{http://webbdcrista1.ehu.es/magndata/index.php?this_label=0.3}{0.3} &$R-3c (167)$&15.89&15.5.96&\href{https://cmpdc.iphy.ac.cn/ssg/ssgs/167.1.2.3.L}{167.1.2.3.L} \\\hline 
\ch{NiCr2O4}&\href{http://webbdcrista1.ehu.es/magndata/index.php?this_label=0.4}{0.4} &$I4_1/amd (141)$&70.530&70.4.619&\href{https://cmpdc.iphy.ac.cn/ssg/ssgs/141.1.1.1.L}{141.1.1.1.L} \\\hline 
\ch{Cr2S3}&\href{http://webbdcrista1.ehu.es/magndata/index.php?this_label=0.5}{0.5} &$R-3 (148)$&2.4&2.1.4&\href{https://cmpdc.iphy.ac.cn/ssg/ssgs/148.1.1.1.L}{148.1.1.1.L} \\\hline 
\ch{YMnO3}&\href{http://webbdcrista1.ehu.es/magndata/index.php?this_label=0.6}{0.6} &$P6_3cm (185)$&185.197&185.1.1429&\href{https://cmpdc.iphy.ac.cn/ssg/ssgs/185.1.12.1.P}{185.1.12.1.P} \\\hline 
\ch{ScMnO3}&\href{http://webbdcrista1.ehu.es/magndata/index.php?this_label=0.7}{0.7} &$P6_3cm (185)$&185.201&185.5.1433&\href{https://cmpdc.iphy.ac.cn/ssg/ssgs/185.1.12.1.P}{185.1.12.1.P} \\\hline 
\ch{ScMnO3}&\href{http://webbdcrista1.ehu.es/magndata/index.php?this_label=0.8}{0.8} &$P6_3cm (185)$&173.129&173.1.1360&\href{https://cmpdc.iphy.ac.cn/ssg/ssgs/185.1.12.1.P}{185.1.12.1.P} \\
\hline 
\hline\hline
\end{longtable}

\normalem

\end{document}